\documentclass[%
preprint,
superscriptaddress,
groupedaddress,
unsortedaddress,
 amsmath,amssymb,
 aps,
showkeys
]{revtex4-2}

\usepackage{graphicx}
\usepackage{dcolumn}
\usepackage{bm}
\usepackage{hyperref}
\usepackage{subcaption}
\usepackage{float}
\usepackage{booktabs} 
\usepackage{placeins}
\usepackage{todonotes}
\usepackage{comment}
\usepackage{adjustbox} 
\usepackage{romannum}
\usepackage{natbib}

\begin{document}


\title{Permuted Charged Lepton Correction in the Framework of Dirac Seesaw}

\author{Sagar Tirtha Goswami}
\email{sagartirtha@gauhati.ac.in}
 \affiliation{Physics Department, Gauhati University, India.}
\author{Subhankar Roy}%
 \email{subhankar@gauhati.ac.in}
\affiliation{Physics Department, Gauhati University, India}%



\date{\today}

\begin{abstract}
A Dirac neutrino mass model is proposed, based on an extended group structure of \(SU(2)_L \otimes U(1)_Y \otimes A_4 \otimes Z_3 \otimes Z_{10}\) with the Dirac Type I seesaw mechanism. This work explores the impact of parametrization and permutation in the charged lepton diagonalizing matrix, 
driven by free parameters in the charged lepton sector, on the predictions of observable parameters.  Some interesting consequences on the neutrino mass hierarchies, mixing angles and the Dirac CP phase are observed.
The framework also finds application in the study of charged lepton flavour violation and dark matter.



\end{abstract}

\maketitle


\section{\label{Section 1}{Introduction}}
The neutrino ($\nu$) is a peculiar particle in the standard model (SM) of particle physics. The SM says that it is massless, because there is no evidence of a right handed $\nu$ ($\nu_R$) in nature. Hence, we can not form a mass term ($ \bar{\nu}_L \nu_R+h.c.$) in the Lagrangian. But the idea of $\nu$ oscillation, first proposed by Bruno Pontecorvo in 1957\,\cite{Pontecorvo:1957cp}, and later confirmed by experiments\,\cite{Davis:1968cp,SNO:2001kpb,Super-Kamiokande:2001ljr,SNO:2002tuh,Bionta:1987qt,Super-Kamiokande:1998kpq,KamLAND:2002uet,KamLAND:2002uet,DayaBay:2012fng,K2K:2002icj,T2K:2011ypd,T2K:2013ppw,Kamiokande-II:1989hkh} suggest that $\nu$ has mass, however tiny it may be. To account for this mass, we might assume the existence of $\nu_R$ and form the mass term. But then we run into a new problem. To explain the tiny mass of $\nu$, we have no other way but to accept an extremely small Yukawa coupling constant ($\lesssim 10^{-11}$). This does not answer why the mass is so small, but rather it enhances the flavour hierarchy problem\,\cite{Xing:2020ijf}. The other alternative is to consider $\nu$ to be a Majorana particle, thanks to its electrical neutrality. In this case, we do not require a right handed counterpart and can form a mass term with simply the left handed part only ($\bar{\nu}_L \nu_L^c+h.c.$). As to why the mass is so small, we have several mechanisms that explain it. Popular among these are seesaw mechanism\,\cite{Minkowski:1977sc,Ramond:1979py,Gell-Mann:1979vob,Yanagida:1979as,Mohapatra:1979ia}, radiative mass mechanism\,\cite{Weinberg:1979sa,Cai:2017jrq} etc. But the catch is that a Majorana mass term violates lepton number and lepton number violation (LNV) has not been observed in nature\,\cite{Barabash:2023dwc}. This fact alone casts doubt on the Majorana paradigm for $\nu$.

So, given this uncertainty, it is reasonable to stick to the Dirac nature of $\nu$'s. If we put aside the issue of non availability of $\nu_R$, the other issue, namely the smallness of $\nu$ mass, can be explained by similar mechanisms that we use in Majorana scenario. For example, seesaw mechanism can be used to explain this smallness\,\cite{Roy:1983be,CentellesChulia:2016rms,CentellesChulia:2017koy,Bonilla:2017ekt,Borah:2018nvu}. In this work, we use Dirac Type \Romannum{1} seesaw mechanism considering $\nu$'s to be Dirac particles. Several works have explored this approach\,\cite{CentellesChulia:2016rms,CentellesChulia:2017koy,Chen:2022bjb,Borah:2017dmk,Mahapatra:2023oyh,Singh:2024imk}. Our work here emphasizes the charged lepton (CL) sector in shaping predictions for observable parameters related to $\nu$.  The impact of CL corrections on these parameters is well documented\,\cite{Dev:2013esa,Antusch:2005kw,Roy:2012ib,Dev:2011bd,King:2017guk}. In this work, however, we introduce a specific \textit{Permuted Charged Lepton Correction} (PCLC), which captures the combined effect of parametrization and permutation of the charged lepton (CL) diagonalizing matrix on the predictions. This is an interesting aspect not often found in literature. PCLC arises due to the presence of free parameters in the CL sector, offering ample flexibility that can be leveraged by linear algebra. We highlight how this flexibility leads to intriguing consequences for observable quantities.


The plan of the work is as follows: in section \,\ref{Section 2}, we outline the framework of our work,  including the effect of linear algebra on the diagonalizing matrices. We then present the numerical analysis in section \,\ref{Section 3}, while section \,\ref{Section 4} explores the application of our framework to charged lepton flavour violation (CLFV) and dark matter (DM). Finally, the paper is concluded with a summary in section \,\ref{Section 5}. 

\section{\label{Section 2}{Framework}}
\subsection{Basic Structure}
The present work is a $\nu$ mass model, considering $\nu$'s as Dirac particles. The underlying mass generation mechanism is Dirac Type I seesaw. The model is based on an extensive group structure of $SU(2)_L \otimes U(1)_Y \otimes A_4 \otimes Z_3 \otimes Z_{10}$, which breaks down to $U(1)_{EM} \otimes Z_3$ after spontaneous symmetry breaking.  Apart from the SM Higgs, $H$, we introduce two other scalar fields, $\phi$ and $\chi$ in the model. The Lagrangian of the model is presented as follows,

\begin{eqnarray}
\label{1}
    -\mathcal{L}&=&\tilde{y}_1\,(\bar{l}_{L}\,H)_1\, l_{1,R}+ \tilde{y}_2\,(\bar{l}_{L}\,H)_{1^{'}}\,l_{2,R} 
    + \tilde{y}_3\,(\bar{l}_{L}\,H)_{1^{''}}\, l_{3,R}+ \tilde{y}_4\,(\bar{l}_{L}\,N_{R})_{3_s}\, \phi 
   + \tilde{y}_5\,(\bar{l}_{L}\,N_{R})_{3_a}\, \phi \nonumber \\
  && + \tilde{y}_6\,(\bar{N}_{L}\,\nu_{R})_{3_s}\, \chi 
   +\tilde{y}_7\,(\bar{N}_{L}\,\nu_{R})_{3_a}\, \chi+ m_N\,(\bar{N}_{L}\,N_{R})_1,
\end{eqnarray}
where the Yukawa coupling constants are complex numbers, as denoted by the sign (\textasciitilde).
The charge assignments of the fields under the said group structure is tabulated in Table \ref{table1a}. The multiplication rules for the groups $A_4$, $Z_3$ and, $Z_{10}$ are shown in Appendixes \ref{A4} and \ref{ZN} respectively.

\begin{table}[h]
\begin{center}
\footnotesize
\begin{tabular*}{\textwidth}{@{\extracolsep{\fill}}c c c c c c c}
\hline
Fields \textbackslash Groups &$U(1)_Y$&$SU(2)_L$&$A_4$&$Z_3\,(q_3)$ & $Z_{10}\,(q_{10})$\\
\hline
$\bar{l}_L$ & 1 & 2 & 3 & 2 & 3  \\
\hline
$l_{i,R}$ & -2 & 1 & $1, 1^{''},1^{'}$ & 1 & 3 \\
\hline
$H$ & 1 &2 & 3 & 0 & 4  \\
\hline
$\bar{N}_L$ & 0 & 1 & 3 & 2 & 4\\
\hline
$N_R$ & 0 &1 & 3 & 1 & 6\\
\hline
$\nu_R$ & 0 &1 & 3 & 1 & 2\\
\hline
$\phi$ & -1 & 2 & 3 & 0 &1 \\
\hline
$\chi$ & 0 & 1 & 3 & 0 &4 \\
\hline
\end{tabular*}
\caption{\footnotesize \label{table1a} The charge assignments of various fields under the group structure of the model.}.
\end{center}
\end{table}

The cyclic groups $Z_3$ and $Z_{10}$ have special roles to play.  The group $Z_3$ prevents a Majarona mass term for $\nu_R$, i.e., $\bar{\nu_R}^c\,\nu_R$ and it remains unbroken at all energy scales.  The $Z_{10}$ group\,\cite{Dey:2024ctx} 
is instrumental in preventing tree level Dirac mass terms like, $\bar{l}_LH\nu_R$,\,$\bar{l}_L \phi\nu_R$,\,$\bar{l}_L \chi \nu_R$ etc.  

With the applications of cyclic groups explained,  we now consider the vacuum expectation values (VEV) of the scalar field, $H$ as $(1,1,1)^T v_H$,  which gives the charged lepton mass matrix, $M_L$ as follows,
\begin{equation}
\label{equation:A}
M_{L} = \begin{pmatrix}
\tilde{m}_{L1} & \tilde{m}_{L2} & \tilde{m}_{L3} \\
 \tilde{m}_{L1} & \omega\, \tilde{m}_{L2} & \omega^2\, \tilde{m}_{L3}\\
 \tilde{m}_{L1} & \omega^2\, \tilde{m}_{L2}&  \omega\, \tilde{m}_{L3}
\end{pmatrix}, 
\end{equation}
where we have  $\omega=(-1+i \sqrt{3})/2$ and $\tilde{m}_{Li}=\tilde{y}_i\,v_H, i=1,2,3$. 
It is seen that $M_L$ is non-diagonal and has three independent complex parameters, viz., $\tilde{m}_{L1},\,\tilde{m}_{L2}$ and $\tilde{m}_{L3}$. The $\nu$ mass matrix, \( M_\nu \), on the other hand, from the Dirac Type \Romannum{1} seesaw mechanism, considering the VEV alignments of the scalar fields \(\phi\) and \(\chi\) as \((0, k_1, k_2)^T v_\phi\) and \((0, f_1, f_2)^T v_\chi\), respectively, with \(k_1, k_2, f_1, f_2, v_\phi, v_\chi \in \mathbb{R}\), is derived as

\begin{equation}
\label{aa}
M_\nu=B\,M^{-1}\,D,
\end{equation}
where we have,
\begin{eqnarray}
\label{ab}
B &=& \begin{pmatrix}
0& (\tilde{y}_4+\tilde{y}_5)k_2 & (\tilde{y}_4-\tilde{y}_5)k_1 \\
 (\tilde{y}_4-\tilde{y}_5)k_2 & 0 & 0\\
 (\tilde{y}_4+\tilde{y}_5)k_1 & 0&  0
\end{pmatrix}v_\phi,\nonumber \\
\label{ac}
M &=& \begin{pmatrix}
m_N & 0 & 0 \\
 0 & m_N & 0\\
 0 & 0&  m_N
\end{pmatrix}, \nonumber \\
\label{ac}
D&=& \begin{pmatrix}
0& (\tilde{y}_6+\tilde{y}_7)f_2 & (\tilde{y}_6-\tilde{y}_7)f_1 \\
(\tilde{y}_6-\tilde{y}_7)f_2  & 0 &0\\
(\tilde{y}_6+\tilde{y}_7)f_1 & 0&  0
\end{pmatrix}v_\chi . 
\end{eqnarray}

Clearly, $M_\nu$ is defined in the symmetry basis,  and it looks like the following,

\begin{equation}
\label{equation:E}
M_{\nu } = \begin{pmatrix}
x_1 e^{i x_2} & 0 & 0 \\
 0 & s_1 e^{i s_2} & z_1 e^{i z_2}\\
 0 & w_1 e^{i w_2} &  \frac{z_1 w_1}{s_1} e^{[i(z_2+w_2-s_2)]}
\end{pmatrix},
\end{equation}
with $x_1=(v_\chi\,v_\phi)|(\tilde{y}_4-\tilde{y}_5)(\tilde{y}_6+\tilde{y}_7)f_1k_1+(\tilde{y}_4+\tilde{y}_5)(\tilde{y}_6-\tilde{y}_7)f_2k_2|/m_N$,\, $s_1=(f_2k_2v_\chi v_\phi)|(\tilde{y}_4-\tilde{y}_5)(\tilde{y}_6+\tilde{y}_7)|/m_N$,\, $z_1=(f_1 k_2 v_\chi v_\phi)|(\tilde{y}_4-\tilde{y}_5)(\tilde{y}_6-\tilde{y}_7)|/m_N$,\, $w_1=(f_2 k_1 v_\chi v_\phi)|(\tilde{y}_4+\tilde{y}_5)(\tilde{y}_6+\tilde{y}_7)|/m_N$, \,$x_2=(v_\chi\,v_\phi)$ arg$[(\tilde{y}_4-\tilde{y}_5)(\tilde{y}_6+\tilde{y}_7)f_1k_1+(\tilde{y}_4+\tilde{y}_5)(\tilde{y}_6-\tilde{y}_7)f_2k_2]/m_N$,\, $s_2=(f_2k_2v_\chi v_\phi)$ arg$[\tilde{y}_4-\tilde{y}_5)(\tilde{y}_6+\tilde{y}_7)]/m_N$, \, $z_2=(f_1 k_2 v_\chi v_\phi)$ arg$[(\tilde{y}_4-\tilde{y}_5)(\tilde{y}_6-\tilde{y}_7)]/m_N$,\, and $w_2=(f_2 k_1 v_\chi v_\phi)$ arg$[(\tilde{y}_4+\tilde{y}_5)(\tilde{y}_6+\tilde{y}_7)]/m_N$.
We see that $M_\nu$ contains eight real free parameters. With $M_\nu$ defined,  we next introduce the hermitian matrix, $H_\nu=M_\nu\,M_\nu^\dagger$, as shown below,

\begin{equation}
\label{equation:F}
H_\nu = \begin{pmatrix}
a  & 0 & 0 \\
 0 & b & \sqrt{bc}\, e^{-i \rho} \\
 0 & \sqrt{bc} \,e^{i \rho}  &  c
\end{pmatrix},
\end{equation}

where $a=x_1^2$,\, $b=s_1^2+z_1^2$,\, $c=w_1^2+(w_1^2 z_1^2)/s_1^2$ and $\rho=w_2-s_2$, with $a,\,b,\,c>0$. 

It is important to note that the VEV alignments of both the scalar fields $\phi$ and $\chi$ are similar in our framework and hence, the structure of $H_\nu$ is also similar to $M_\nu$.  Had we taken different VEV alignments, we would have obtained an $H_\nu$ structurally different from $M_\nu$.  For example,  for VEV alignments of $\phi$ and $\chi$ as $(k_1,0,k_2)^Tv_\phi$ and $(0,f_1,f_2)^T v_\chi$ respectively,  the structures of the obtained $M_\nu$ and $H_\nu$ are as follows,
 \begin{eqnarray}
 M_\nu^ \prime=\begin{pmatrix}
\checkmark & 0 & 0 \\
 \checkmark & \checkmark & \checkmark\\
 \checkmark & 0 & 0
\end{pmatrix}, \quad
H_\nu^\prime=\begin{pmatrix}
\checkmark & \checkmark & \checkmark \\
 \checkmark & \checkmark & \checkmark \\
 \checkmark & \checkmark &  \checkmark
\end{pmatrix},
 \end{eqnarray}
 
 where the $\checkmark$ indicates a non zero entry.  
 
 Such an $H_\nu$ would, of course, lead to different predictions, which will be explored in our future works. With $H_\nu$ defined, the neutrino masses can now be expressed in terms of $a$, $b$, and $c$. Having $M_L$ and $M_\nu$ at hand, we proceed to determine their respective diagonalizing matrices in the subsequent sections.


\subsection{\label{Section 2(b)}{Permuted Charged Lepton Correction }}
As discussed in the earlier section, the three complex free parameters that $M_L$ has, viz., $\tilde{m}_{L1}, \,\tilde{m}_{L2}$ and $\tilde{m}_{L3}$ (Eq.\,(\ref{equation:A})), play an important part in our work.  These free parameters provide the necessary flexibility for the CL diagonalizing matrices ($U_L$ and $U_R$) to take different forms, mainly through two ways, \textit{parametrization} and \textit{permutation}. In this section, we discuss them one by one, starting with parametrization.

Let us then diagonalize $M_L$ in the following way,
\begin{eqnarray}
\label{equation:B}
M_L^{diag}=U_L^\dagger\, M_L \,U_R,
\end{eqnarray}

with $U_L$ and $U_R$ as,
\begin{eqnarray}
\label{ba}
U_{L} &=&\frac{1}{\sqrt{3}} \begin{pmatrix}
1 &  \omega & \omega^2 \\
 1& \omega^2 & \omega\\
1& 1&  1
\end{pmatrix}, \quad
\label{bb}
U_R = \begin{pmatrix}
1 & 0& 0 \\
 0 & 1 & 0\\
 0 & 0&  1
\end{pmatrix}.
\end{eqnarray}
 
 The diagonal matrix is found to be,
 \begin{equation}
 \label{bc}
 M_L^{diag} = \begin{pmatrix}
\sqrt{3}\,\tilde{m}_{L1} & 0& 0 \\
 0 & \sqrt{3}\,\omega^2\,\tilde{m}_{L2} & 0\\
 0 & 0&  \sqrt{3}\,\omega\,\tilde{m}_{L3}
\end{pmatrix}.
 \end{equation}

 The above matrix (Eq.\,(\ref{bc})) can also be expressed, with $\tilde{m}_{Li}=m_{Li}\,\exp(i\,\kappa_i), i=1,2,3$, as,
\begin{equation}
 \label{bd}
 M_L^{diag} = \begin{pmatrix}
m_{L1}\,e^{i\,\kappa_1} & 0& 0 \\
 0 & m_{L2} \,e^{i\,\kappa_2}& 0\\
 0 & 0& m_{L3}\,e^{i\,\kappa_3}
\end{pmatrix},
 \end{equation}

where $m_{Li}$'s and $\kappa_i$'s are real free parameters.
To absorb the unphysical phases from $ M_L^{diag} $ (Eq.\,(\ref{bd})), we can redefine either $U_L$ or $U_R$ or both in several possible ways.  As an example, we can redefine them in the following way, 
\begin{eqnarray}
\label{be}
U_L &\longrightarrow&  \begin{pmatrix}
e^{i\,\kappa_1} & 0& 0 \\
 0 &  e^{i\,\kappa_2} & 0\\
 0 & 0&  e^{i\,\kappa_3}
\end{pmatrix} U_L, \quad
\label{bf}
U_R \longrightarrow  U_R.
\end{eqnarray}

In this parametrization, the phases remain with $U_L$ leaving $U_R$ independent of $\kappa_i$'s.  Similarly, we can also make $U_L$ completely independent of the phases using the following redefinition,
\begin{eqnarray}
\label{bg}
U_L &\longrightarrow& U_L,\quad
\label{bh}
 U_R \longrightarrow \begin{pmatrix}
e^{-i\,\kappa_1} & 0& 0 \\
 0 &  e^{-i\,\kappa_2}& 0\\
 0 & 0&  e^{-i\,\kappa_3}
\end{pmatrix} U_R.
\end{eqnarray}

Besides the parametrizations shown in Eqs.\,(\ref{bf}) and (\ref{bh}), many other parametrization schemes for the diagonalizing matrices can be adopted based on specific requirements. 

 Apart from the parametrization of the diagonalizing matrices, there is another way they can be redefined and that is through permutation of rows and columns. These transformations correspond to basis changes and effectively permute the diagonal elements, which is allowed as they are all free parameters.  For example, if we take the column permutation, $C_1 \longleftrightarrow C_2$ on both $U_L$ and $U_R$, the new set of $U_L$ and $U_R$ takes the following form,
 
 \begin{eqnarray}
 U_{L}^\prime &=&\frac{1}{\sqrt{3}} \begin{pmatrix}
  \omega &1 & \omega^2 \\
  \omega^2 & 1& \omega\\
1& 1&  1
\end{pmatrix}, \quad
U_R^\prime = \begin{pmatrix}
0 & 1& 0 \\
 1 & 0 & 0\\
 0 & 0&  1
\end{pmatrix},\nonumber
 \end{eqnarray}
 
with $M_L^{diag}$ as $\text{diag}\, \left[ m_{L2}\,\exp(i\, \kappa_2), \,m_{L1}\,\exp(i\,\kappa_1),  \nonumber \\
 \, m_{L3}\,\exp(i\,\kappa_3) \right]$.  In this new $M_L^{diag}$, we see that the masses of electron ($m_e$), muon ($m_\mu$) and tauon ($m_\tau$) are $m_{L2},\,m_{L1}$ and $m_{L3}$ respectively, a departure from $M_L^{diag}$ in Eq.\,(\ref{bd}), where $m_e,\,m_\mu$ and $m_\tau$ are identified with $m_{L1},\,m_{L2}$ and $m_{L3}$ respectively.  This is possible only because $m_{L i}$'s are free parameters.
 Similarly, if we do a row operation, $R_1 \longleftrightarrow R_2$ on $U_L$ and another, $R_2 \longleftrightarrow R_3$ on $U_R$, the new set of $U_L$ and $U_R$ is given below,
 
 \begin{eqnarray}
 U_{L}^{\prime \prime} &=&\frac{1}{\sqrt{3}} \begin{pmatrix}
1 &  \omega^2  & \omega \\
 1 & \omega & \omega^2\\
1& 1&  1
\end{pmatrix}, \quad
U_R^{\prime \prime} = \begin{pmatrix}
1 & 0& 0 \\
 0 & 0 & 1\\
 0 & 1&  0
\end{pmatrix}. \nonumber
 \end{eqnarray}
 
 They would give $M_L^{diag}$ as $\text{diag}\, \left[m_{L1}\,\exp(i\,\kappa_1), \, m_{L3}\,\exp(i\,\kappa_3),\, m_{L2}\,\exp(i\,\kappa_2) \right]$, yet another different positioning of the eigenvalues, with $m_{L1},\,m_{L3}$ and $m_{L2}$ denoting $m_e,\,m_\mu$ and $m_\tau$ respectively.  
 

The discussion of how we can redefine the diagonalizing matrices  is important because $U_L$ is involved in the Pontecorvo-Maki-Nakagawa-Sakata (PMNS) matrix\,\cite{Maki:1962mu,Pontecorvo:1957qd}, which is defined in the flavour basis. As PMNS matrix contains the information of various observable parameters, these parameters are expected to be affected by the form of $U_L$, whether in terms of the unobserved phases, $\kappa_i$'s or a particular permuted variant of $U_L$. In fact,  there are 18 distinct \( U_L \)'s that can be obtained through row and column transformations, as illustrated in Table \ref{table2a}. The effect of parametrization and permutation of $U_L$ on the predictions of observable parameters is discussed in detail in sections \ref{Section 2(c)} and \ref{Section 3} respectively.

\begin{table}
\begin{center}
\footnotesize
\begin{tabular*}{\textwidth}{@{\extracolsep{\fill}}c c c c c c}
\hline
 $\mathbf{U_{L1}}$ & $\mathbf{U_{L2}}$ & $\mathbf{U_{L3}}$ & $\mathbf{U_{L4}}$ & $\mathbf{U_{L5}}$ & $\mathbf{U_{L6}}$ \\
\hline
 $\frac{1}{\sqrt{3}}  \begin{pmatrix}
1 &  \omega & \omega^2 \\
 1& \omega^2 & \omega\\
1& 1&  1
\end{pmatrix}$ & $\frac{1}{\sqrt{3}} \begin{pmatrix}
\omega &  1 & \omega^2 \\
 \omega^2& 1 & \omega\\
1& 1&  1
\end{pmatrix}$ & $\frac{1}{\sqrt{3}} \begin{pmatrix}
\omega^2 &  \omega & 1 \\
 \omega & \omega^2 & 1\\
1& 1&  1
\end{pmatrix}$ & $\frac{1}{\sqrt{3}} \begin{pmatrix}
1 &  \omega^2 & \omega \\
 1& \omega & \omega^2\\
1& 1&  1
\end{pmatrix}$ & $\frac{1}{\sqrt{3}} \begin{pmatrix}
\omega &  \omega^2 & 1 \\
 \omega^2 & \omega & 1 \\
1& 1&  1
\end{pmatrix}$ & $\frac{1}{\sqrt{3}} \begin{pmatrix}
\omega^2 &  1 & \omega \\
 \omega& 1 & \omega^2 \\
1& 1&  1
\end{pmatrix}$ \\
\hline
\hline
  $\mathbf{U_{L7}}$ & $\mathbf{U_{L8}}$ & $\mathbf{U_{L9}}$ & $\mathbf{U_{L10}}$ & $\mathbf{U_{L11}}$ & $\mathbf{U_{L12}}$ \\
\hline
 $\frac{1}{\sqrt{3}} \begin{pmatrix}
1 &  1 & 1 \\
 1& \omega & \omega^2 \\
1& \omega^2&  \omega
\end{pmatrix}$ & $\frac{1}{\sqrt{3}} \begin{pmatrix}
1 &  1 & 1 \\
 \omega & 1 & \omega^2 \\
\omega^2 & 1 &  \omega
\end{pmatrix}$ & $\frac{1}{\sqrt{3}} \begin{pmatrix}
1 &  1 & 1 \\
 \omega^2 & \omega & 1 \\
\omega & \omega^2 &  1
\end{pmatrix}$ & $\frac{1}{\sqrt{3}} \begin{pmatrix}
1 &  1 & 1 \\
 1 & \omega^2 & \omega \\
1 & \omega &  \omega^2
\end{pmatrix}$ & $\frac{1}{\sqrt{3}} \begin{pmatrix}
1 &  1 & 1 \\
 \omega & \omega^2 & 1 \\
\omega^2 & \omega &  1
\end{pmatrix}$ & $\frac{1}{\sqrt{3}} \begin{pmatrix}
1 &  1 & 1 \\
 \omega^2 & 1 & \omega \\
\omega & 1 &  \omega^2
\end{pmatrix}$ \\
\hline
\hline
 $\mathbf{U_{L13}}$ & $\mathbf{U_{L14}}$ & $\mathbf{U_{L15}}$ & $\mathbf{U_{L16}}$ & $\mathbf{U_{L17}}$ & $\mathbf{U_{L18}}$ \\
\hline
 $\frac{1}{\sqrt{3}} \begin{pmatrix}
1 &  \omega & \omega^2 \\
 1 & 1 & 1 \\
1 & \omega^2 &  \omega
\end{pmatrix}$ & $\frac{1}{\sqrt{3}} \begin{pmatrix}
\omega^2 &  1 & \omega \\
 1 & 1 & 1 \\
1 & \omega^2 &  \omega
\end{pmatrix}$ & $\frac{1}{\sqrt{3}} \begin{pmatrix}
\omega^2 &  \omega & 1 \\
 1 & 1 & 1 \\
\omega & \omega^2 &  1
\end{pmatrix}$ & $\frac{1}{\sqrt{3}} \begin{pmatrix}
1 &  \omega^2 & \omega \\
 1 & 1 & 1 \\
1 & \omega &  \omega^2
\end{pmatrix}$ & $\frac{1}{\sqrt{3}} \begin{pmatrix}
\omega &  \omega^2 & 1 \\
 1 & 1 & 1 \\
\omega^2 & \omega &  1
\end{pmatrix}$ & $\frac{1}{\sqrt{3}} \begin{pmatrix}
\omega^2 &  1 & \omega \\
 1 & 1 & 1 \\
\omega & 1 &  \omega^2
\end{pmatrix}$ \\
\hline
\end{tabular*}
 \caption{\footnotesize \label{table2a} The 18 distinct permuted variants of $U_L$.}
\end{center}
\end{table}

\subsection{\label{Section 2(c)}{Permuted Variants of Neutrino Diagonalizing Matrix}}
In the preceding section, we have discussed PCLC extensively, i.e., how the presence of free parameters results in many different forms of $U_L$ through parametrization and permutation. The same holds true for the $\nu$ sector too. 

 $M_\nu$, like $M_L$, is diagonalized as,
\begin{equation}
\label{bi}
M_\nu^{diag}=U_\nu^\dagger\,M_\nu\,V_\nu,
\end{equation}

where $U_\nu$ and $V_\nu$ are $\nu$ diagonalizing matrices. But unlike in the CL sector, we will not focus on the parametrizations of $U_\nu$ and $V_\nu$; instead,  we will try to uncover the possible permuted variants of $U_\nu$. The reason is that only $U_\nu$ is present in the PMNS matrix along with $U_L$ (see Eq.\,(\ref{ca})) and even if we parametrize $U_\nu$ with the unphysical phases from the $\nu$ mass eigenvalues, they will just form linear combinations with those from $U_L$, effectively losing their individual identities. 
Following this reasoning, with $H_\nu$ at our disposal,  the diagonalizing matrix, $U_\nu$ is easily found to be,
\begin{eqnarray}
\label{bj}
U_{\nu} &=&\begin{pmatrix}
0  & 1 & 0 \\
 -\sqrt{\frac{c}{b+c}}e^{-i \rho} & 0 &  \sqrt{\frac{b}{b+c}}e^{-i \rho}  \\
   \sqrt{\frac{b}{b+c}} & 0  &   \sqrt{\frac{c}{b+c}}
\end{pmatrix} .
\end{eqnarray}

It is found that there are actually six different permuted variants of $U_\nu$, corresponding to six column transformations. Taking the one given in Eq.\,(\ref{bj}) as $U_{\nu 1}$, the list of the various $U_\nu$'s is given in Table \ref{tin(a)} ,  

\begin{table}
\begin{center}
\footnotesize
\begin{tabular*}{\textwidth}{@{\extracolsep{\fill}}ccc}
\hline
$\mathbf{U_{\nu 1}}$ & $\mathbf{U_{\nu 2}}$ & $\mathbf{U_{\nu 3}}$  \\
\hline
$\begin{pmatrix}
0  & 1 & 0 \\
 -\sqrt{\frac{c}{b+c}}e^{-i \rho} & 0 &  \sqrt{\frac{b}{b+c}}e^{-i \rho}  \\
   \sqrt{\frac{b}{b+c}} & 0  &   \sqrt{\frac{c}{b+c}}
\end{pmatrix}$ & $\begin{pmatrix}
1  & 0 & 0 \\
 0&  -\sqrt{\frac{c}{b+c}}e^{-i \rho}  &  \sqrt{\frac{b}{b+c}}e^{-i \rho}  \\
0 &  \sqrt{\frac{b}{b+c}} &  \sqrt{\frac{c}{b+c}}
\end{pmatrix}$ & $\begin{pmatrix}
0  & 1 & 0 \\
  \sqrt{\frac{b}{b+c}}e^{-i \rho} & 0 &  -\sqrt{\frac{c}{b+c}}e^{-i \rho}   \\
 \sqrt{\frac{c}{b+c}} & 0  &   \sqrt{\frac{b}{b+c}}
\end{pmatrix}$ \\ 
\hline
\hline
$\mathbf{U_{\nu 4}}$ & $\mathbf{U_{\nu 5}}$ & $\mathbf{U_{\nu 6}}$  \\
\hline
$\begin{pmatrix}
0  & 0& 1 \\
-\sqrt{\frac{c}{b+c}}e^{-i \rho}  & \sqrt{\frac{b}{b+c}}e^{-i \rho}  & 0 \\
\sqrt{\frac{b}{b+c}}  & \sqrt{\frac{c}{b+c}}  &   0
\end{pmatrix}$ & $\begin{pmatrix}
1  & 0 & 0 \\
0 &  \sqrt{\frac{b}{b+c}}e^{-i \rho}   & -\sqrt{\frac{c}{b+c}}e^{-i \rho}  \\
0& \sqrt{\frac{c}{b+c}}   &  \sqrt{\frac{b}{b+c}} 
\end{pmatrix}$ & $\begin{pmatrix}
0  & 0& 1 \\
 \sqrt{\frac{b}{b+c}}e^{-i \rho}  &  -\sqrt{\frac{c}{b+c}}e^{-i \rho}   & 0 \\
 \sqrt{\frac{c}{b+c}}  &  \sqrt{\frac{b}{b+c}}  &   0
\end{pmatrix}$ \\
\hline
\end{tabular*}
 \caption{\footnotesize \label{tin(a)} The six distinct permuted variants of $U_\nu$.}
\end{center}
\end{table}

Of the six $U_\nu$'s, only three survive, viz., $U_{\nu\,1}$,\,$U_{\nu\,3}$ and $U_{\nu\,5}$. The reasons for the exclusion of the others are either one or both of the following:
\begin{itemize}
\item  $\Delta\,m_{21}^2$ is found to be less than zero.
\item $\sin^2\,\theta_{13}$ is found to equal $1/3$, where $\theta_{13}$ is the reactor mixing angle (see Sect.\ref{Section 2(c)}).
\end{itemize}

Upon further examination, it is also found that for $U_{\nu\,5}$, the values of $\sin^2\,\theta_{12}$, $\theta_{12}$ being the solar angle (see Sect.\ref{Section 2(c)}), fall outside the experimental bounds\,\cite{Esteban:2020cvm,Esteban:2024eli}. Hence,  we are left with only two feasible $U_\nu$'s, namely $U_{\nu\,1}$ and $U_{\nu\,3}$. 

If we diagonalize $M_\nu$ with $U_{\nu\,1}$, we get a strict normal hierarchy (NH) of neutrino masses, with the masses  given below,

\begin{eqnarray}
\label{bq}
m_1 &=& 0,\quad
\label{br}
m_2 = \sqrt{a},\quad
\label{bs}
m_3 = \sqrt{b+c}.
\end{eqnarray}

On the other hand, diagonalizing $M_\nu$ with $U_{\nu\,3}$ gives rise to a strictly inverted hierarchy (IH) of neutrino mass,  as shown below,

\begin{eqnarray}
\label{bv}
m_1 &=& \sqrt{b+c},\quad
\label{bw}
m_2 = \sqrt{a},\quad
\label{bx}
m_3 = 0.
\end{eqnarray}

Thus we see that the number of feasible permuted variants of $U_\nu$ is smaller than that of $U_L$. It is because, unlike in the CL sector, the model parameters in the $\nu$ sector are subject to various constraints like mass squared differences, mixing angles,  CP phases etc. In the next section, we will see analytical expressions of these observable parameters in terms of the model parameters in $H_\nu$ (Eq.\,(\ref{equation:F})).


\subsection{\label{Section 2(c)}{Standard Parametrization of the Lepton Mixing Matrix}}

Given $U_L$'s and $U_\nu$'s,  we proceed to find out the lepton mixing matrix, $U$ as,
\begin{equation}
\label{ca}
U=U_L^\dagger\,U_\nu.
\end{equation}
  $U$, like any ordinary $3 \times 3$ unitary matrix, has nine free parameters. However neutrino oscillation experiments measure only four of them, viz., three mixing angles: the solar mixing angle, $\theta_{12}$, the atmospheric mixing angle, $\theta_{23}$, the reactor mixing angle, $\theta_{13}$ and the Dirac charge-parity (CP) violating phase, $\delta$. The remaining five parameters remain unobserved or should we say,  unphysical.  To make sure that these unphysical phases do not create complexities in the calculations of observable parameters, we prevent them from appearing in the calculations by transforming $U$ to a standardized form as given below,
\begin{equation}
\label{cx}
U_{std}=U_{\eta_{123}}^*\,U\,U_{\eta_{45}}^*,
\end{equation}
where,
\begin{eqnarray}
\label{cy}
U_{\eta_{123}} &=& \begin{pmatrix}
e^{i\,\eta_1} & 0& 0 \\
 0 &  e^{i\,\eta_2}& 0\\
 0 & 0&  e^{i\,\eta_3}\\
 \end{pmatrix},\quad
 \label{cz}
 U_{\eta_{45}} = \begin{pmatrix}
e^{i\,\eta_4} & 0& 0 \\
 0 &  e^{i\,\eta_5}& 0\\
 0 & 0&  1\\
\end{pmatrix},
\end{eqnarray}

so that $U_{std}$ resembles the one as per PDG convention\,\cite{ParticleDataGroup:2024cfk},
\begin{equation}
\label{da}
U_{std}=\begin{pmatrix}
c_{12}\,c_{13} & c_{13}\,s_{12} & e^{-i\,\delta}\,s_{13} \\
-c_{23}s_{12}-e^{i\,\delta}c_{12}s_{13}s_{23} & c_{12}c_{23}-e^{i\,\delta}s_{12}s_{23} & c_{13}s_{23}\\
s_{12}s_{23}-e^{i\,\delta}c_{12}c_{23}s_{13} & -c_{12}s_{23}-e^{i\,\delta}c_{23}s_{12}s_{13} & c_{13}c_{23}
\end{pmatrix},
\end{equation}
with $c_{ij}=\cos\,\theta_{ij}$ and $s_{ij}=\sin\,\theta_{ij}$.

The three mixing angles depend on the parameters $a,\,b, \,c$ and $\rho$. In this work, we take as our inputs four parameters:  three observable parameters,viz.,  $\sin^2\,\theta_{13}$,\,$\Delta m^2_{21} $, \,$\Delta m^2_{31} $, and a model parameter, $\rho$. Our predictions include the parameters $a,\,b, \,c$,  $\sin^2\,\theta_{12}$,\, $\sin^2\,\theta_{23}$,\, $\delta$ and the three $\nu$ masses (Eqs.\,(\ref{bs}) and (\ref{bx})).  If we choose the parametrization scheme given by Eq.\,(\ref{bh}) and $U_{L}$ as $U_{L1}$ (see Table \ref{table2a}), the analytical expressions of the predicted parameters, for NH case, in terms of the input parameters are given below,
 
\vspace*{-\baselineskip}
\begin{eqnarray}
\label{cb}
a &=& \Delta m^2_{21},\nonumber \\
b &=& \frac{1}{2}\left(\Delta m^2_{31}+\frac{\sqrt{\Delta m^4_{31} \cos^2\,\rho(\cos^2\,\rho-9\sin^4\,\theta_{13}+6\sin^2\,\theta_{13}-1)}}{\cos^2\,\rho} \right), \nonumber \\
c &=& \frac{1}{2}\left(\Delta m^2_{31}-\frac{\sqrt{\Delta m^4_{31} \cos^2\,\rho(\cos^2\,\rho-9\sin^4\,\theta_{13}+6\sin^2\,\theta_{13}-1)}}{\cos^2\,\rho} \right), \nonumber \\
\sin^2\,\theta_{12} &=& \frac{\Delta m^2_{31}\cos^2\,\rho}{2\Delta m^2_{31}\sqrt{\cos^2\,\rho}-\cos\,\rho\sqrt{\Delta m^4_{31}(3\sin^2\,\theta_{13}-1)^2}},\nonumber \\
\sin^2\,\theta_{23} &=& \frac{2\Delta m^2_{31}\sqrt{\cos^2\,\rho}+(\sqrt{3-3\cos^2\,\rho}-\cos\,\rho)\sqrt{\Delta m^4_{31}(1-3\sin^2\,\theta_{13})^2}}{4\Delta m^2_{31}\sqrt{\cos^2\,\rho}-2\cos\,\rho \sqrt{\Delta m^4_{31}(3\sin^2\,\theta_{13}-1})},\nonumber \\
\delta &=& -arg \left[ \frac{e^{-i \eta_1}(\sqrt{b}-\sqrt{c}\,e^{-i \rho})}{\sqrt{3(b+c)}} \right].
\end{eqnarray}




 For IH case, on the other hand, we have
 \begin{eqnarray}
\label{ei1}
a &=& \Delta m^2_{21}-\Delta m^2_{31},\nonumber \\
b &=& \frac{1}{2}\left(-\Delta m^2_{31}+\frac{\sqrt{\Delta m^4_{31} \cos^2\,\rho(\cos^2\,\rho-9\sin^4\,\theta_{13}+6\sin^2\,\theta_{13}-1)}}{\cos^2\,\rho} \right), \nonumber \\
c &=& -\frac{1}{2}\left(\Delta m^2_{31}+\frac{\sqrt{\Delta m^4_{31} \cos^2\,\rho(\cos^2\,\rho-9\sin^4\,\theta_{13}+6\sin^2\,\theta_{13}-1)}}{\cos^2\,\rho} \right), \nonumber \\
\sin^2\,\theta_{12} &=& \frac{\Delta m^2_{31}\cos^2\,\rho}{2\Delta m^2_{31}\sqrt{\cos^2\,\rho}-\cos\,\rho\sqrt{\Delta m^4_{31}(3\sin^2\,\theta_{13}-1)^2}},\nonumber \\
\sin^2\,\theta_{23} &=& \frac{2\Delta m^2_{31}\sqrt{\cos^2\,\rho}+(\sqrt{3-3\cos^2\,\rho}-\cos\,\rho)\sqrt{\Delta m^4_{31}(1-3\sin^2\,\theta_{13})^2}}{4\Delta m^2_{31}\sqrt{\cos^2\,\rho}-2\cos\,\rho \sqrt{\Delta m^4_{31}(3\sin^2\,\theta_{13}-1})},\nonumber \\
\delta &=& -arg \left[ \frac{e^{i\eta_1}(\sqrt{b}-\sqrt{c}\,e^{-i\rho})}{\sqrt{3(b+c)}} \right].
\end{eqnarray}

 It seems that even though we have standardized $U$, $\delta$ in both NH and IH cases still depends on an unphysical phase, $\eta_1$. But it turns out that $\eta_1$ is not an independent parameter, rather a function of the input parameters. This is assured by the following five equations, known as the standard parametrization conditions,

For NH,
\begin{flushleft}
\begin{align}
\textbf{(A)} & \quad \sqrt{c}\sin\,(\eta_4+\rho+\eta_1)-\sqrt{b}\sin\,(\eta_1+\eta_4)=0, \nonumber \\
\textbf{(B)} & \quad \sin\,(\eta_1 + \eta_5) = 0, \nonumber\\
\textbf{(C)} & \quad 3\,\sqrt{b} \cos\,(\rho + \eta_2)-2\sqrt{3c}\sin\,\eta_2 + \sqrt{3b} \sin\,(\rho+\eta_2)=0, \nonumber \\
\textbf{(D)} & \quad 3\,\sqrt{b} \cos\,(\rho + \eta_3)+2\sqrt{3c}\sin\,\eta_3 - \sqrt{3b} \sin\,(\rho+\eta_3)=0, \nonumber \\
\label{db5}
\textbf{(E)} & \quad 2\sqrt{3b(b+c)}\sin\,(\eta_2+\eta_4)-\sqrt{b+c-2 \sqrt{bc}\cos\,\rho}\,(3\cos\,(\eta_2+\eta_5)-\sqrt{3}\sin\,(\eta_2+\eta_5)) \nonumber \\ 
& \quad -\sqrt{c}(3\cos\,(\rho+\eta_2+\eta_4)+\sqrt{3}\sin\,(\rho+\eta_2+\eta_4))=0,
\end{align}
\end{flushleft}
and for IH, 
\begin{flushleft}
\begin{align}
\textbf{(A)} & \quad \sqrt{c}\,\sin\,(\eta_1+\eta_4)+\sqrt{b}\,\sin\,(\rho+\eta_1+\eta_4)=0, \nonumber \\
\textbf{(B)} & \quad \sin\,(\eta_1+\eta_5)=0, \nonumber \\
\textbf{(C)} & \quad 3\sqrt{c}\,\cos\,(\rho+\eta_2)+2\sqrt{3b}\,\sin\,\eta_2+\sqrt{3c}\,\sin\,(\rho+\eta_2)=0, \nonumber \\
\textbf{(D)} & \quad 3\sqrt{c}\,\cos\,(\rho+\eta_3)-2\sqrt{3b}\,\sin\,\eta_3-\sqrt{3c}\,\sin\,(\rho+\eta_3)=0, \nonumber \\
\label{ej5}
\textbf{(E)} & \quad \sqrt{b+c+2\sqrt{bc}\,\cos\rho}\,(3\cos\,(\eta_2+\eta_5)-\sqrt{3}\,\sin\,(\eta_2+\eta_5)) +\sqrt{b}\,(3\cos\,(\rho+\eta_2+\eta_4) \nonumber \\
& \quad +\sqrt{3}\sin\,(\rho+\eta_2+\eta_4))-2\sqrt{3c}\,\sin\,(\eta_2+\eta_4)=0.
\end{align}
\end{flushleft}

Now, let us see the effects of different parametrizations of $U_L$ on the expressions of observable parameters and standard parametrization conditions. If we had used the parametrization scheme given by Eq.\,(\ref{bf}), the forms of the expressions and the standard parametrization conditions for both NH and IH would have remained the same with the following redefinition of parameters,
\begin{eqnarray}
\label{dn}
\rho &\longrightarrow & \rho+\kappa_2-\kappa_3,\nonumber \\
\label{do}
\eta_1 &\longrightarrow & \eta_1+\kappa_3,\nonumber \\
\label{dp}
\eta_2 &\longrightarrow & \eta_3+\kappa_3,\nonumber \\
\label{dq}
\eta_3 &\longrightarrow & \eta_3+\kappa_3,\nonumber \\
\label{dr}
\eta_5 &\longrightarrow & \eta_5+\kappa_1.
\end{eqnarray}

Likewise,  had we implemented another scheme defined by $U_L \longrightarrow \text{diag}\,[\exp(i\,\kappa_1),\,\exp(i\,\kappa_2),1]\,U_L$ and $U_R \longrightarrow \text{diag}\,[1,1,\exp(-i\,\kappa_3)]\,U_R$, then also the forms of the equations would have remained unchanged, with the following redefinition of parameters,
\begin{eqnarray}
\label{ds}
\rho &\longrightarrow & \rho + \kappa_2,\nonumber \\
\label{dt}
\eta_5 &\longrightarrow & \eta_5+\kappa_1.
\end{eqnarray}

In all the cases of parametrization schemes discussed, we see that the redefinition of parameters is possible because $\kappa_1,\,\kappa_2$ and $\kappa_3$ are present as linear combinations with the other parameters, $\rho$ and $\eta_i$'s and can be redefined as they are free parameters.  Although we have used $U_{L1}$ in the example, the same is true for all the other 17 $U_L$'s. As such, the presence of the $\kappa_i$'s in $U_L$ do not have much impact on the predictions of the model. But this may not be always so, especially if they were not present as linear combinations with $\rho$ and $\eta_i$'s.  This is a topic left for our future studies. 

In the next section, we will see, on the other hand, the effect of the permuted variants of $U_L$  on the predictions.




\section{\label{Section 3}{Numerical Analysis}}
In the preceding sections, we have seen that the proposed framework allows both NH and IH of $\nu$ masses. We have also shown that PCLC allows 18 possible $U_L$'s through permutation. 
Now, we define a model of a given type for each hierarchy. For NH, a \textit{Type k} model, with $k=1,2,..18$, is defined by a combination of \( U_{Lk} \) and \( U_{\nu 1} \), while for IH, it corresponds to a combination of \( U_{Lk} \) and \( U_{\nu 3} \). Upon further analysis, it is observed that not all 18 models yield experimentally valid predictions for \( \sin^2\theta_{12} \) for a given hierarchy. Only 12 models are valid for each hierarchy. Notably, a given model does not necessarily involve the same \( U_L \) for both NH and IH. The list of valid models and their combinations for NH and IH is provided in Table \ref{table4a}.

\begin{table}
\centering
\footnotesize
\begin{tabular}{ccccccccccccc}
\hline
\textbf{NH} & Type 1 & Type 2 & Type 3 & Type 4 & Type 5& Type 6 & Type 7& Type 8&Type 9& Type 10& Type 11& Type 12 \\
\hline
$\mathbf{U_L}$ & $U_{L1}$ & $U_{L3}$ & $U_{L4}$ & $U_{L6}$ & $U_{L7}$ & $U_{L9}$ & $U_{L10}$ & $U_{L12}$ & $U_{L16}$ & $U_{L17}$ & $U_{L13}$& $U_{L14}$ \\
\hline
$\mathbf{U_\nu}$ & $U_{\nu 1}$ & $U_{\nu 1}$ & $U_{\nu 1}$ & $U_{\nu 1}$ & $U_{\nu 1}$ & $U_{\nu 1}$ & $U_{\nu 1}$ & $U_{\nu 1}$ & $U_{\nu 1}$ & $U_{\nu 1}$ & $U_{\nu 1}$ & $U_{\nu 1}$ \\
\hline
\hline
\textbf{IH} & Type 1 & Type 2 & Type 3 & Type 4 & Type 5& Type 6 & Type 7& Type 8&Type 9& Type 10& Type 11& Type 12 \\
\hline
$\mathbf{U_L}$ & $U_{L1}$ & $U_{L2}$ & $U_{L4}$ & $U_{L5}$ & $U_{L7}$ & $U_{L8}$ & $U_{L10}$ & $U_{L11}$ & $U_{L16}$ & $U_{L18}$ & $U_{L13}$& $U_{L15}$ \\
\hline
$\mathbf{U_\nu}$ & $U_{\nu 3}$ & $U_{\nu 3}$ & $U_{\nu 3}$ & $U_{\nu 3}$ & $U_{\nu 3}$ & $U_{\nu 3}$ & $U_{\nu 3}$ & $U_{\nu 3}$ & $U_{\nu 3}$ & $U_{\nu 3}$ & $U_{\nu 3}$ & $U_{\nu 3}$ \\
\hline
\end{tabular}
\footnotesize
\caption{\footnotesize  \label{table4a} Various types of valid models under NH and IH.}
\end{table}

\begin{figure}
    \centering
    \begin{subfigure}{0.45\textwidth}
        \includegraphics[width=\textwidth]{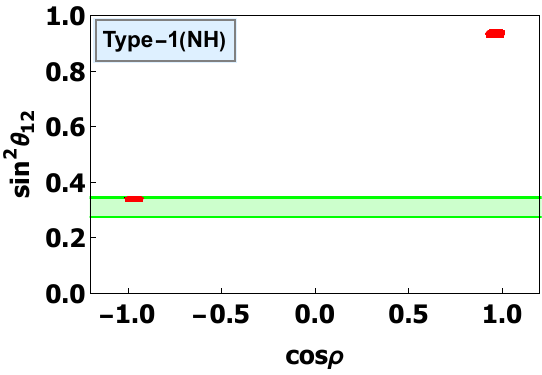}
        \caption{ Type 1 Model}
        \label{a}
    \end{subfigure}
    \hfill
    \begin{subfigure}{0.45\textwidth}
        \includegraphics[width=\textwidth]{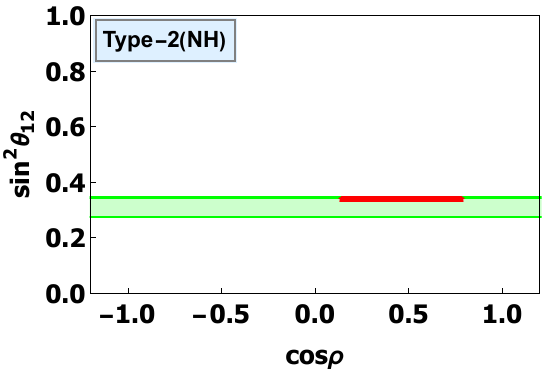}
        \caption{Type 2 Model}
    \end{subfigure}
   \caption{\footnotesize \label{1n} The correlation plots of $\sin^2\,\theta_{12}$ vs $\cos\,\rho$ for normal hierarchy. All the 12 types of models show either of the two plots.  Types 1, 3,5,7, 9 and 11 have the one similar plot and types 2, 4,6,8,10 and 12 have the other one. }
\end{figure}

\begin{figure}
    \centering
    \begin{subfigure}{0.32\textwidth}
        \includegraphics[width=\textwidth]{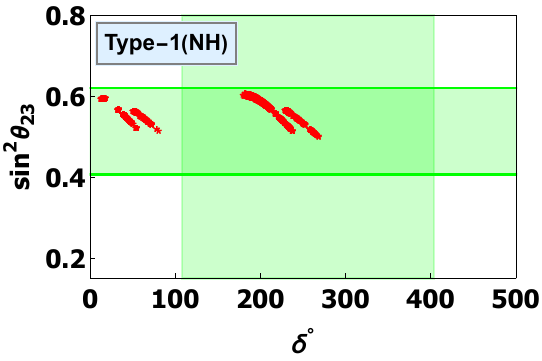}
        \caption{ Type 1 Model}
    \end{subfigure}
    \hfill
    \begin{subfigure}{0.32\textwidth}
        \includegraphics[width=\textwidth]{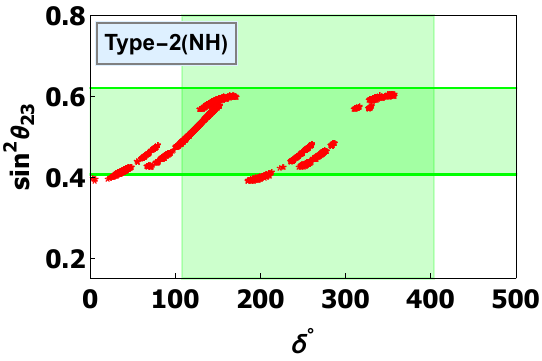}
        \caption{Type 2 Model}
    \end{subfigure}
    \hfill
    \begin{subfigure}{0.32\textwidth}
        \includegraphics[width=\textwidth]{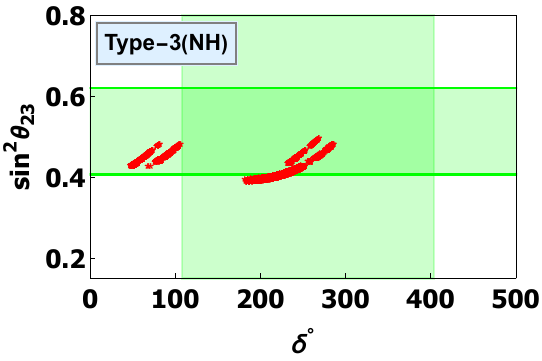}
        \caption{ Type 3 Model}
    \end{subfigure}

    \vspace{10pt} 

    \begin{subfigure}{0.3\textwidth}
        \includegraphics[width=\textwidth]{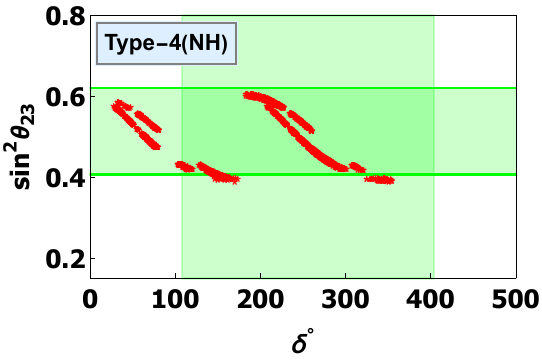}
        \caption{Type 4 Model}
    \end{subfigure}
    \hfill
    \begin{subfigure}{0.3\textwidth}
        \includegraphics[width=\textwidth]{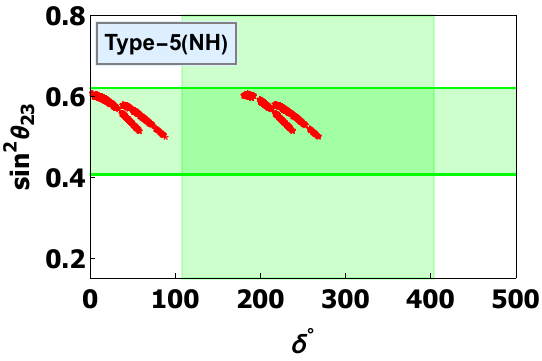}
        \caption{Type 5 Model}
    \end{subfigure}
    \hfill
    \begin{subfigure}{0.3\textwidth}
        \includegraphics[width=\textwidth]{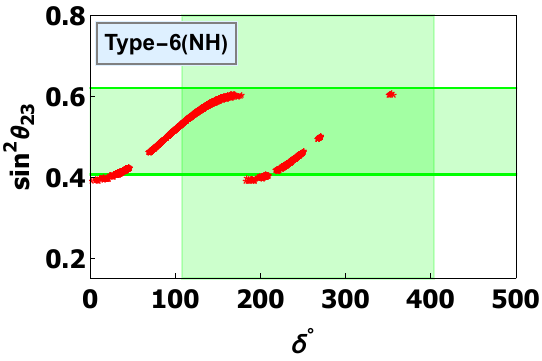}
        \caption{Type 6 Model}
    \end{subfigure}

    \vspace{10pt} 

    \begin{subfigure}{0.3\textwidth}
        \includegraphics[width=\textwidth]{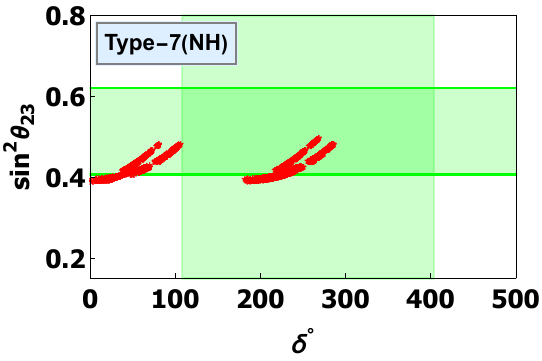}
        \caption{Type 7 Model}
    \end{subfigure}
    \hfill
    \begin{subfigure}{0.3\textwidth}
        \includegraphics[width=\textwidth]{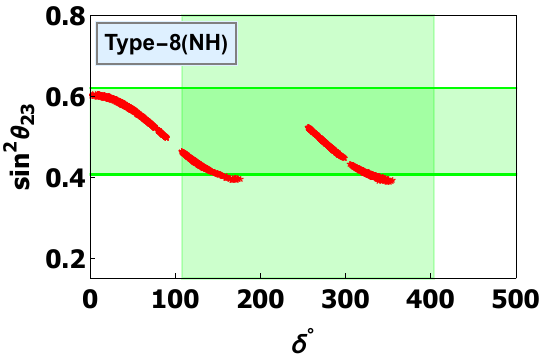}
        \caption{Type 8 Model}
    \end{subfigure}
    \hfill
    \begin{subfigure}{0.3\textwidth}
        \includegraphics[width=\textwidth]{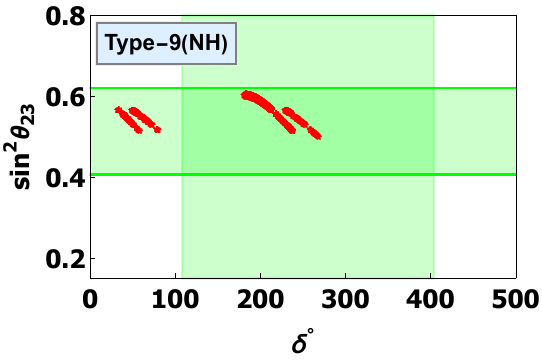}
        \caption{Type 9 Model}
    \end{subfigure}

    \vspace{10pt} 

    \begin{subfigure}{0.3\textwidth}
        \includegraphics[width=\textwidth]{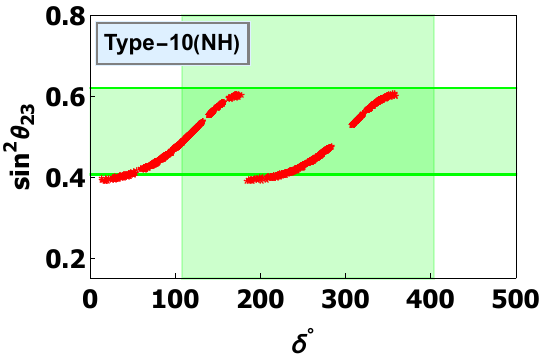}
        \caption{Type 10 Model}
    \end{subfigure}
    \hfill
    \begin{subfigure}{0.3\textwidth}
        \includegraphics[width=\textwidth]{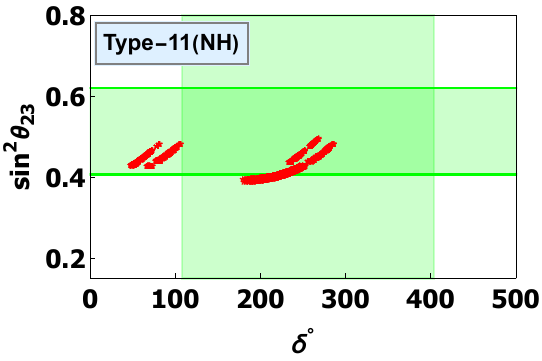}
        \caption{ Type 11 Model }
    \end{subfigure}
    \hfill
    \begin{subfigure}{0.3\textwidth}
        \includegraphics[width=\textwidth]{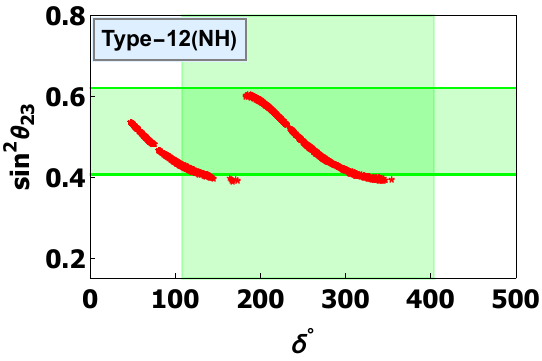}
        \caption{Type 12 Model}
    \end{subfigure}

    \caption{\footnotesize \label{2n} The correlation plots of $\sin^2\,\theta_{23}$ vs $\delta^\circ$ for the 12 models under normal hierarchy.}
\end{figure}

\begin{figure}
\centering
 \begin{subfigure}{0.45\textwidth}
        \includegraphics[width=\textwidth]{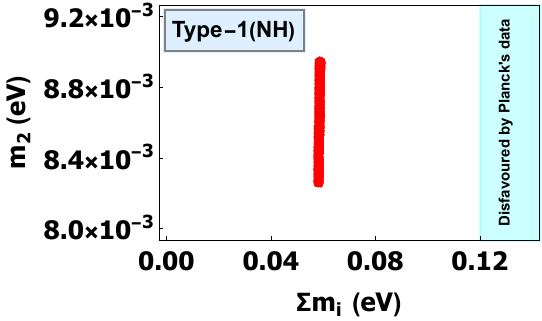}
        \caption{}
        \label{a3}
    \end{subfigure}
    \hfill
    \begin{subfigure}{0.40\textwidth}
        \includegraphics[width=\textwidth]{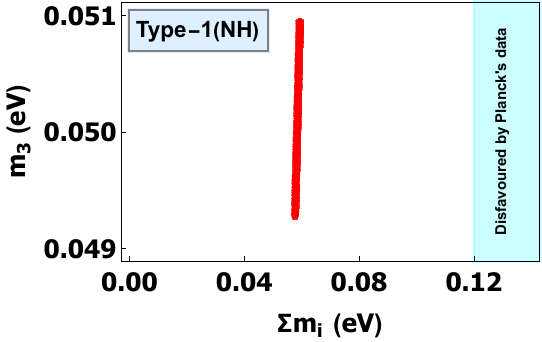}
        \caption{}
        \label{a4}
    \end{subfigure}

    \vspace{10pt} 

    \begin{subfigure}{0.48\textwidth}
        \includegraphics[width=\textwidth]{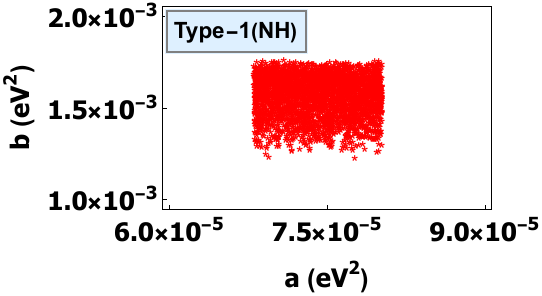}
        \caption{}
        \label{a5}
    \end{subfigure}
    \hfill
    \begin{subfigure}{0.48\textwidth}
        \includegraphics[width=\textwidth]{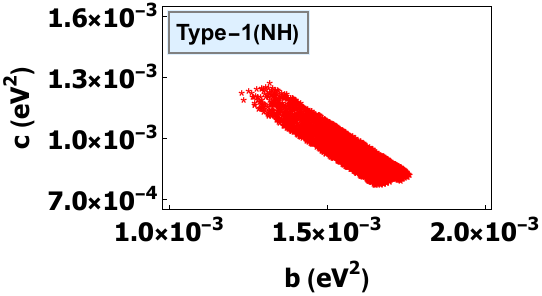}
        \caption{}
        \label{a6}
    \end{subfigure}
 \caption{\footnotesize \label{Cn} The correlation plot of (a) $m_2$ vs $\Sigma m_i$, (b) $m_3$ vs $\Sigma m_i$,  (c) $b$ vs $a$ and (d) $c$ vs $b$ for normal hierarchy. All types of models have similar plots.}
\end{figure}

\begin{figure}
    \centering
    \begin{subfigure}{0.45\textwidth}
        \includegraphics[width=\textwidth]{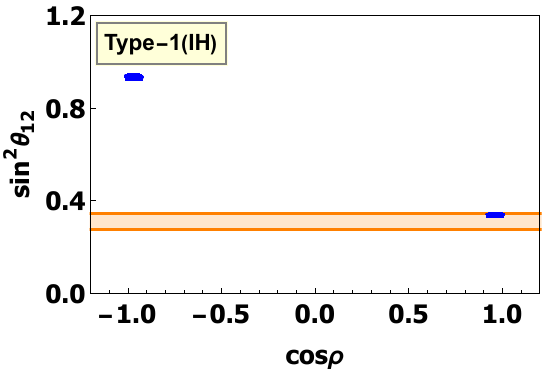}
        \caption{Type 1 Model}
        \label{aii}
    \end{subfigure}
    \hfill
    \begin{subfigure}{0.45\textwidth}
        \includegraphics[width=\textwidth]{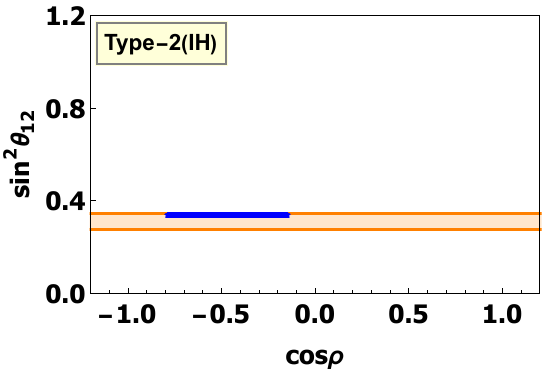}
        \caption{Type 2 Model}
    \end{subfigure}

    \caption{\footnotesize \label{1i} The correlation plots of $\sin^2\,\theta_{12}$ vs $\cos\,\rho$ for inverted hierarchy case. All the 12 types showing either of the following two plots.  Types 1, 3,5,7, 9 and 11 have a similar plot, while types 2, 4,6,8,10 and 12 have another similar plot. }
\end{figure}

\begin{figure}
    \centering
    \begin{subfigure}{0.32\textwidth}
        \includegraphics[width=\textwidth]{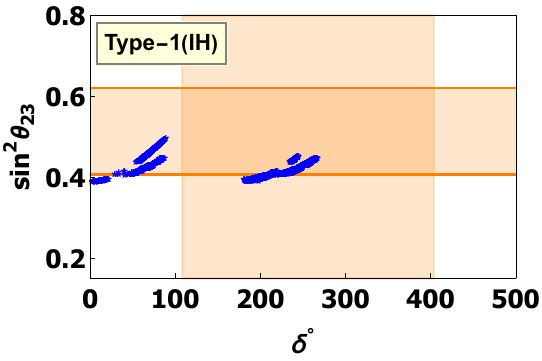}
        \caption{Type 1 Model}
    \end{subfigure}
    \hfill
    \begin{subfigure}{0.32\textwidth}
        \includegraphics[width=\textwidth]{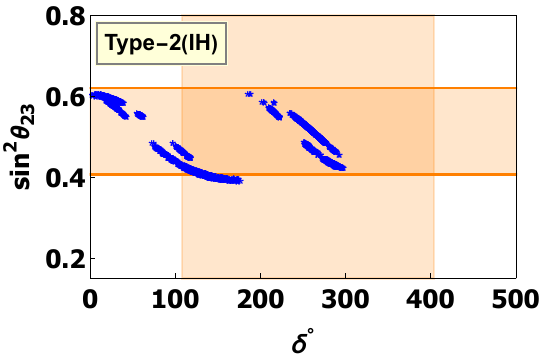}
        \caption{Type 2 Model}
    \end{subfigure}
    \hfill
    \begin{subfigure}{0.32\textwidth}
        \includegraphics[width=\textwidth]{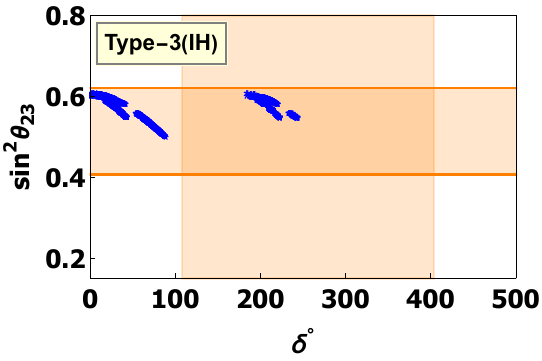}
        \caption{Type 3 Model}
    \end{subfigure}

    \vspace{10pt} 

    \begin{subfigure}{0.3\textwidth}
        \includegraphics[width=\textwidth]{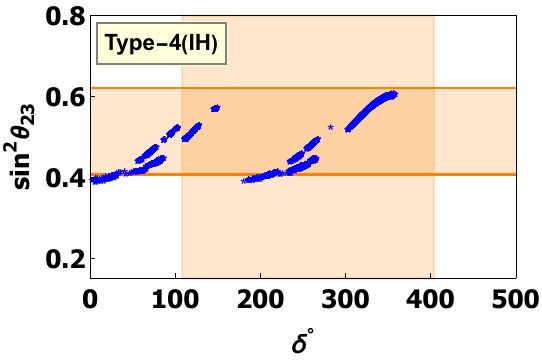}
        \caption{Type 4 Model}
    \end{subfigure}
    \hfill
    \begin{subfigure}{0.3\textwidth}
        \includegraphics[width=\textwidth]{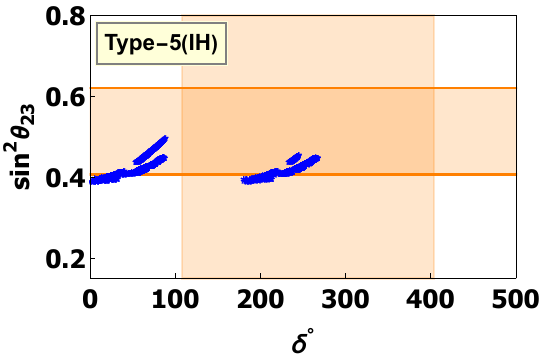}
        \caption{Type 5 Model}
    \end{subfigure}
    \hfill
    \begin{subfigure}{0.3\textwidth}
        \includegraphics[width=\textwidth]{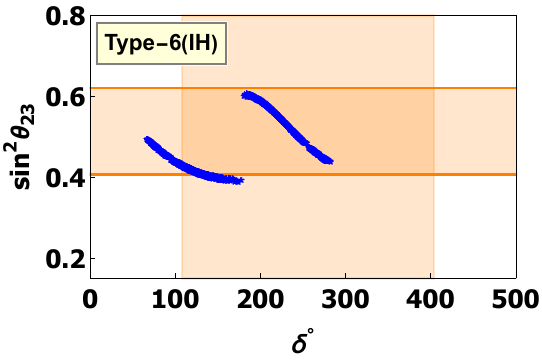}
        \caption{Type 6 Model}
    \end{subfigure}

    \vspace{10pt} 

    \begin{subfigure}{0.3\textwidth}
        \includegraphics[width=\textwidth]{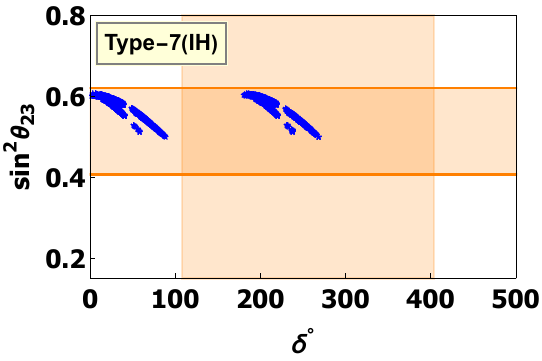}
        \caption{Type 7 Model}
    \end{subfigure}
    \hfill
    \begin{subfigure}{0.3\textwidth}
        \includegraphics[width=\textwidth]{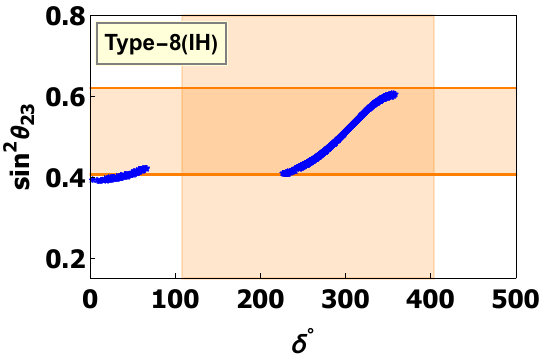}
        \caption{Type 8 Model}
    \end{subfigure}
    \hfill
    \begin{subfigure}{0.3\textwidth}
        \includegraphics[width=\textwidth]{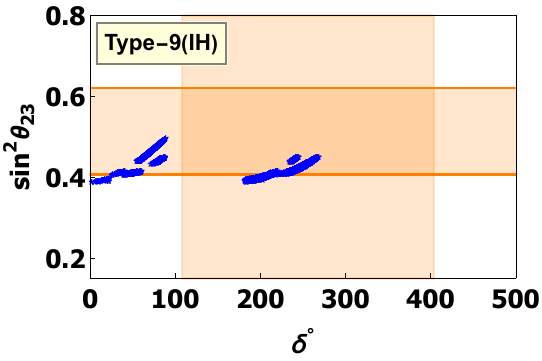}
        \caption{Type 9 Model}
    \end{subfigure}

    \vspace{10pt} 

    \begin{subfigure}{0.3\textwidth}
        \includegraphics[width=\textwidth]{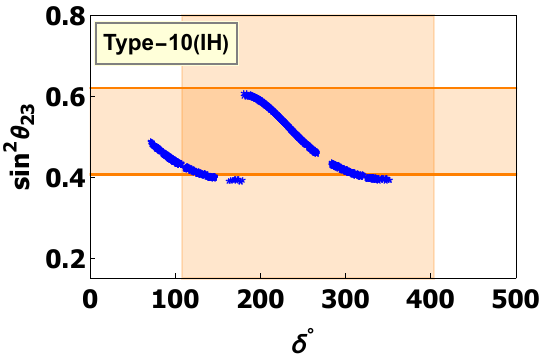}
        \caption{Type 10 Model}
    \end{subfigure}
    \hfill
    \begin{subfigure}{0.3\textwidth}
        \includegraphics[width=\textwidth]{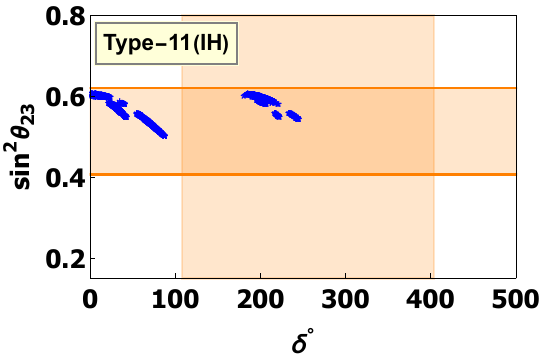}
        \caption{Type 11 Model}
    \end{subfigure}
    \hfill
    \begin{subfigure}{0.3\textwidth}
        \includegraphics[width=\textwidth]{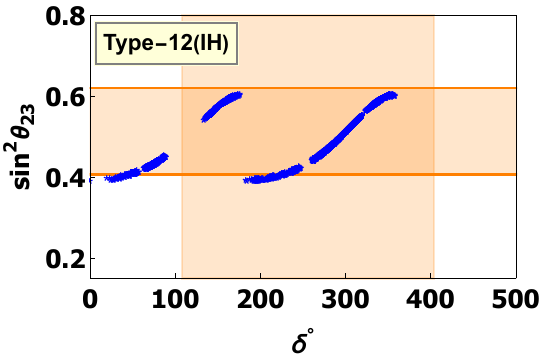}
        \caption{Type 12 Model}
    \end{subfigure}

    \caption{\footnotesize \label{2i}The correlation plots of $\sin^2\,\theta_{23}$ vs $\delta$ for the 12 models under inverted hierarchy. }
\end{figure}

\begin{figure}
\centering
 \begin{subfigure}{0.40\textwidth}
        \includegraphics[width=\textwidth]{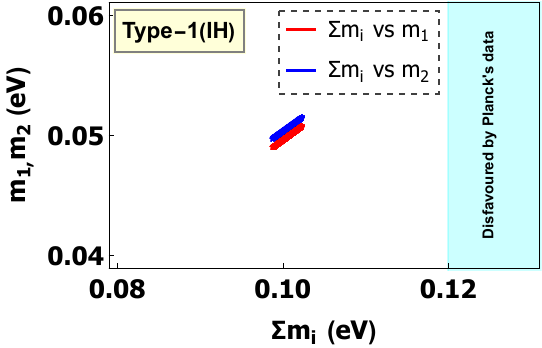}
        \caption{}
        \label{b3}
    \end{subfigure}
    \hfill
    \begin{subfigure}{0.40\textwidth}
        \includegraphics[width=\textwidth]{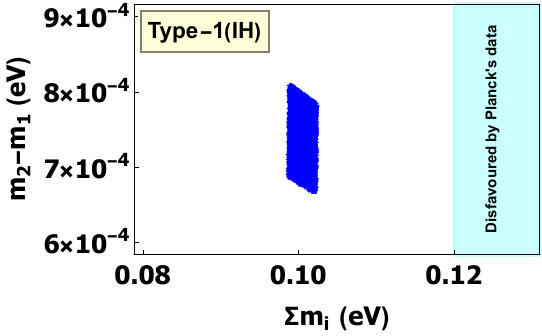}
        \caption{}
        \label{b4}
    \end{subfigure}

    \vspace{10pt} 

    \begin{subfigure}{0.45\textwidth}
        \includegraphics[width=\textwidth]{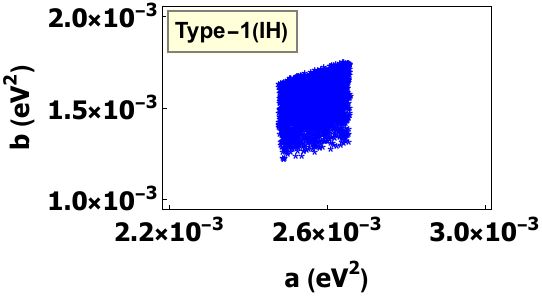}
        \caption{}
        \label{b5}
    \end{subfigure}
    \hfill
    \begin{subfigure}{0.45\textwidth}
        \includegraphics[width=\textwidth]{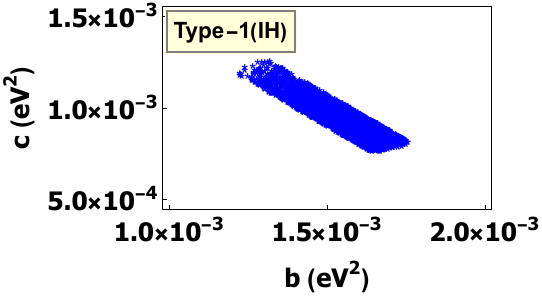}
        \caption{}
        \label{b6}
    \end{subfigure}
 \caption{\footnotesize \label{Ci} The correlation plot of (a) $m_1,\,m_2$ vs $\Sigma m_i$, (b) $m_2-m_1$ vs $\Sigma m_i$,  (c) $b$ vs $a$ and (d) $c$ vs $b$ for inverted hierarchy. All types of models have similar plots.}
\end{figure}

 As mentioned in the previous section, we have four input parameters, viz.,  $\Delta m^2_{21}$, $\Delta m^2_{31}$, $\sin^2\,\theta_{13}$ and $\rho$. Now, instead of fixing them with single values, we have taken ranges of values, with $3 \sigma$ bounds for the observational parameters, and a range of $[-1,1]$ for $\cos\,\rho$. To observe the predictions, we use Eqs.\,(\ref{cb}) and (\ref{db5}) for NH and Eqs.\,(\ref{ei1}) and (\ref{ej5}) for IH, which happen to be the equations of Type 1 models for NH and IH respectively. Since, we have 12 models each, the equations for the rest are given in Appendix \ref{A}.  Some special features of the results are mentioned below,
 
 \begin{itemize}
 \item For both NH and IH, our work predicts a very sharp range of values for $\sin^2\,\theta_{12}$.  We have found, for all types of models under NH and IH,  the minimum value at $0.340$ approximately and the maximum value at $0.341$ approximately.  It is important to note that the correlation plots of $\sin^2\theta_{12}$ versus $\cos\rho$ exhibit two distinct patterns, as illustrated in Figs.,(\ref{1n}) and (\ref{1i}). From Figs.,(\ref{a}) and (\ref{aii}), it can be observed that certain values of $\sin^2\theta_{12}$ lie outside the experimental bounds (experimental bounds denoted by coloured patches) for specific values of $\cos\rho$. For models with similar correlation patterns, the data points corresponding to $\cos\rho$ that yield $\sin^2\theta_{12}$ outside the experimental range are excluded from subsequent plots.
 \item For both NH and IH,  different models show different predictions for the octant of $\theta_{23}$. A list is provided in Table \ref{table5a}, detailing the maximum and minimum values of $\theta_{23}$ for different types of models with favoured octants. The correlation plots of $\sin^2\,\theta_{23}$ vs $\delta$ are shown in Figs.\,(\ref{2n}) and (\ref{2i}).
 The values of $\delta$ show some interesting patterns too. In some cases,they show something similar to forbidden gaps extending from region outside the experimental bounds to experimentally valid regions.  
 \item The work also predicts absolute NH as well as IH for $\nu$ masses.  For each hierarchy, the respective models show approximately the same upper and lower bounds on mass eigenvalues. For NH,  we have approximately $0.0082 \text{eV}<m2<0.009 \text{eV}\,, 0.049 \text{eV}<m3<0.051 \text{eV}$, while the IH predicts, approximately, $0.049 \text{eV}<m1<0.051 \text{eV}, 0.050 \text{eV}<m2<0.052 \text{eV}$. All these values are valid based on cosmological upper bounds on sum of $\nu$ masses\,\cite{ParticleDataGroup:2024cfk} (refer to Figs.\,(\ref{Cn}) and (\ref{Ci})). Since, the values of $m_1$ and $m_2$ are almost similar for IH, we have also shown the difference of them, $m_2-m_1$.
 \end{itemize}
 
\begin{figure}
    \centering
    \begin{subfigure}{0.32\textwidth}
        \includegraphics[width=\textwidth]{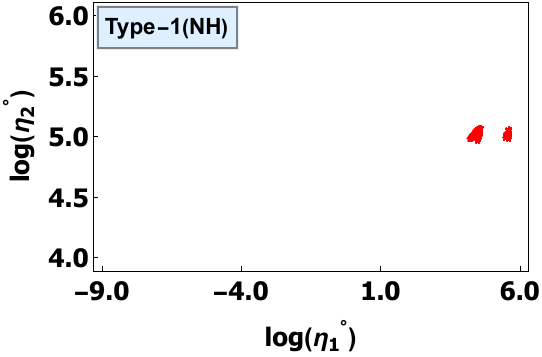}
        \caption{Type 1 Model}
    \end{subfigure}
    \hfill
    \begin{subfigure}{0.32\textwidth}
        \includegraphics[width=\textwidth]{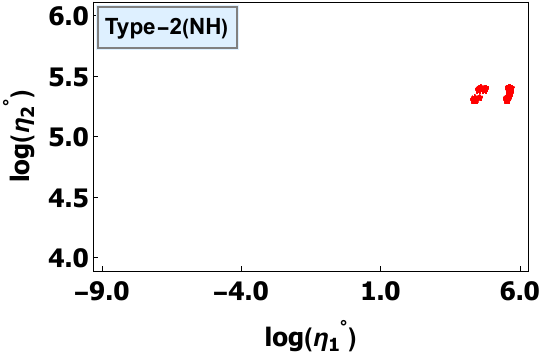}
        \caption{Type 2 Model}
    \end{subfigure}
    \hfill
    \begin{subfigure}{0.32\textwidth}
        \includegraphics[width=\textwidth]{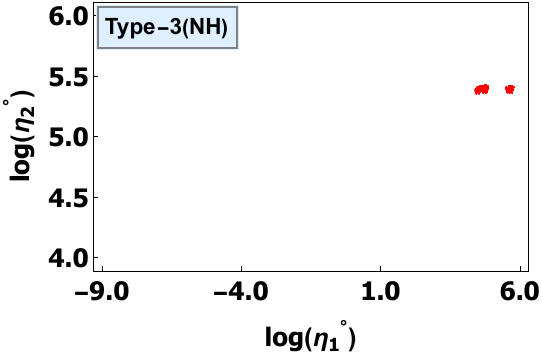}
        \caption{Type 3 Model}
    \end{subfigure}

    \vspace{10pt} 

    \begin{subfigure}{0.3\textwidth}
        \includegraphics[width=\textwidth]{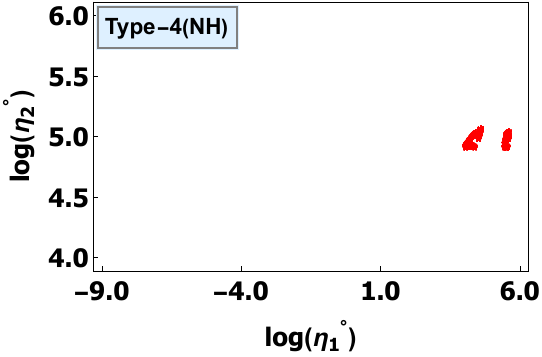}
        \caption{Type 4 Model}
    \end{subfigure}
    \hfill
    \begin{subfigure}{0.3\textwidth}
        \includegraphics[width=\textwidth]{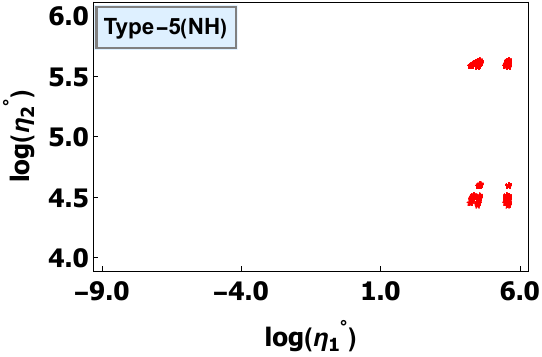}
        \caption{Type 5 Model}
    \end{subfigure}
    \hfill
    \begin{subfigure}{0.3\textwidth}
        \includegraphics[width=\textwidth]{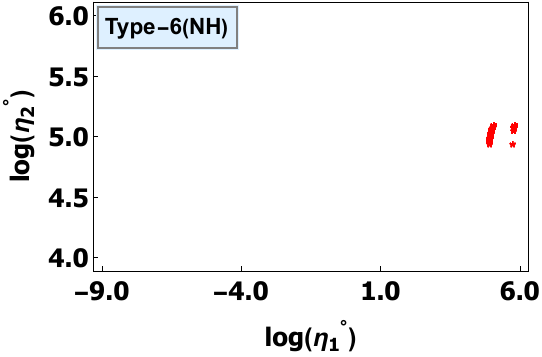}
        \caption{Type 6 Model}
    \end{subfigure}

    \vspace{10pt} 

    \begin{subfigure}{0.3\textwidth}
        \includegraphics[width=\textwidth]{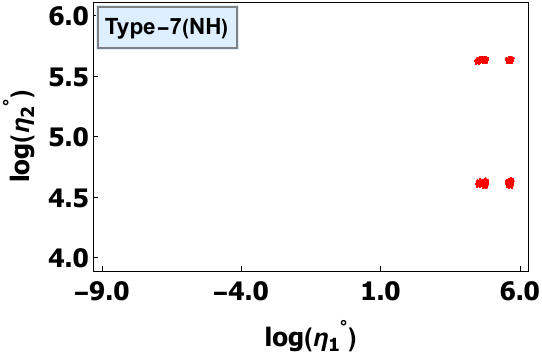}
        \caption{Type 7 Model}
    \end{subfigure}
    \hfill
    \begin{subfigure}{0.3\textwidth}
        \includegraphics[width=\textwidth]{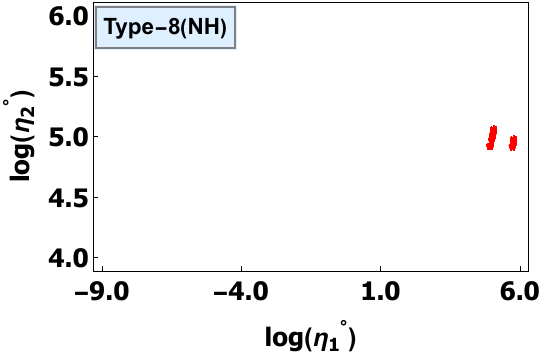}
        \caption{Type 8 Model}
    \end{subfigure}
    \hfill
    \begin{subfigure}{0.3\textwidth}
        \includegraphics[width=\textwidth]{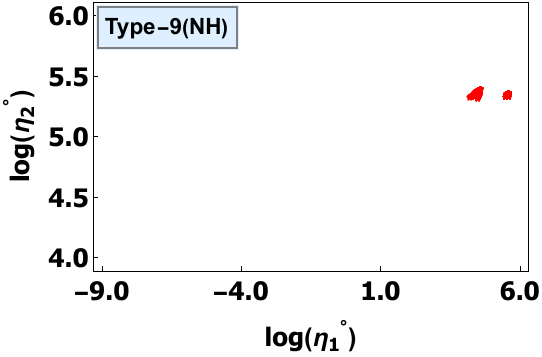}
        \caption{Type 9 Model}
    \end{subfigure}

    \vspace{10pt} 

    \begin{subfigure}{0.3\textwidth}
        \includegraphics[width=\textwidth]{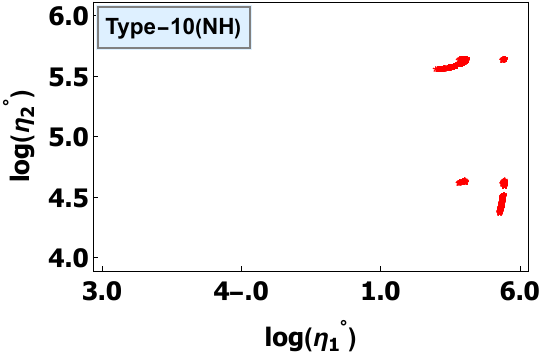}
        \caption{Type 10 Model}
    \end{subfigure}
    \hfill
    \begin{subfigure}{0.3\textwidth}
        \includegraphics[width=\textwidth]{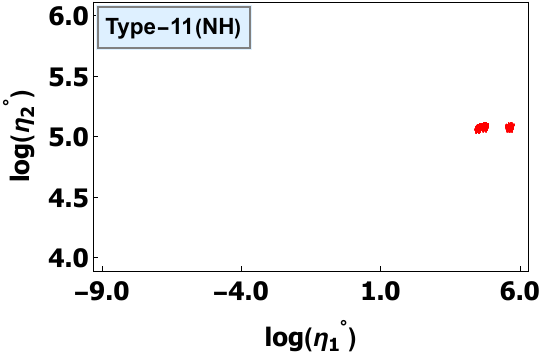}
        \caption{Type 11 Model}
    \end{subfigure}
    \hfill
    \begin{subfigure}{0.3\textwidth}
        \includegraphics[width=\textwidth]{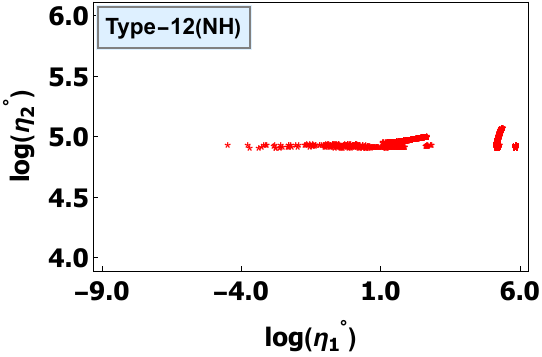}
        \caption{Type 12 Model}
    \end{subfigure}

    \caption{\footnotesize \label{7n}The correlation plots of $\log(\eta_2^\circ)$ and $\log(\eta_1^\circ)$ for the 12 models under normal hierarchy. }
\end{figure}

\begin{figure}
    \centering
    \begin{subfigure}{0.32\textwidth}
        \includegraphics[width=\textwidth]{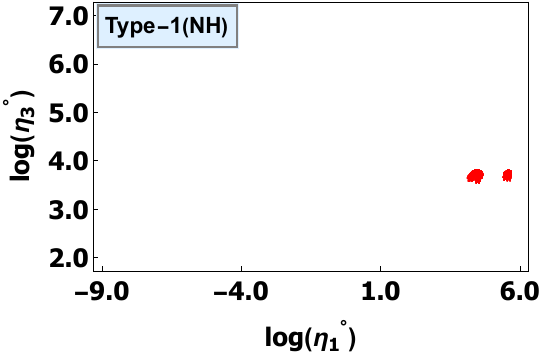}
        \caption{Type 1 Model}
    \end{subfigure}
    \hfill
    \begin{subfigure}{0.32\textwidth}
        \includegraphics[width=\textwidth]{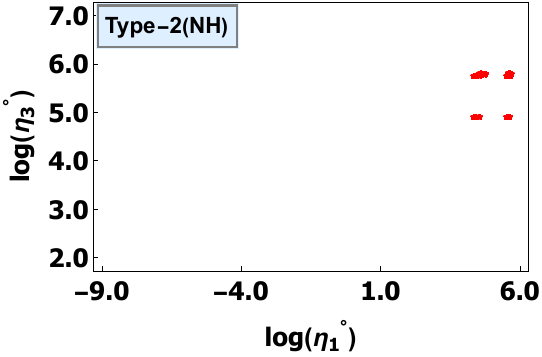}
        \caption{Type 2 Model}
    \end{subfigure}
    \hfill
    \begin{subfigure}{0.32\textwidth}
        \includegraphics[width=\textwidth]{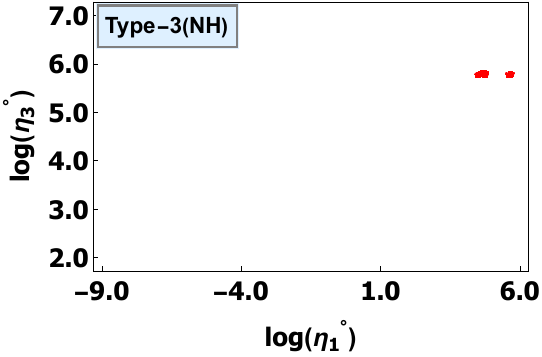}
        \caption{Type 3 Model}
    \end{subfigure}

    \vspace{10pt} 

    \begin{subfigure}{0.3\textwidth}
        \includegraphics[width=\textwidth]{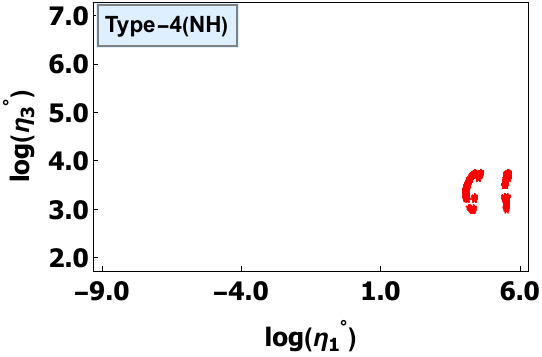}
        \caption{Type 4 Model}
    \end{subfigure}
    \hfill
    \begin{subfigure}{0.3\textwidth}
        \includegraphics[width=\textwidth]{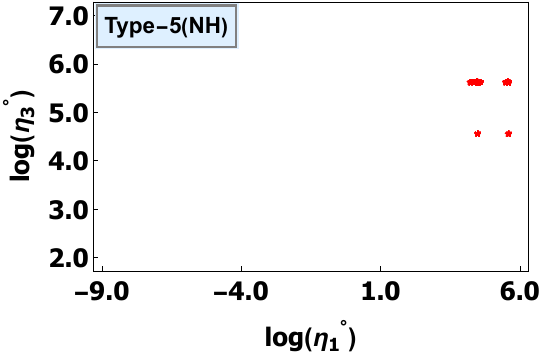}
        \caption{Type 5 Model}
    \end{subfigure}
    \hfill
    \begin{subfigure}{0.3\textwidth}
        \includegraphics[width=\textwidth]{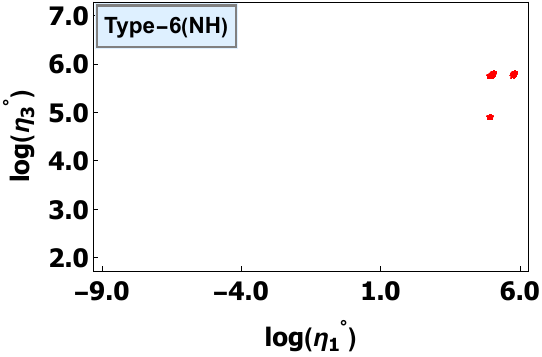}
        \caption{Type 6 Model}
    \end{subfigure}

    \vspace{10pt} 

    \begin{subfigure}{0.3\textwidth}
        \includegraphics[width=\textwidth]{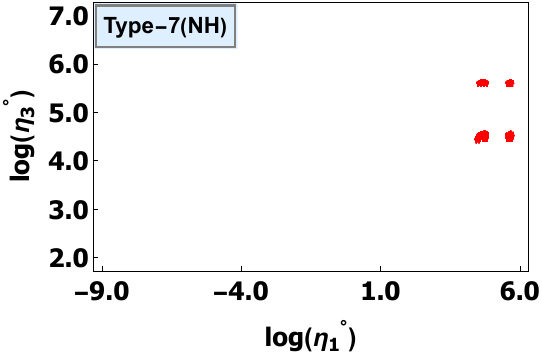}
        \caption{Type 7 Model}
    \end{subfigure}
    \hfill
    \begin{subfigure}{0.3\textwidth}
        \includegraphics[width=\textwidth]{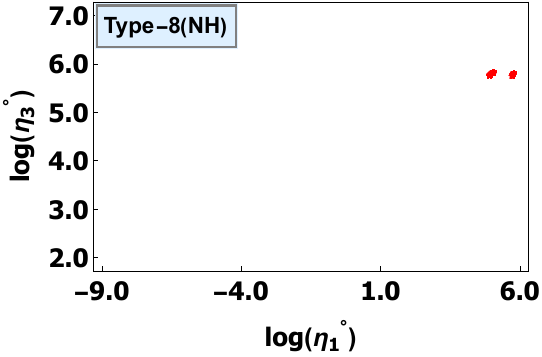}
        \caption{Type 8 Model}
    \end{subfigure}
    \hfill
    \begin{subfigure}{0.3\textwidth}
        \includegraphics[width=\textwidth]{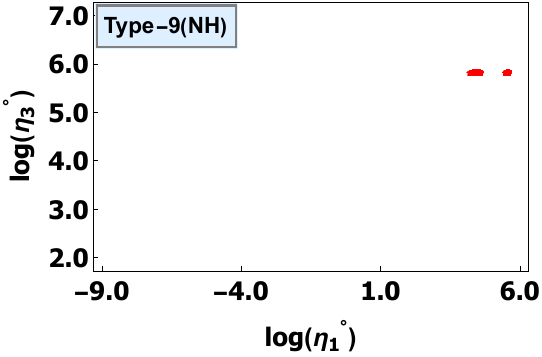}
        \caption{Type 9 Model}
    \end{subfigure}

    \vspace{10pt} 

    \begin{subfigure}{0.3\textwidth}
        \includegraphics[width=\textwidth]{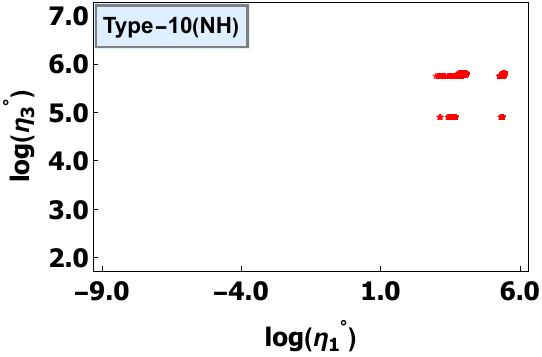}
        \caption{Type 10 Model}
    \end{subfigure}
    \hfill
    \begin{subfigure}{0.3\textwidth}
        \includegraphics[width=\textwidth]{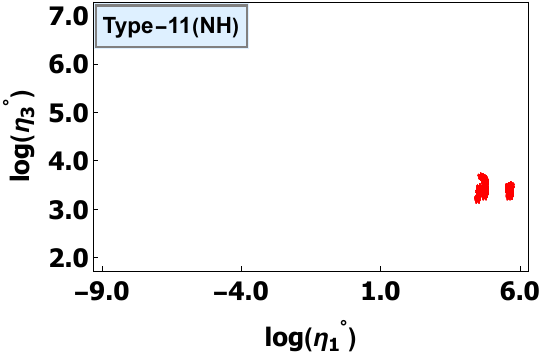}
        \caption{Type 11 Model}
    \end{subfigure}
    \hfill
    \begin{subfigure}{0.3\textwidth}
        \includegraphics[width=\textwidth]{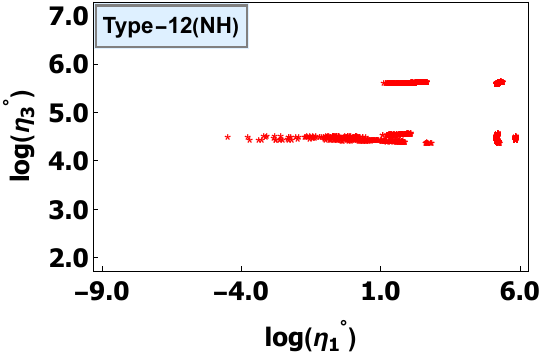}
        \caption{Type 12 Model}
    \end{subfigure}

    \caption{\footnotesize \label{8n}The correlation plots of $\log(\eta_3^\circ)$ vs $\log(\eta_1^\circ)$ for the 12 models under normal hierarchy. }
\end{figure}

\begin{figure}
    \centering
    \begin{subfigure}{0.32\textwidth}
        \includegraphics[width=\textwidth]{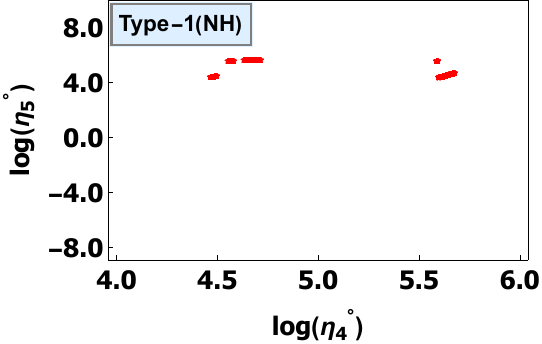}
        \caption{Type 1 Model}
    \end{subfigure}
    \hfill
    \begin{subfigure}{0.32\textwidth}
        \includegraphics[width=\textwidth]{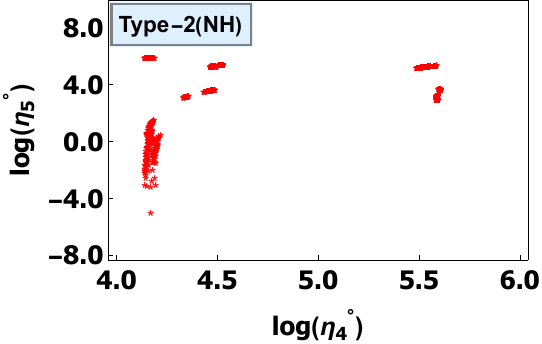}
        \caption{Type 2 Model}
    \end{subfigure}
    \hfill
    \begin{subfigure}{0.32\textwidth}
        \includegraphics[width=\textwidth]{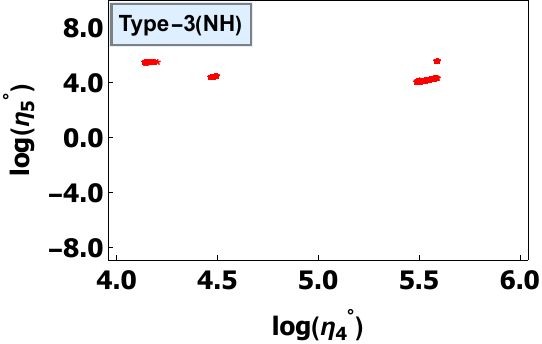}
        \caption{Type 3 Model}
    \end{subfigure}

    \vspace{10pt} 

    \begin{subfigure}{0.3\textwidth}
        \includegraphics[width=\textwidth]{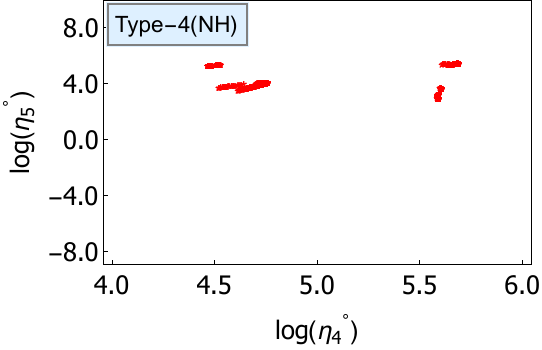}
        \caption{Type 4 Model}
    \end{subfigure}
    \hfill
    \begin{subfigure}{0.3\textwidth}
        \includegraphics[width=\textwidth]{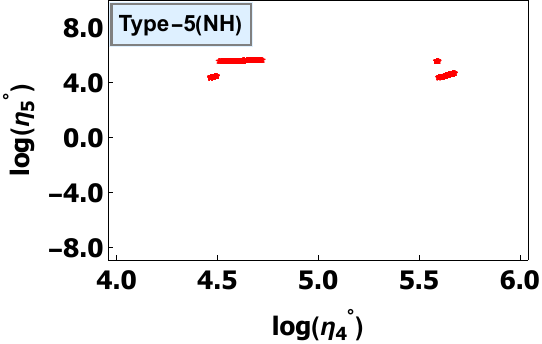}
        \caption{Type 5 Model}
    \end{subfigure}
    \hfill
    \begin{subfigure}{0.3\textwidth}
        \includegraphics[width=\textwidth]{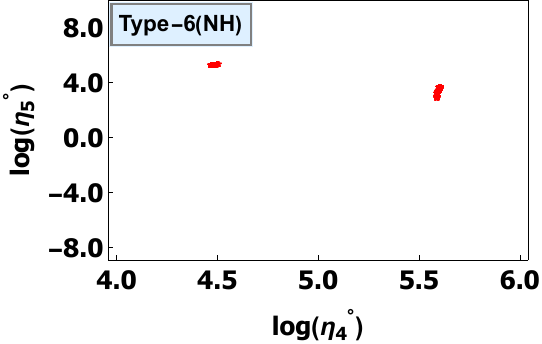}
        \caption{Type 6 Model}
    \end{subfigure}

    \vspace{10pt} 

    \begin{subfigure}{0.3\textwidth}
        \includegraphics[width=\textwidth]{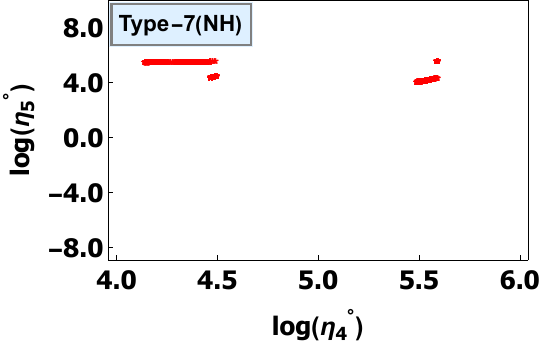}
        \caption{Type 7 Model}
    \end{subfigure}
    \hfill
    \begin{subfigure}{0.3\textwidth}
        \includegraphics[width=\textwidth]{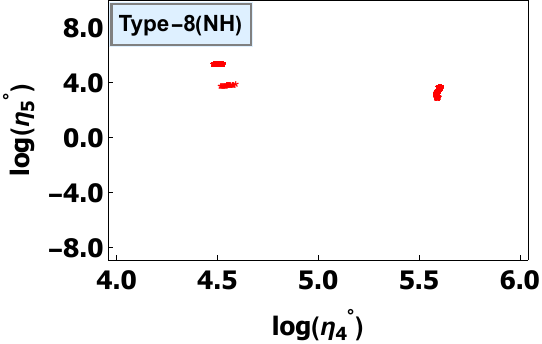}
        \caption{Type 8 Model}
    \end{subfigure}
    \hfill
    \begin{subfigure}{0.3\textwidth}
        \includegraphics[width=\textwidth]{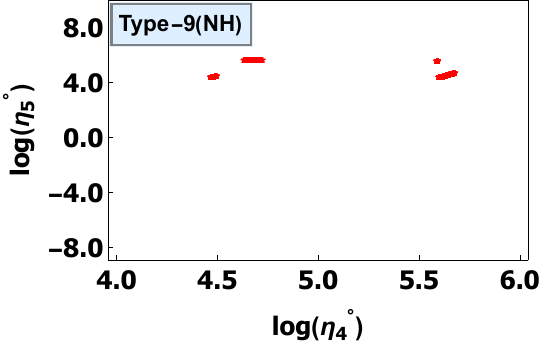}
        \caption{Type 9 Model}
    \end{subfigure}

    \vspace{10pt} 

    \begin{subfigure}{0.3\textwidth}
        \includegraphics[width=\textwidth]{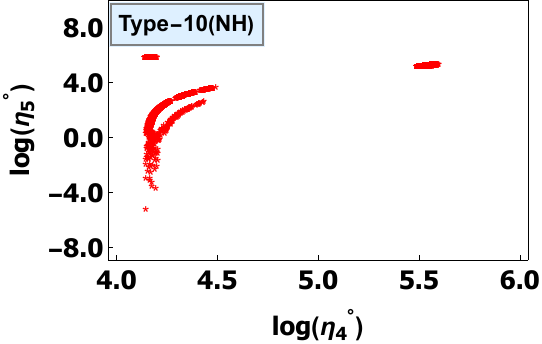}
        \caption{Type 10 Model}
    \end{subfigure}
    \hfill
    \begin{subfigure}{0.3\textwidth}
        \includegraphics[width=\textwidth]{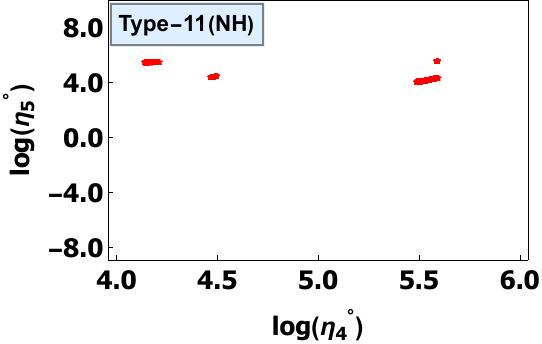}
        \caption{Type 11 Model}
    \end{subfigure}
    \hfill
    \begin{subfigure}{0.3\textwidth}
        \includegraphics[width=\textwidth]{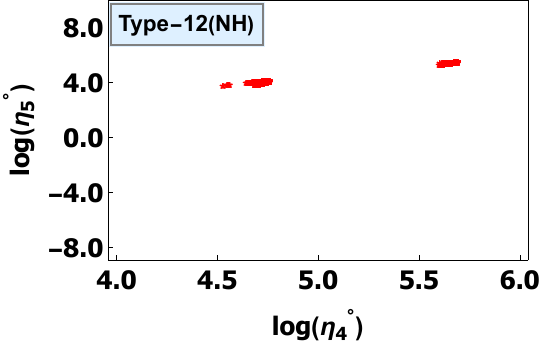}
        \caption{Type 12 Model}
    \end{subfigure}

    \caption{\footnotesize \label{9n}The correlation plots of $\log(\eta_5^\circ)$ vs $\log(\eta_4^\circ)$ for the 12 models normal hierarchy. }
\end{figure}

\begin{figure}
    \centering
    \begin{subfigure}{0.32\textwidth}
        \includegraphics[width=\textwidth]{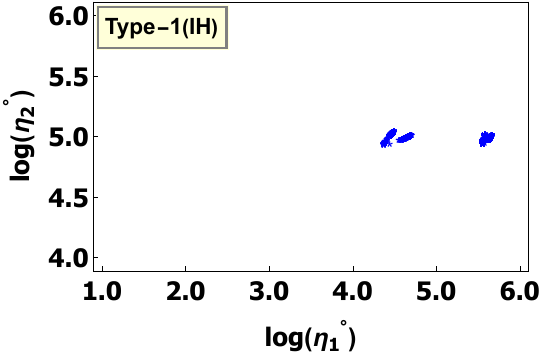}
        \caption{Type 1 Model}
    \end{subfigure}
    \hfill
    \begin{subfigure}{0.32\textwidth}
        \includegraphics[width=\textwidth]{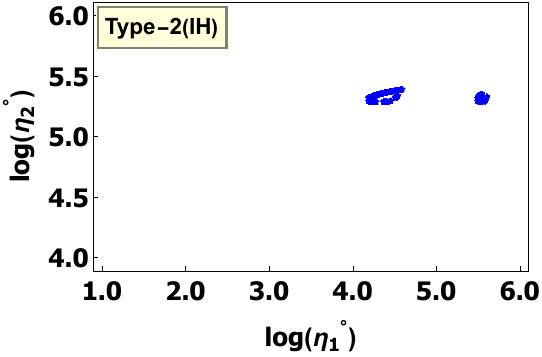}
        \caption{Type 2 Model}
    \end{subfigure}
    \hfill
    \begin{subfigure}{0.32\textwidth}
        \includegraphics[width=\textwidth]{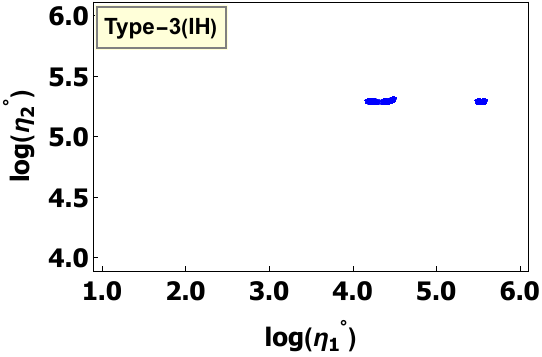}
        \caption{Type 3 Model}
    \end{subfigure}

    \vspace{10pt} 

    \begin{subfigure}{0.3\textwidth}
        \includegraphics[width=\textwidth]{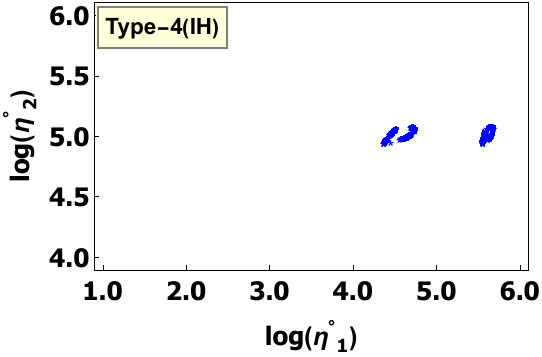}
        \caption{Type 4 Model}
    \end{subfigure}
    \hfill
    \begin{subfigure}{0.3\textwidth}
        \includegraphics[width=\textwidth]{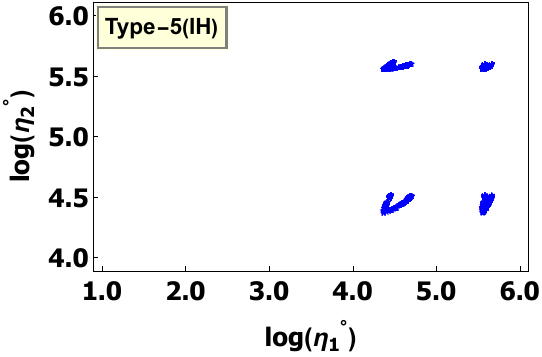}
        \caption{Type 5 Model}
    \end{subfigure}
    \hfill
    \begin{subfigure}{0.3\textwidth}
        \includegraphics[width=\textwidth]{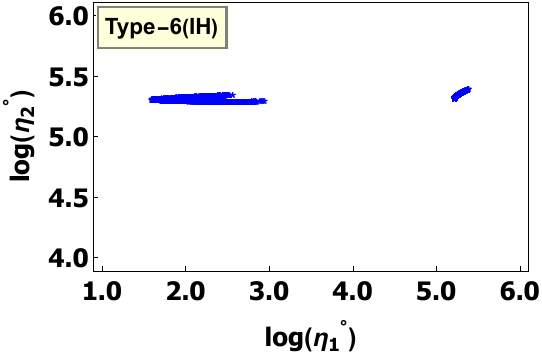}
        \caption{Type 6 Model}
    \end{subfigure}

    \vspace{10pt} 

    \begin{subfigure}{0.3\textwidth}
        \includegraphics[width=\textwidth]{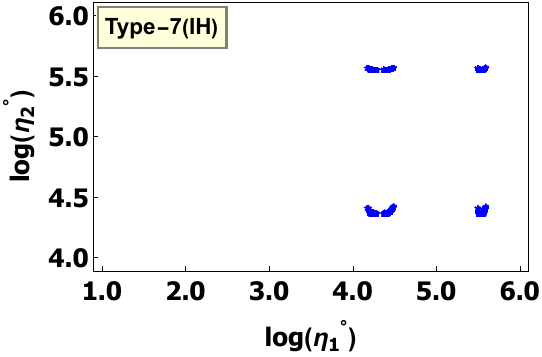}
        \caption{Type 7 Model}
    \end{subfigure}
    \hfill
    \begin{subfigure}{0.3\textwidth}
        \includegraphics[width=\textwidth]{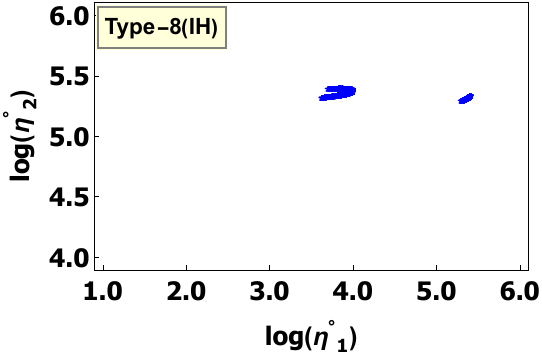}
        \caption{Type 8 Model}
    \end{subfigure}
    \hfill
    \begin{subfigure}{0.3\textwidth}
        \includegraphics[width=\textwidth]{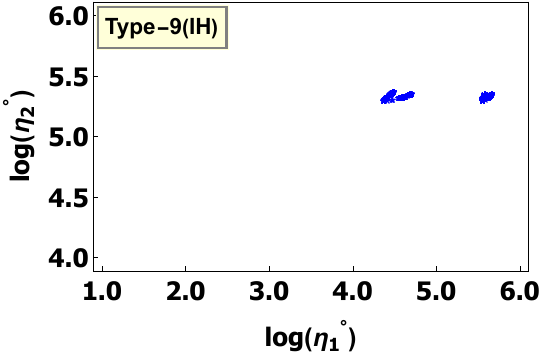}
        \caption{Type 9 Model}
    \end{subfigure}

    \vspace{10pt} 

    \begin{subfigure}{0.3\textwidth}
        \includegraphics[width=\textwidth]{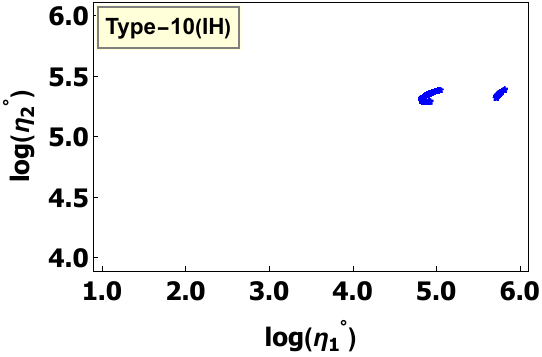}
        \caption{Type 10 Model}
    \end{subfigure}
    \hfill
    \begin{subfigure}{0.3\textwidth}
        \includegraphics[width=\textwidth]{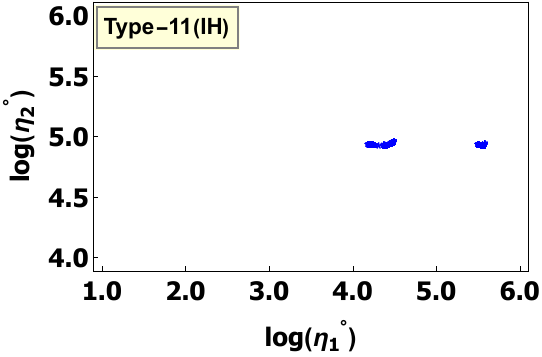}
        \caption{Type 11 Model}
    \end{subfigure}
    \hfill
    \begin{subfigure}{0.3\textwidth}
        \includegraphics[width=\textwidth]{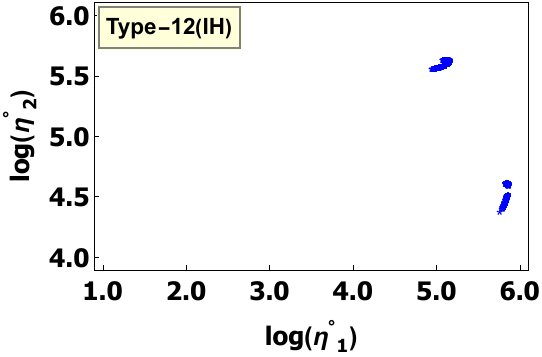}
        \caption{Type 12 Model}
    \end{subfigure}

    \caption{\footnotesize \label{7i}The correlation plots of $\log(\eta_2^\circ)$ vs $\log(\eta_1^\circ)$ for the 12 models under inverted hierarchy. }
\end{figure}

\begin{figure}
    \centering
    \begin{subfigure}{0.32\textwidth}
        \includegraphics[width=\textwidth]{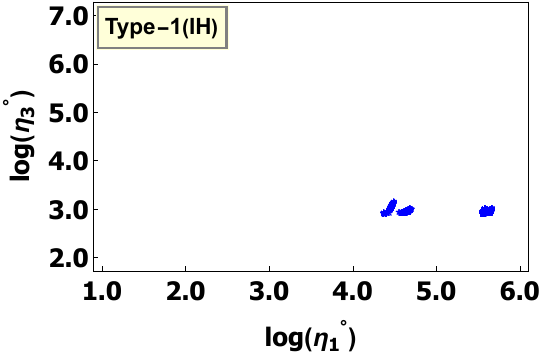}
        \caption{Type 1 Model}
    \end{subfigure}
    \hfill
    \begin{subfigure}{0.32\textwidth}
        \includegraphics[width=\textwidth]{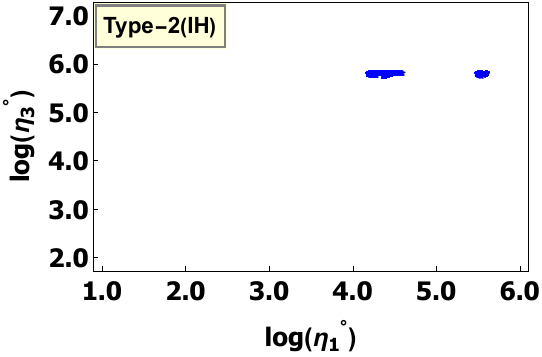}
        \caption{Type 2 Model}
    \end{subfigure}
    \hfill
    \begin{subfigure}{0.32\textwidth}
        \includegraphics[width=\textwidth]{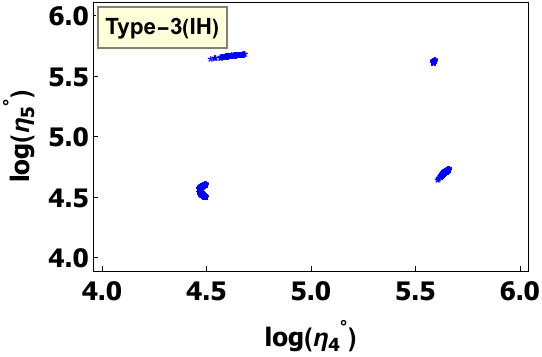}
        \caption{Type 3 Model}
    \end{subfigure}

    \vspace{10pt} 

    \begin{subfigure}{0.3\textwidth}
        \includegraphics[width=\textwidth]{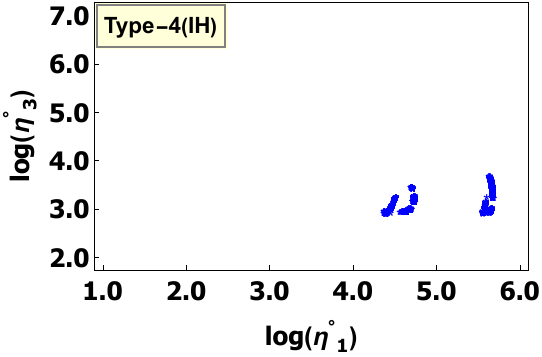}
        \caption{Type 4 Model}
    \end{subfigure}
    \hfill
    \begin{subfigure}{0.3\textwidth}
        \includegraphics[width=\textwidth]{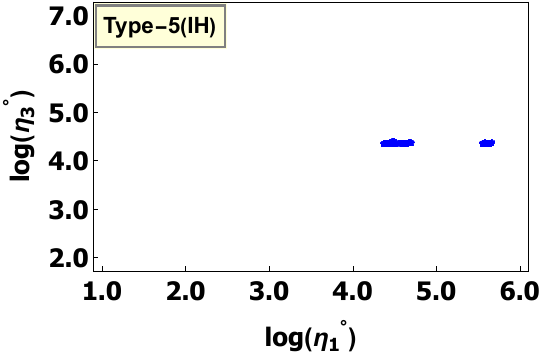}
        \caption{Type 5 Model}
    \end{subfigure}
    \hfill
    \begin{subfigure}{0.3\textwidth}
        \includegraphics[width=\textwidth]{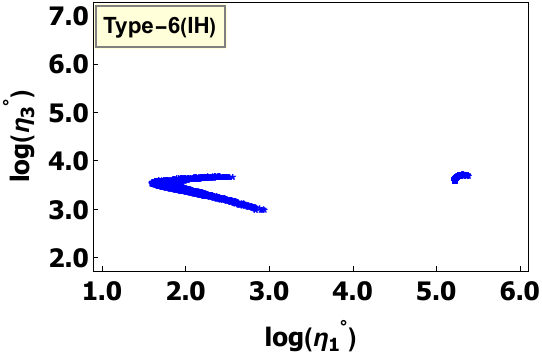}
        \caption{Type 6 Model}
    \end{subfigure}

    \vspace{10pt} 

    \begin{subfigure}{0.3\textwidth}
        \includegraphics[width=\textwidth]{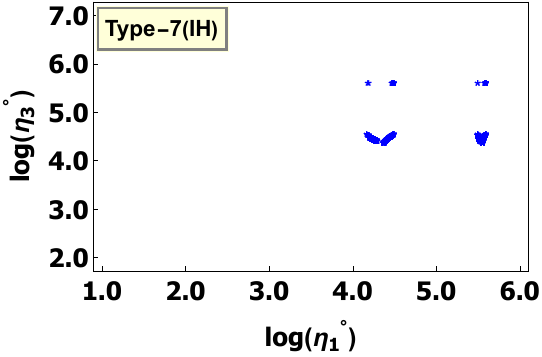}
        \caption{Type 7 Model}
    \end{subfigure}
    \hfill
    \begin{subfigure}{0.3\textwidth}
        \includegraphics[width=\textwidth]{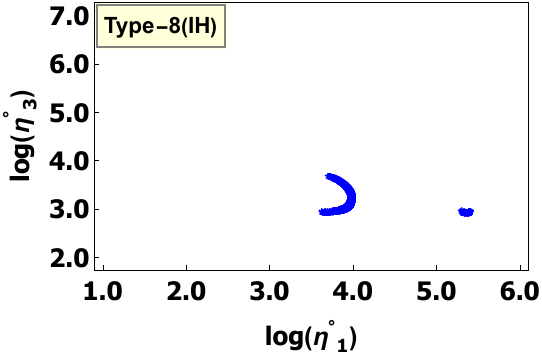}
        \caption{Type 8 Model}
    \end{subfigure}
    \hfill
    \begin{subfigure}{0.3\textwidth}
        \includegraphics[width=\textwidth]{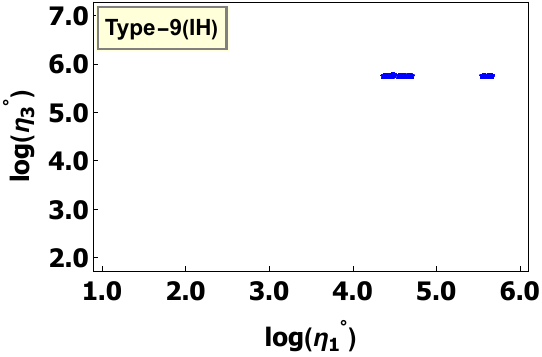}
        \caption{Type 9 Model}
    \end{subfigure}

    \vspace{10pt} 

    \begin{subfigure}{0.3\textwidth}
        \includegraphics[width=\textwidth]{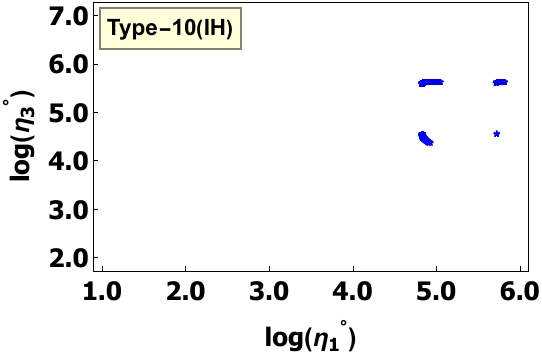}
        \caption{Type 10 Model}
    \end{subfigure}
    \hfill
    \begin{subfigure}{0.3\textwidth}
        \includegraphics[width=\textwidth]{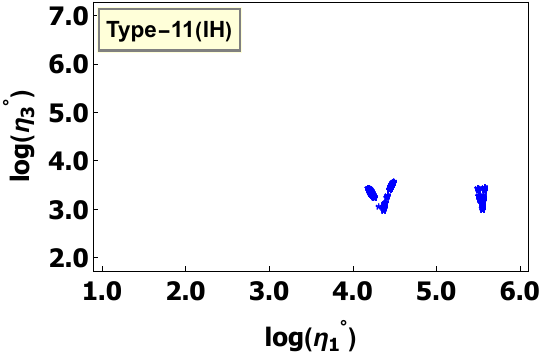}
        \caption{Type 11 Model}
    \end{subfigure}
    \hfill
    \begin{subfigure}{0.3\textwidth}
        \includegraphics[width=\textwidth]{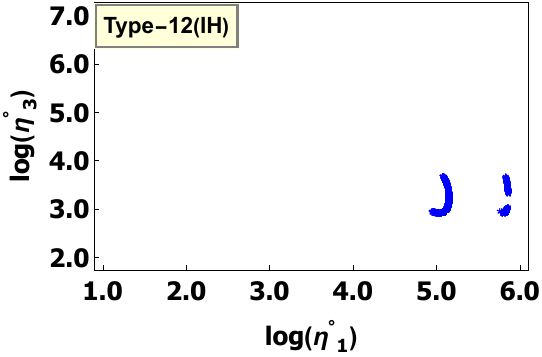}
        \caption{Type 12 Model}
    \end{subfigure}

    \caption{\footnotesize \label{8i}The correlation plots of $\log(\eta_3^\circ)$ vs $\log(\eta_1^\circ)$ for the 12 models under inverted hierarchy.}
\end{figure}

\begin{figure}
    \centering
    \begin{subfigure}{0.32\textwidth}
        \includegraphics[width=\textwidth]{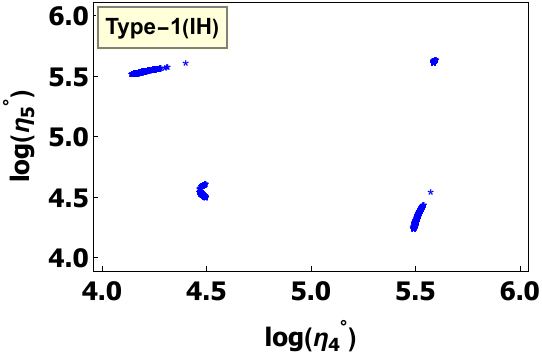}
        \caption{Type 1 Model}
    \end{subfigure}
    \hfill
    \begin{subfigure}{0.32\textwidth}
        \includegraphics[width=\textwidth]{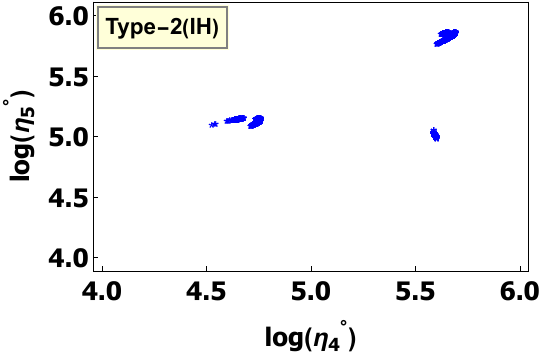}
        \caption{Type 2 Model}
    \end{subfigure}
    \hfill
    \begin{subfigure}{0.32\textwidth}
        \includegraphics[width=\textwidth]{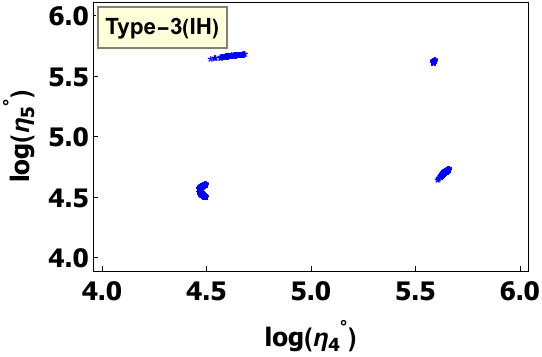}
        \caption{Type 3 Model}
    \end{subfigure}

    \vspace{10pt} 

    \begin{subfigure}{0.3\textwidth}
        \includegraphics[width=\textwidth]{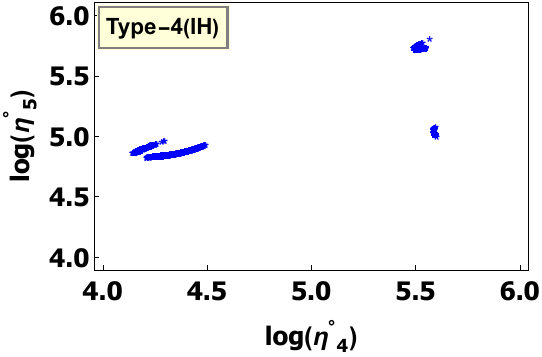}
        \caption{Type 4 Model}
    \end{subfigure}
    \hfill
    \begin{subfigure}{0.3\textwidth}
        \includegraphics[width=\textwidth]{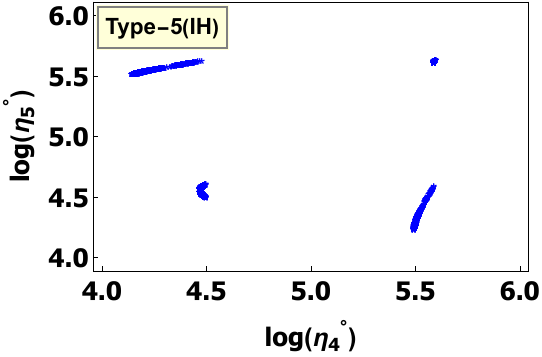}
        \caption{Type 5 Model}
    \end{subfigure}
    \hfill
    \begin{subfigure}{0.3\textwidth}
        \includegraphics[width=\textwidth]{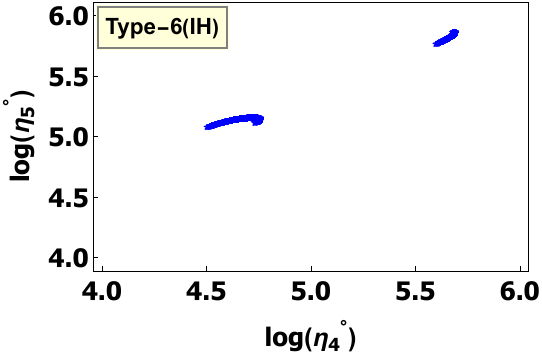}
        \caption{Type 6 Model}
    \end{subfigure}

    \vspace{10pt} 

    \begin{subfigure}{0.3\textwidth}
        \includegraphics[width=\textwidth]{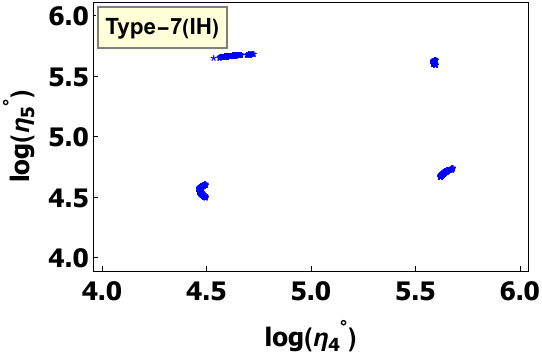}
        \caption{Type 7 Model}
    \end{subfigure}
    \hfill
    \begin{subfigure}{0.3\textwidth}
        \includegraphics[width=\textwidth]{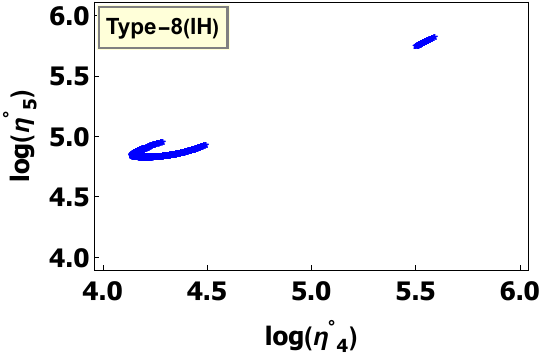}
        \caption{Type 8 Model}
    \end{subfigure}
    \hfill
    \begin{subfigure}{0.3\textwidth}
        \includegraphics[width=\textwidth]{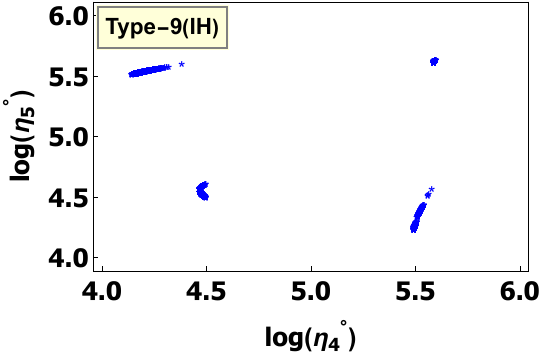}
        \caption{Type 9 Model}
    \end{subfigure}

    \vspace{10pt} 

    \begin{subfigure}{0.3\textwidth}
        \includegraphics[width=\textwidth]{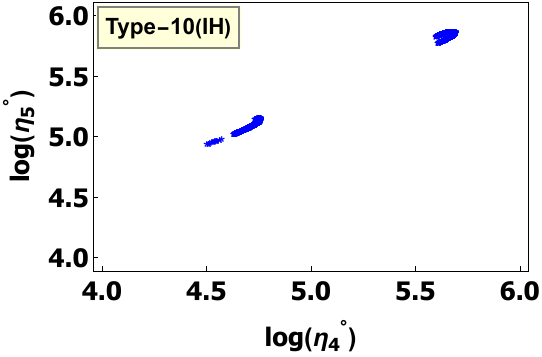}
        \caption{Type 10 Model}
    \end{subfigure}
    \hfill
    \begin{subfigure}{0.3\textwidth}
        \includegraphics[width=\textwidth]{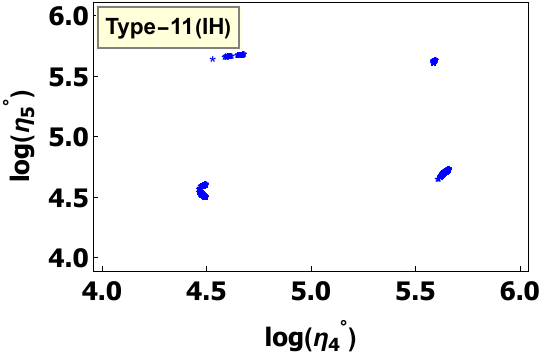}
        \caption{Type 11 Model}
    \end{subfigure}
    \hfill
    \begin{subfigure}{0.3\textwidth}
        \includegraphics[width=\textwidth]{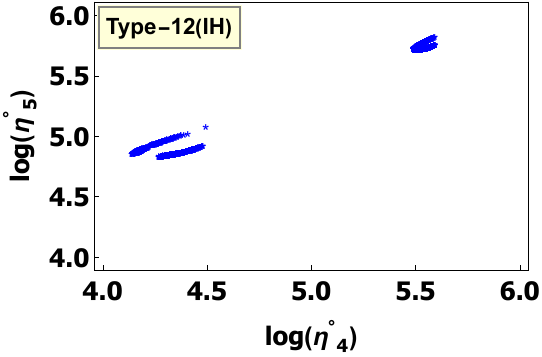}
        \caption{Type 12 Model}
    \end{subfigure}

    \caption{\footnotesize \label{9i}The correlation plots of $\log(\eta_5^\circ)$ vs $\log(\eta_4^\circ)$ for the 12 models under inverted hierarchy. }
\end{figure}

 Besides, we also have done correlation plots among the model parameters (Figs.\,(\ref{Cn}) and (\ref{Ci})) and the unphysical phases (Figs.\,(\ref{7n})-(\ref{9n}) and (\ref{7i})-(\ref{9i})).
 
 \begin{table}
\centering
\footnotesize
\begin{tabular*}{\textwidth}{@{\extracolsep{\fill}}ccc||ccc}
\hline
 \multicolumn{3}{c||}{\textbf{NH}} & \multicolumn{3}{c}{\textbf{IH}} \\
  \hline
\textbf{Model Type} & $\mathbf{\theta_{23}}$ & \textbf{Favoured Octant} & \textbf{Model Type} & $\mathbf{\theta_{23}}$&  \textbf{Favoured Octant}  \\ 
  \hline
 Type 1&($44.94^\circ$, $51.88^\circ$)&  Predominantly HO &Type 1&($39.91^\circ$, $43.20^\circ$) & LO  \\
 \hline
 Type 2&($39.64^\circ$, $51.82^\circ$)& Both &Type 2&($39.91^\circ$, $52.01^\circ$) & Both\\
 \hline
 Type 3&($39.64^\circ$, $45.68^\circ$)& Predominantly LO &Type 3&($47.51^\circ$, $52.01^\circ$) & HO \\
 \hline
 Type 4&($39.64^\circ$, $51.88^\circ$) & Both &Type 4&($39.91^\circ$, $52.01^\circ$) &  Both\\
 \hline
 Type 5&($44.77^\circ$, $51.88^\circ$)& Predominantly HO &Type 5&($39.91^\circ$, $42.83^\circ$) & LO \\
 \hline
 Type 6&($39.64^\circ$, $51.88^\circ$)& Both & Type 6&($39.91^\circ$, $52.01^\circ$) & Both \\
 \hline
 Type 7&($39.64^\circ$, $45.68^\circ$)& Predominantly LO &Type 7&($44.82^\circ$, $51.81^\circ$) & Predominantly HO \\
 \hline
 Type 8&($39.64^\circ$, $47.18^\circ$)& Both &Type 8&($39.91^\circ$, $52.01^\circ$) & Both \\
 \hline
 Type 9&($44.76^\circ$, $51.88^\circ$)& Predominantly HO &Type 9&($39.91^\circ$, $42.70^\circ$) & LO  \\
 \hline
 Type 10&($39.64^\circ$, $51.94^\circ$)& Both &Type 10&($39.91^\circ$, $52.01^\circ$) & Both \\
 \hline
 Type 11&($39.64^\circ$, $45.57^\circ$)& Predominantly LO &Type 11&($47.33^\circ$, $51.64^\circ$) & HO \\
 \hline
 Type 12&($39.64^\circ$, $51.94^\circ$)& Both & Type 12&($39.91^\circ$, $52.01^\circ$) & Both\\
 \hline
\end{tabular*}
\footnotesize
\caption{\footnotesize \label{table5a} The ranges of $\theta_{23}$ under various types of models under  NH and IH. The LO and HO mean lower and higher octants respectively.}
\end{table}

\section{\label{Section 4}{Charged Lepton Flavour Violation and Dark Matter}}
Besides the prediction of the observable parameters,  the proposed framework is also capable of  accommodating various CLFV  and DM annihilation processes,  and making predictions for their branching ratio (BR) and annihilation cross section respectively. In this section, we study how the model predicts those values, particularly in the light of PCLC. We start with CLFV and then move to DM.

In our model, CLFV processes like $\mu \longrightarrow e\,\gamma$, \, $\tau \longrightarrow e\,\gamma$,\, $\tau \longrightarrow \mu\,\gamma$ etc., arise from the Lagrangian terms  $\tilde{y}_4\,(\bar{l}_{L}\,N_{R})_{3_s}\, \phi $ and $\tilde{y}_5\,(\bar{l}_{L}\,N_{R})_{3_a}\, \phi,$ where they are associated with Yukawa couplings $\tilde{y}_4$ and $\tilde{y}_5$. These processes are mediated by the heavy neutral fermion singlet, $N_R$ and  the complex scalar field, $\phi$. 

The formula\,\cite{Guo:2020qin} for the BR of a CLFV process like $l_\alpha \longrightarrow \beta\,\gamma$ is given by

\begin{eqnarray}
BR(l_\alpha \longrightarrow l_\beta\,\gamma)= \frac{3\,\alpha_{em}\, BR(l_\alpha \longrightarrow l_\beta\,\nu_\alpha\,\bar{\nu}_\beta)}{16\,\pi\, G^2} \left|\sum_i \frac{(Y_\phi)_{\beta\,i}\,(Y^*_\phi)_{\alpha\,i}}{m_\phi^2}\,j\left( \frac{m^2_{N_i}}{m^2_\phi} \right) \right|^2,
\end{eqnarray}
with

\begin{equation}
j(r)=\frac{1-6r+3r^2+2r^3-6r^2\,\ln r}{12(1-r)^4},\nonumber
\end{equation}

where $Y_\phi$ is the Yukawa matrix in flavour basis corresponding to the Yukawa couplings that mediate the cLFV process. In our work,  $Y_\phi$ is given by 

\begin{eqnarray}
\label{cd2}
Y_\phi = U_L\,Y_{\phi_s},
\end{eqnarray}
where $Y_{\phi_s}$ is the Yukawa matrix for the $\phi$ channel in symmetry basis and is given by
\begin{eqnarray}
Y_{\phi_s}=\begin{pmatrix}
0 & \tilde{y}_{\phi_s}k_2 & \tilde{y}_{\phi_a}k_1 \\
 \tilde{y}_{\phi_a}k_2 & 0 & 0\\
 \tilde{y}_{\phi_s}k_1 & 0&0
\end{pmatrix}, \nonumber
\end{eqnarray}
with $\tilde{y}_{\phi_s}=\tilde{y}_4+\tilde{y}_5=|\tilde{y}_{\phi_s} | \exp(i \phi_s)$ and $\tilde{y}_{\phi_a}=\tilde{y}_4-\tilde{y}_5=|\tilde{y}_{\phi_a} | \exp(i \phi_a)$.  The presence of $U_L$ in Eq.\,(\ref{cd2}) gives us the opportunity to see how PCLC might affect the BR of the CLFV processes.

If we use parametrization of $U_L$, as given by Eq.\,(\ref{bf}), we find that the expressions of BR do not change, provided we make the following redefinition of the phases,
\begin{eqnarray}
\label{cd3}
\theta_a \longrightarrow \theta_a + \kappa_i, \nonumber \\
\phi_s \longrightarrow \phi_s + \kappa_i, 
\end{eqnarray}
with $i=1,2,3$. Since all these phases are free parameters, the parametrization of $U_L$ does not have much impact on the result as they can be absorbed by redefinition of $\theta_a$ and $\phi_s$.  This is true for all other parametrizations of $U_L$ too. 

\begin{figure}
    \begin{subfigure}{0.32\textwidth}
        \includegraphics[width=\textwidth]{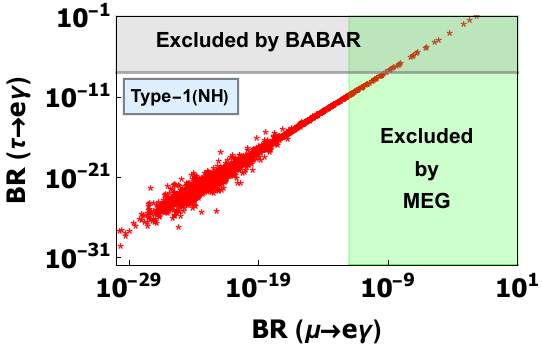}
        \caption{ }
    \end{subfigure}
    \hfill
    \begin{subfigure}{0.32\textwidth}
        \includegraphics[width=\textwidth]{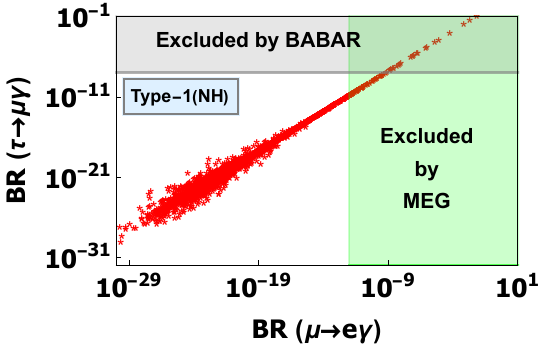}
        \caption{}
    \end{subfigure}
    \hfill
    \begin{subfigure}{0.32\textwidth}
        \includegraphics[width=\textwidth]{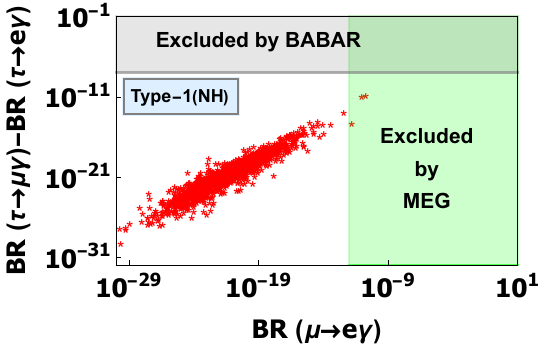}
        \caption{}
    \end{subfigure}

    \vspace{10pt} 

    \begin{subfigure}{0.32\textwidth}
        \includegraphics[width=\textwidth]{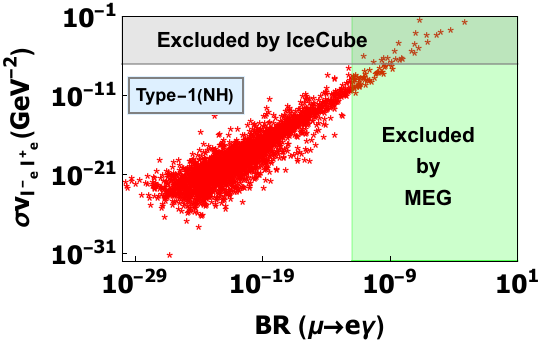}
        \caption{}
    \end{subfigure}
    \hfill
    \begin{subfigure}{0.32\textwidth}
        \includegraphics[width=\textwidth]{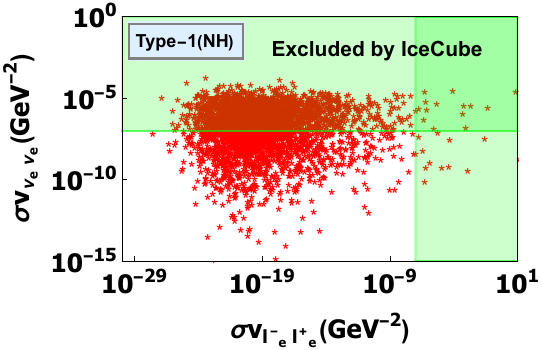}
        \caption{}
    \end{subfigure}
    \hfill
     \begin{subfigure}{0.32\textwidth}
        \includegraphics[width=\textwidth]{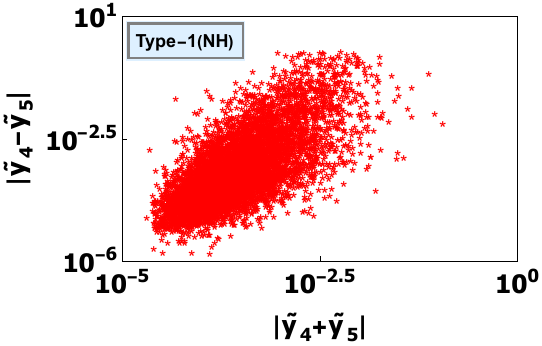}
        \caption{}
    \end{subfigure}

    \begin{flushleft}
    \caption{\label{cd} \footnotesize The correlation plots of (a) the branching ratios of the processes $\tau \longrightarrow e\,\gamma$ and $\mu \longrightarrow e\,\gamma$, (b) the branching ratios of the processes $\tau \longrightarrow \mu\,\gamma$ and $\mu \longrightarrow e\,\gamma$, (c) the difference between the branching ratios of the processes $\tau \longrightarrow \mu\,\gamma$ and $\tau \longrightarrow e\,\gamma$ against $\mu \longrightarrow e\,\gamma$, (d) the cross section of the DM annihilation process $N_1 \bar{N}_1 \longrightarrow l_e^- l_e^+$ ($\phi$ channel) and branching ratio of the process $\mu \longrightarrow e\,\gamma$, (e) the cross sections of DM annihilation processes $N_1 N_1 \longrightarrow \nu_{eR} \nu_{eR}$ ($\chi$ channel) and $N_1 \bar{N}_1 \longrightarrow l_e^- l_e^+$ ($\phi$ channel), and (f) $| \tilde{y}_4-\tilde{y}_5|$ and $|\tilde{y}_4+\tilde{y}_5 |$ . All models under both NH and IH show similar plots. For DM annihilation cross section, we have $1 \text{GeV}^{-2} \sim 1.2 \times 10^{-17} \text{cm}^3 \text{s}^{-1}$}
    \end{flushleft}
\end{figure}

Likewise, permutation of $U_L$ (see Table\,\ref{table4a}) also does not have significant effect on the results. When we use permutation of $U_L$, we find that the all the models under both NH and IH give almost similar results for the BR's. So, we have shown only the Type 1 model under NH in Fig.\,(\ref{cd}).

Apart from CLFV, the framework also contains the provision for the discussion of  DM.
We consider
the lightest of the neutral fermion triplet, i.e., $N_1$ to be a DM candidate. The framework provides two channels for $N_1$ to annihilate: the $\phi$ channel mediated by $\phi$ and the $\chi$ channel mediated by $\chi$.

The annihilation cross sections\,\cite{Guo:2020qin} of $N_1$ for $\phi$ channel is given by
\begin{eqnarray}
\sigma(N_1\,\bar{N}_1 \longrightarrow e_L^-\,e_L^+)v_{\text{rel}} = \sigma(N_1\,\bar{N}_1 \longrightarrow \nu_{e_L}\,\bar{\nu}_{e_L})v_{\text{rel}} =\frac{\left|(Y_\phi)_{11} (Y^*_\phi)_{11} \right|^2}{32 \pi} \frac{m_N^2}{(m_N^2+m_\phi^2)^2},
\end{eqnarray}

and, the same for $\chi$ channel is given by
\begin{eqnarray}
\label{chi1}
\sigma(N_1\,\bar{N}_1 \longrightarrow \nu_{e_R}\,\bar{\nu}_{e_R})v_{\text{rel}} &=& \frac{\left|(Y_\chi)_{11} (Y^*_\chi)_{11} \right|^2}{32 \pi} \frac{m_N^2}{(m_N^2+m_\chi^2)^2},\nonumber \\
\sigma(N_1\,N_1 \longrightarrow \nu_{e_R}\,\nu_{e_R})v_{\text{rel}} &=& \frac{\left|(Y_\chi)_{11} (Y_\chi)_{11} \right|^2}{16 \pi} \frac{m_N^2}{(m_N^2+m_\chi^2)^2}.
\end{eqnarray}

In Eq.\,(\ref{chi1}), $Y_\chi$ is the corresponding Yukawa matrix in the flavour basis comprising coupling constants $\tilde{y}_6$ and $\tilde{y}_7$. It is given by
\begin{eqnarray}
Y_\chi = U_L\,Y_{\chi_s},
\end{eqnarray}
where $Y_{\chi_s}$ is the Yukawa matrix for the $\chi$ channel in symmetry basis as given below,
\begin{eqnarray}
Y_{\chi_s}=\begin{pmatrix}
0 & \tilde{y}_{\chi_s}f_2 & \tilde{y}_{\chi_a}f_1 \\
 \tilde{y}_{\chi_a}f_2 & 0 & 0\\
 \tilde{y}_{\chi_s}f_1 & 0&0
\end{pmatrix},
\end{eqnarray}
with $\tilde{y}_{\chi_s}=\tilde{y}_6+\tilde{y}_7=|\tilde{y}_{\chi_s} | \exp(i \chi_s)$ and $\tilde{y}_{\chi_a}=\tilde{y}_6-\tilde{y}_7=|\tilde{y}_{\chi_a} | \exp(i \chi_a)$. As we have seen in the case of $Y_{\phi}$ (Eq.\,(\ref{cd2})), PCLC does not have appreciable consequences on the results in this case also.  Here, the parametrization phases $\kappa_i$'s can be absorbed into $\chi_s$ and $\chi_a$ in the same way as Eq.\,(\ref{cd3}). As for the permutation of $U_L$, we have the same situation as in the case of CLFV, namely all the models showing similar patterns of results (see Fig.\,(\ref{cd})).

As we see in both the CLFV and DM sectors,  there are a lot of free parameters in terms of Yukawa coupling constants, phases and masses of the mediating scalars and the DM candidate.
But we can express at least two of them, $\tilde{y}_{\phi_s}$ and $\tilde{y}_{\phi_a}$ in terms of $a$ and $c$ using Eqs.\,(\ref{equation:E}) and (\ref{equation:F}). They are given as,

\begin{eqnarray}
\label{cd4}
\tilde{y}_{\phi_s}&=&\frac{\sqrt{c}\,e^{i\,\phi_s}}{\frac{v_\phi v_\chi}{m_N}\sqrt{(f_2\,k_1\,|\tilde{y}_{\chi_s}|)^2+(f_1\,k_1\,|\tilde{y}_{\chi_a}|)^2}},\nonumber \\
\tilde{y}_{\phi_a}&=&\frac{1}{\frac{v_\phi v_\chi}{m_N}\,|\tilde{y}_{\chi_s}|\,e^{i\,\chi_s}} \left[\frac{\sqrt{a}\,e^{i\,\theta_a}}{f_1\,k_1} -\frac{k_2\,f_2\,|\tilde{y}_{\chi_a}|\,e^{i(\phi_s+\chi_a)}\,\sqrt{c}}{k_1^2\,f_1^2\,\sqrt{|\tilde{y}_{\chi_a}|^2+\frac{|\tilde{y}_{\chi_s}|^2\,f_2^2}{f_1^2}}} \right].
\end{eqnarray}
For the remaining parameters, we choose some suitable ranges as shown in the Table \ref{table6a}.

\begin{table}
\begin{center}
\footnotesize
\begin{tabular*}{\textwidth}{@{\extracolsep{\fill}}cccccccc}
\hline
Parameters & $\chi_s$ & $\chi_a$ & $\phi_s$ & $\theta_a$ & $f_1$ & $f_2$ & $k_1$ \\
\hline
Ranges & $(0,2\pi)$ & $(0,2\pi)$ & $(0,2\pi)$ & $(0,2\pi)$ & $(0.1,9)$ & $(0.1,9)$ & $(0.1,9)$ \\
\hline
\hline
Parameters & $k_2$ & $|\tilde{y}_{\chi_s}|$ & $|\tilde{y}_{\chi_a} |$ & $v_\phi v_\chi/m_N$ (GeV) & $m_N$ (GeV) & $m_\phi$ (GeV) & $m_\chi$ (GeV) \\
\hline
Ranges & $(0.1,9)$ & $(0,2)$ & $(0,2)$ & $(1,10)$ & $(10^3, 1.5 \times 10^3)$  & $(10^3, 1.5 \times 10^3)$  & $(10^3, 1.5 \times 10^3)$  \\
\hline
\end{tabular*}
\caption{\footnotesize \label{table6a} The ranges of the free parameters present in the expressions of BR's of CLFV processes, DM annihilation cross sections and the Yukawa coupling constants, $\tilde{y}_4\, \text{and}\, \tilde{y}_5$.}
\end{center}
\end{table}

The Fig.\,(\ref{cd}) shows the graphical representations of BR's of the CLFV processes, annihilation cross sections of the DM and the absolute values of $\tilde{y}_4+\tilde{y}_5\, (\tilde{y}_{\phi_s})$ and $\tilde{y}_4-\tilde{y}_5 \,(\tilde{y}_{\phi_a})$ for the Type 1 model under NH. It is worth mentioning that we choose the mass range of the DM candidate and the scalars to be between $1000$ GeV and $1500$ GeV. The results are also compared with those reported by several recent and ongoing experiments\,\cite{IceCube:2023ies,Mori:2016vwi,Venturini:2024dvx,Venturini:2024keu,BaBar:2009hkt}. Some important features of the results are discussed below,

\begin{itemize}
\item The annihilation cross sections for the processes for $\chi$ channel, $N_1\,N_1 \longrightarrow \nu_{eR}\,\nu_{eR}$ and $N_1\,\bar{N}_1 \longrightarrow \nu_{eR}\,\bar{\nu}_{eR}$ are the same.
\item The work also predicts new results for the BR's that can be explored with upcoming experiments. The current limits for the BR's for the processes $\mu \longrightarrow e\,\gamma$, $\tau \longrightarrow e\,\gamma$ and $\tau \longrightarrow \mu\,\gamma$ are respectively $3.1 \times 10^{-13}$\,\cite{Venturini:2024dvx} ,$3.3 \times 10^{-8}$ and $4.2 \times 10^{-8}$\,\cite{BaBar:2009hkt}.  We see that our proposed framework predicts BR's for these processes over a long range, right from the current upper bounds to as low as to the order of $10^{-29}$ (for $\mu \longrightarrow e\,\gamma$) and $10^{-28}$(for $\tau \longrightarrow e/\mu\,\gamma$). Although currently beyond experimental capacity, at least some values are expected to be probed by upcoming experiments, e.g.,MEG-II\,\cite{Venturini:2024dvx} and future experiments.
\item The framework also predicts a wide range of values for DM annihilation cross sections for both the $\phi$ and $\chi$ channels. For the chosen mass ranges of the mediating scalars and the DM candidate (refer to Table \ref{table6a}),  the annihilation cross section for the $\phi$ channel ranges from the current limit of $10^{-24} \text{cm}^3\, \text{s}^{-1}$\,\cite{IceCube:2023ies} to $10^{-46} \text{cm}^3\, \text{s}^{-1}$. For the $\chi$ channel, the annihilation cross section covers a range from the current limit to $10^{-32} \text{cm}^3\, \text{s}^{-1}$. It is notable that the lower limit for $\phi$ channel is far lower than that of $\chi$ channel.  We expect the results to be probed with increased sensitivities of the experiments.
\item For the Yukawa coupling constants, $\tilde{y}_4$ and $\tilde{y}_5$, related to $\phi$ channel, our proposed model predicts $|\tilde{y}_4+\tilde{y}_5 |$ and $|\tilde{y}_4-\tilde{y}_5 |$. The values of these coupling constants are also found across a large range, with approximately $10^{-5} < |\tilde{y}_4+\tilde{y}_5 | < 1$ and $10^{-6} < |\tilde{y}_4-\tilde{y}_5 | < 1$, as seen in Fig.\,(\ref{cd}).
\end{itemize}

\section{\label{Section 5} {Summary and Discussion}}


In this work, we have introduced a framework to explain the masses of $\nu$'s using Dirac Type \Romannum{1} seesaw mechanism under the assumption that neutrinos are Dirac particles. 
A significant feature of our work lies in augmenting the Standard Model (SM) group structure with three discrete symmetry groups: one non Abelian group, $A_4$ and two cyclic groups, \(Z_3\) and \(Z_{10}\). The \(Z_3\) symmetry is instrumental in forbidding a Majorana mass term for $\nu$'s, ensuring the Dirac nature of their mass, while \(Z_{10}\) plays a crucial role in preventing Dirac mass terms
at the tree level. The $A_4$ group, in contrast,  enables a predictive structure for the $\nu$ mass matrix.

Beyond the SM Higgs, we have introduced two additional scalar fields, $\phi$ and $\chi$, whose VEV alignments play a pivotal role in shaping the structure of $M_\nu$. Notably, similar VEV alignments for these fields yield analogous structures for both $M_\nu$ and $M_\nu M_\nu^\dagger$. However, differing VEV alignments lead to distinct structures, a feature that might influence the predictions derived from the texture.

We have also studied PCLC, i.e., how the free parameters in the CL sector affects the predictions related to $\nu$ by moulding $U_L$ into various forms through parametrization and linear algebraic operations like permutation of columns and rows.  We have seen that both NH and IH of $\nu$ masses can be accommodated in the same framework by merely applying a particular column transformation to $U_\nu$.  It is also notable that the model predicts strict NH and IH, i.e., $m_1=0$ for the former, and, $m_3=0$ for the latter.

Under both NH and IH, our framework offers a sharp ranges of values for the solar mixing angle $\theta_{12}$.  Furthermore, under different permutations of the diagonalizing matrices, the models predict distinct octant preferences for the atmospheric mixing angle \(\theta_{23}\), revealing an interesting connection between theoretical structures and experimental observables.  The Dirac CP phase, $\delta$ also shows some interesting patterns in various types of models.

Beyond neutrino mass and mixing, our framework extends to address CLFV, predicting branching ratios for various processes. These predictions range from the current experimental limits to new values that can be tested in future experiments.
Moreover, the framework has implications for DM, offering new values for DM annihilation cross-sections. These predictions provide an exciting opportunity to test the model in future DM detection experiments, offering a bridge between neutrino physics and cosmology. The model also predicts the Yukawa coupling constants ($\tilde{y}_4$ and $\tilde{y}_5$) related to CLFV and one channel of DM annihilation.

Although the proposed framework is based on Type \Romannum{1} mechanism for $\nu$ mass generation, loop contributions might arise from quartic coupling terms, such as $\phi^\dagger \phi \chi^\dagger \chi$ in the scalar potential (Appendix \ref{Section 8}).  Such contributions are neglected in this work for simplicity.

In the field of phenomenology, our proposed work attempts to underline the power of free parameters in shaping predictions of the observable parameters.  We believe that it could open up new avenues in model building, relating results to intricacies of linear algebra.
\section*{Acknowledgement}
The authors thank Dr Salvador Centelles Chuli\'{a}, Heidelberg, Max Planck Institute and Dr Shu-Yan Guo, Yantai University for important discussions during the preparation of the manuscript. STG also thanks Pralay Chakraborty and Manansh Dey, Gauhati University for their help in some calculations and plots in the manuscript.
\newpage
\appendix
\section{\label{A4} $A_4$ Group}
In this section, we discuss the multiplication rules for $A_4$ group we have used in the manuscript based on a particular basis for representation.

All elements of the group $A_4$ can be written as products of the generators, $s$ and $t$, which satisfy the following relation\,\cite{Ishimori:2012zz},

\begin{equation}
s^2=t^3=(s t)^3=e. \nonumber
\end{equation}

If the we take $s$ and $t$ as,

\begin{equation}
s=\begin{pmatrix}
1 & 0 & 0\\
0 & -1 & 0 \\
0 & 0 & -1
\end{pmatrix}, \quad t=\begin{pmatrix}
0 & 0 & 1\\
1 & 0 & 0\\
0 & 1 & 0
\end{pmatrix}, \nonumber
\end{equation}

then the multiplication rules in terms of the irreducible representations are given as,

\begin{eqnarray}
\begin{pmatrix}
a_1 \\
a_2 \\
a_3
\end{pmatrix}_3 \otimes \begin{pmatrix}
b_1 \\
b_2 \\
b_3
\end{pmatrix}_3 &=& (a_1 b_1+a_2 b_2+ a_3 b_3)_1 \oplus (a_1 b_1+\omega a_2 b_2+ \omega^2 a_3 b_3)_{1^\prime} \oplus (a_1 b_1+\omega^2 a_2 b_2+ \omega a_3 b_3)_{1^{\prime \prime}} \nonumber \\
&&\oplus \begin{pmatrix}
a_2 b_3 + a_3 b_2\\
a_3 b_1 + a_1 b_3\\
a_1 b_2 + a_2 b_1
\end{pmatrix}_{3s} \oplus \begin{pmatrix}
a_2 b_3 - a_3 b_2\\
a_3 b_1 - a_1 b_3\\
a_1 b_2 - a_2 b_1
\end{pmatrix}_{3a}. \nonumber
\end{eqnarray}
\section{\label{ZN} $Z_N$ Group}
The $Z_N$ group represents a symmetry that operates within the integers modulo $N$; it involves the numbers $0$ to $(N-1)$. Suppose $n_1$ and $n_2$ are two group elements of $Z_N$, then we have,
\begin{equation}
    n_1 \times n_2 = (n_1 + n_2) \mod N.
\end{equation}
The irreducible representation of a general group element ($q_n$) is given by 
\begin{equation}
    \exp\left(\frac{2 \pi i q_n}{N}\right),
\end{equation}
where $q_n$ can take values from $0$ to $N-1$. 

The multiplication tables for the groups $Z_3$ and $Z_{10}$ are given in Tables \ref{t3} and \ref{t10}, respectively.

\begin{table}[H]
\begin{center}
\begin{tabular*}{0.5\textwidth}{@{\extracolsep{\fill}}c c c c}
\hline
$q_3$ & 0 & 1 & 2\\
\hline
0 & 0 & 1 & 2 \\
\hline
1 & 1 & 2 & 0\\
\hline
2 & 2 & 0 & 1\\
\hline
\end{tabular*}
\caption{\footnotesize 
\label{t3}
Multiplication Table for $Z_3$ Group.}
\end{center}
\end{table} \quad

\begin{table}[H]
\begin{center}
\begin{tabular*}{\textwidth}{@{\extracolsep{\fill}}c c c c c c c c c c c}
\hline
$q_{10}$ & 0 & 1 & 2 & 3 & 4 & 5 & 6 & 7 & 8 & 9\\
\hline
0 & 0 & 1 & 2 & 3 & 4 & 5 & 6 & 7 & 8 & 9 \\
\hline
1 & 1 & 2 & 3 & 4 & 5 & 6 & 7 & 8 & 9 & 0\\
\hline
2 & 2 & 3 & 4 & 5 & 6 & 7 & 8 & 9 & 0 & 1\\
\hline
3 & 3 & 4 & 5 & 6 & 7 & 8 &9 & 0 & 1 & 2\\
\hline
4 & 4 & 5 & 6 & 7 & 8 & 9 & 0 & 1 & 2 & 3 \\
\hline 
5 & 5 & 6 & 7 & 8 & 9 & 0 & 1 & 2 & 3 & 4\\
\hline
6 & 6 & 7 & 8 & 9 & 0 &1 & 2 & 3 & 4 & 5 \\
\hline
7 & 7 & 8 & 9 & 0 & 1 & 2 & 3 & 4 & 5&6\\
\hline
8&8&9&0&1&2&3&4&5&6&7\\
\hline
9&9&0&1&2&3&4&5&6&7&8\\
\hline
\end{tabular*}
\caption{\footnotesize 
\label{t10}
Multiplication Table for $Z_{10}$ Group.}
\end{center}
\end{table}

\newpage
\section{\label{Section 8} {Scalar Potential of Our Framework}}
The $SU(2)_L \otimes U(1)_Y \otimes A_4 \otimes Z_3 \otimes Z_{10}$ invariant scalar potential of our framework is given below,
\begin{eqnarray}
V=V(H)+V(\phi)+V(\chi)+V(\phi,H)+V(\phi,\chi)+V(\chi,H),
\end{eqnarray}

where,

\begin{eqnarray}
V(H)&=&\mu_1^2(H^\dagger H)_1+\lambda_1^H (H^\dagger H)_1 (H^\dagger H)_1+\lambda_2^H (H^\dagger H)_{1^{'}} (H^\dagger H)_{1^{''}}+\lambda_3^H (H^\dagger H)_{3s} (H^\dagger H)_{3s} \nonumber \\
&&+\lambda_4^H (H^\dagger H)_{3a} (H^\dagger H)_{3a} +\lambda_5^H (H^\dagger H)_{3s} (H^\dagger H)_{3a},\nonumber \\
V(\phi)&=&V(H \longrightarrow \phi, \mu_1 \longrightarrow \mu_2,\lambda_i^H \longrightarrow \lambda_i^\phi),\nonumber \\
V(\chi)&=&V(H \longrightarrow \chi, \mu_1 \longrightarrow \mu_3,\lambda_i^H \longrightarrow \lambda_i^\chi),\nonumber \\
V(\phi,H) &=& \lambda_1^{\phi H}(\phi^\dagger \phi)_1 (H^\dagger H)_1+\lambda_2^{\phi H}(\phi^\dagger \phi)_{1^{'}} (H^\dagger H)_{1^{''}}+\lambda_3^{\phi H}(\phi^\dagger \phi)_{1^{''}} (H^\dagger H)_{1^{'}}+\lambda_4^{\phi H}(\phi^\dagger \phi)_{3s} (H^\dagger H)_{3s} \nonumber \\
&&+\lambda_5^{\phi H}(\phi^\dagger \phi)_{3a} (H^\dagger H)_{3a}+\lambda_6^{\phi H}(\phi^\dagger \phi)_{3s} (H^\dagger H)_{3a} +\lambda_7^{\phi H}(\phi^\dagger \phi)_{3a} (H^\dagger H)_{3s}+\lambda_8^{\phi H}(\phi^\dagger H)_{1} (H^\dagger \phi)_{1} \nonumber \\
&&+\lambda_9^{\phi H}(\phi^\dagger H)_{1^{'}} (H^\dagger \phi)_{1^{''}}+\lambda_{10}^{\phi H}(\phi^\dagger H)_{1^{''}} (H^\dagger \phi)_{1^{'}}+\lambda_{11}^{\phi H}(\phi^\dagger H)_{3s} (H^\dagger \phi)_{3s}+\lambda_{12}^{\phi H}(\phi^\dagger H)_{3a} (H^\dagger \phi)_{3a} \nonumber \\
&&+\lambda_{13}^{\phi H}(\phi^\dagger H)_{3s} (H^\dagger \phi)_{3a}+\lambda_{14}^{\phi H}(\phi^\dagger H)_{3a} (H^\dagger \phi)_{3s},\nonumber \\
V(\phi,\chi) &=& V(\phi, H \longrightarrow \phi, \chi, \lambda_i^{\phi H} \longrightarrow \lambda_i^{\phi \chi}),\nonumber \\
V(\chi,H) &=& V(\phi, H \longrightarrow \chi, H, \lambda_i^{\phi H} \longrightarrow \lambda_i^{\chi H}).
\end{eqnarray}

Now, for the VEV assignments of the scalar fields $\phi$ and $\chi$ to be $(0,k_1,k_2)^Tv_\phi$ and $(0,f_1,f_2)^Tv_\chi$,  we have to satisfy the following minimization conditions,

\begin{eqnarray}
\frac{\partial V}{\partial H_1} &=&v_h \bigg( f_1^2 v_\chi^2 \lambda_{11}^{\chi H} 
+ f_2^2 v_\chi^2 \lambda_{11}^{\chi H} 
- f_1^2 v_\chi^2 \lambda_{12}^{\chi H} 
- f_2^2 v_\chi^2 \lambda_{12}^{\chi H} 
+ f_1^2 v_\chi^2 \lambda_{13}^{\chi H} 
- f_2^2 v_\chi^2 \lambda_{13}^{\chi H} 
- f_1^2 v_\chi^2 \lambda_{14}^{\chi H} \nonumber \\
&&+ f_2^2 v_\chi^2 \lambda_{14}^{\chi H} 
+ 6 v_h^2 \lambda_1^H 
+ f_1^2 v_\chi^2 \lambda_1^{\chi H} 
+ f_2^2 v_\chi^2 \lambda_1^{\chi H} 
+ 2 v_h^2 \lambda_2^H 
+ 2 v_h^2 \omega \lambda_2^H 
+ 2 v_h^2 \omega^2 \lambda_2^H 
+ f_1^2 v_\chi^2 \omega \lambda_2^{\chi H} \nonumber \\
&&+ f_2^2 v_\chi^2 \omega^2 \lambda_2^{\chi H} 
+ 8 v_h^2 \lambda_3^H 
+ k_2^2 v_\phi^2 \big( \lambda_{11}^{\phi H} 
- \lambda_{12}^{\phi H} 
- \lambda_{13}^{\phi H} 
+ \lambda_{14}^{\phi H} 
+ \lambda_1^{\phi H} 
+ \omega^2 \lambda_2^{\phi H} 
+ \omega \lambda_3^{\phi H} \big) \nonumber \\
&&+ k_1^2 v_\phi^2 \big( \lambda_{11}^{\phi H} 
- \lambda_{12}^{\phi H} 
+ \lambda_{13}^{\phi H} 
- \lambda_{14}^{\phi H} 
+ \lambda_1^{\phi H} 
+ \omega \lambda_2^{\phi H} 
+ \omega^2 \lambda_3^{\phi H} \big) 
+ f_2^2 v_\chi^2 \omega \lambda_3^{\chi H} 
+ f_1^2 v_\chi^2 \omega^2 \lambda_3^{\chi H} \nonumber \\
&&+ \mu_1^2 
\bigg)=0
,\nonumber \\
\frac{\partial V}{\partial H_2} &=& v_h \bigg( 
    k_2^2 v_\phi^2 \lambda_{11}^{\phi H} 
    - k_2^2 v_\phi^2 \lambda_{12}^{\phi H} 
    + k_2^2 v_\phi^2 \lambda_{13}^{\phi H} 
    - k_2^2 v_\phi^2 \lambda_{14}^{\phi H} 
    + k_2^2 v_\phi^2 \lambda_1^{\phi H} 
    + f_2^2 v_\chi^2 \big( \lambda_{11}^{\chi H} 
    - \lambda_{12}^{\chi H} 
    + \lambda_{13}^{\chi H} \nonumber \\
  &&  - \lambda_{14}^{\chi H} 
    + \lambda_1^{\chi H} \big) 
    + k_2^2 v_\phi^2 \omega^4 \lambda_2^{\phi H} 
    + f_2^2 v_\chi^2 \omega^4 \lambda_2^{\chi H} 
    + v_h^2 \big( 
        6 \lambda_1^H 
        + \omega (1 + \omega)(1 + \omega + \omega^2) \lambda_2^H 
        + 8 \lambda_3^H \big) \nonumber \\
    &&+ k_2^2 v_\phi^2 \omega^2 \lambda_3^{\phi H} 
    + f_2^2 v_\chi^2 \omega^2 \lambda_3^{\chi H} 
    + k_1 k_2 v_\phi^2 \big( 
        \omega^4 \lambda_{10}^{\phi H} 
        + \lambda_{11}^{\phi H} 
        + \lambda_{12}^{\phi H} 
        - \lambda_{13}^{\phi H} 
        - \lambda_{14}^{\phi H} \nonumber \\
    &&+ 2 \lambda_4^{\phi H} 
        - 2 \lambda_6^{\phi H} 
        + \lambda_8^{\phi H} 
        + \omega^2 \lambda_9^{\phi H} \big) 
    + k_1^2 v_\phi^2 \big( 
        \lambda_1^{\phi H} 
        + \lambda_8^{\phi H} 
        + \omega^3 ( \lambda_{10}^{\phi H} 
        + \lambda_2^{\phi H} 
        + \lambda_3^{\phi H} 
        + \lambda_9^{\phi H} ) \big) \nonumber \\
    &&+ f_1 f_2 v_\chi^2 \big( 
        \omega^4 \lambda_{10}^{\chi H} 
        + \lambda_{11}^{\chi H} 
        + \lambda_{12}^{\chi H} 
        - \lambda_{13}^{\chi H} 
        - \lambda_{14}^{\chi H} 
        + 2 \lambda_4^{\chi H} 
        - 2 \lambda_6^{\chi H} 
        + \lambda_8^{\chi H} 
        + \omega^2 \lambda_9^{\chi H} \big) \nonumber \\
    &&+ f_1^2 v_\chi^2 \big( 
        \lambda_1^{\chi H} 
        + \lambda_8^{\chi H} 
        + \omega^3 ( \lambda_{10}^{\chi H} 
        + \lambda_2^{\chi H} 
        + \lambda_3^{\chi H} 
        + \lambda_9^{\chi H} ) \big) 
    + \mu_1^2 
\bigg)
 = 0
,\nonumber \\
\frac{\partial V}{\partial H_2^\dagger} &=& v_h \bigg( 
k_2^2 v_\phi^2 \lambda_{11}^{\phi H} 
+ f_2^2 v_\chi^2 \lambda_{11}^{\chi H} 
- k_2^2 v_\phi^2 \lambda_{12}^{\phi H} 
- f_2^2 v_\chi^2 \lambda_{12}^{\chi H} 
+ k_2^2 v_\phi^2 \lambda_{13}^{\phi H} 
+ f_2^2 v_\chi^2 \lambda_{13}^{\chi H} 
- k_2^2 v_\phi^2 \lambda_{14}^{\phi H} \nonumber \\
&&- f_2^2 v_\chi^2 \lambda_{14}^{\chi H} 
+ 6 v_h^2 \lambda_1^H 
+ k_2^2 v_\phi^2 \lambda_1^{\phi H} 
+ f_2^2 v_\chi^2 \lambda_1^{\chi H} 
+ v_h^2 \omega \lambda_2^H 
+ 2 v_h^2 \omega^2 \lambda_2^H 
+ 2 v_h^2 \omega^3 \lambda_2^H 
+ v_h^2 \omega^4 \lambda_2^H \nonumber \\
&&+ k_2^2 v_\phi^2 \omega^4 \lambda_2^{\phi H} 
+ f_2^2 v_\chi^2 \omega^4 \lambda_2^{\chi H} 
+ 8 v_h^2 \lambda_3^H 
+ k_2^2 v_\phi^2 \omega^2 \lambda_3^{\phi H} 
+ f_2^2 v_\chi^2 \omega^2 \lambda_3^{\chi H} 
+ k_1 k_2 v_\phi^2 \big( 
\omega^2 \lambda_{10}^{\phi H} 
+ \lambda_{11}^{\phi H} \nonumber \\
&&+ \lambda_{12}^{\phi H} 
+ \lambda_{13}^{\phi H} 
+ \lambda_{14}^{\phi H} 
+ 2 \lambda_4^{\phi H} 
+ 2 \lambda_6^{\phi H} 
+ \lambda_8^{\phi H} 
+ \omega^4 \lambda_9^{\phi H} 
\big) 
+ k_1^2 v_\phi^2 \big( 
\lambda_1^{\phi H} 
+ \lambda_8^{\phi H} 
+ \omega^3 \big( 
\lambda_{10}^{\phi H} 
+ \lambda_2^{\phi H} \nonumber \\
&&+ \lambda_3^{\phi H} 
+ \lambda_9^{\phi H} 
\big) 
\big) 
+ f_1 f_2 v_\chi^2 \big( 
\omega^2 \lambda_{10}^{\chi H} 
+ \lambda_{11}^{\chi H} 
+ \lambda_{12}^{\chi H} 
+ \lambda_{13}^{\chi H} 
+ \lambda_{14}^{\chi H} 
+ 2 \lambda_4^{\chi H} 
+ 2 \lambda_6^{\chi H} 
+ \lambda_8^{\chi H} \nonumber \\
&&+ \omega^4 \lambda_9^{\chi H} 
\big) 
+ f_1^2 v_\chi^2 \big( 
\lambda_1^{\chi H} 
+ \lambda_8^{\chi H} 
+ \omega^3 \big( 
\lambda_{10}^{\chi H} 
+ \lambda_2^{\chi H} 
+ \lambda_3^{\chi H} 
+ \lambda_9^{\chi H} 
\big) 
\big) 
+ \mu_{1}^2
\bigg)
=0,\nonumber \\
\frac{\partial V}{\partial H_3} &=& v_h \bigg(
f_1 f_2 v_{\chi}^2 \omega^2 \lambda_{10}^{\chi H} 
+ f_2^2 v_{\chi}^2 \omega^3 \lambda_{10}^{\chi H} 
+ f_1^2 v_{\chi}^2 \lambda_{11}^{\chi H} 
+ f_1 f_2 v_{\chi}^2 \lambda_{11}^{\chi H} 
- f_1^2 v_{\chi}^2 \lambda_{12}^{\chi H} 
+ f_1 f_2 v_{\chi}^2 \lambda_{12}^{\chi H} \nonumber \\
&&- f_1^2 v_{\chi}^2 \lambda_{13}^{\chi H} 
+ f_1 f_2 v_{\chi}^2 \lambda_{13}^{\chi H} 
+ f_1^2 v_{\chi}^2 \lambda_{14}^{\chi H} 
+ f_1 f_2 v_{\chi}^2 \lambda_{14}^{\chi H} 
+ 6 v_h^2 \lambda_{1}^{H} 
+ f_1^2 v_{\chi}^2 \lambda_{1}^{\chi H} 
+ f_2^2 v_{\chi}^2 \lambda_{1}^{\chi H} 
+ v_h^2 \omega \lambda_{2}^{H} \nonumber \\
&&+ 2 v_h^2 \omega^2 \lambda_{2}^{H} 
+ 2 v_h^2 \omega^3 \lambda_{2}^{H} 
+ v_h^2 \omega^4 \lambda_{2}^{H} 
+ f_1^2 v_{\chi}^2 \omega^2 \lambda_{2}^{\chi H} 
+ f_2^2 v_{\chi}^2 \omega^3 \lambda_{2}^{\chi H} 
+ 8 v_h^2 \lambda_{3}^{H} 
+ k_1^2 v_{\phi}^2 
\big(
\lambda_{11}^{\phi H} - \lambda_{12}^{\phi H} \nonumber \\
&&- \lambda_{13}^{\phi H} 
+ \lambda_{14}^{\phi H} + \lambda_{1}^{\phi H} 
+ \omega^2 \lambda_{2}^{\phi H} + \omega^4 \lambda_{3}^{\phi H} 
\big) 
+ f_2^2 v_{\chi}^2 \omega^3 \lambda_{3}^{\chi H} 
+ f_1^2 v_{\chi}^2 \omega^4 \lambda_{3}^{\chi H} 
+ 2 f_1 f_2 v_{\chi}^2 \lambda_{4}^{\chi H} \nonumber \\
&&+ 2 f_1 f_2 v_{\chi}^2 \lambda_{6}^{\chi H} 
+ f_1 f_2 v_{\chi}^2 \lambda_{8}^{\chi H} 
+ f_2^2 v_{\chi}^2 \lambda_{8}^{\chi H} 
+ k_1 k_2 v_{\phi}^2 
\big(
\omega^2 \lambda_{10}^{\phi H} + \lambda_{11}^{\phi H} + \lambda_{12}^{\phi H} + \lambda_{13}^{\phi H} + \lambda_{14}^{\phi H} + 2 (\lambda_{4}^{\phi H} \nonumber \\
&&+ \lambda_{6}^{\phi H}) + \lambda_{8}^{\phi H}  
+ \omega^4 \lambda_{9}^{\phi H}
\big) 
+ k_2^2 v_{\phi}^2 
\big(
\lambda_{1}^{\phi H} + \lambda_{8}^{\phi H} + \omega^3 (\lambda_{10}^{\phi H} + \lambda_{2}^{\phi H} + \lambda_{3}^{\phi H} + \lambda_{9}^{\phi H})
\big) 
+ f_2 v_{\chi}^2 \omega^3 (f_2 \nonumber \\
&&+ f_1 \omega) \lambda_{9}^{\chi H} 
+ \mu_{1}^2
\bigg)
= 0,\nonumber \\
\end{eqnarray}
\begin{eqnarray}
\frac{\partial V}{\partial H_3^\dagger} &=& v_h \bigg(
k_1^2 v_{\phi}^2 \lambda_{11}^{\phi H} 
- k_1^2 v_{\phi}^2 \lambda_{12}^{\phi H} 
- k_1^2 v_{\phi}^2 \lambda_{13}^{\phi H} 
+ k_1^2 v_{\phi}^2 \lambda_{14}^{\phi H} 
+ k_1^2 v_{\phi}^2 \lambda_{1}^{\phi H} 
+ f_1^2 v_{\chi}^2 
\big(
\lambda_{11}^{\chi H} - \lambda_{12}^{\chi H} - \lambda_{13}^{\chi H} \nonumber \\
&&+ \lambda_{14}^{\chi H} 
+ \lambda_{1}^{\chi H}
\big) 
+ k_1^2 v_{\phi}^2 \omega^2 \lambda_{2}^{\phi H} 
+ f_1^2 v_{\chi}^2 \omega^2 \lambda_{2}^{\chi H} 
+ v_h^2 
\big(
6 \lambda_{1}^H 
+ \omega (1 + \omega) (1 + \omega + \omega^2) \lambda_{2}^H 
+ 8 \lambda_{3}^H
\big) \nonumber \\
&&+ k_1^2 v_{\phi}^2 \omega^4 \lambda_{3}^{\phi H} 
+ f_1^2 v_{\chi}^2 \omega^4 \lambda_{3}^{\chi H} 
+ k_1 k_2 v_{\phi}^2 
\big(
\omega^4 \lambda_{10}^{\phi H} + \lambda_{11}^{\phi H} + \lambda_{12}^{\phi H} 
- \lambda_{13}^{\phi H} - \lambda_{14}^{\phi H} 
+ 2 \lambda_{4}^{\phi H} - 2 \lambda_{6}^{\phi H} \nonumber \\
&&+ \lambda_{8}^{\phi H} + \omega^2 \lambda_{9}^{\phi H}
\big) 
+ k_2^2 v_{\phi}^2 
\big(
\lambda_{1}^{\phi H} + \lambda_{8}^{\phi H} 
+ \omega^3 (\lambda_{10}^{\phi H} + \lambda_{2}^{\phi H} + \lambda_{3}^{\phi H} + \lambda_{9}^{\phi H})
\big) 
+ f_1 f_2 v_{\chi}^2 
\big(
\omega^4 \lambda_{10}^{\chi H} \nonumber \\
&&+ \lambda_{11}^{\chi H} + \lambda_{12}^{\chi H} 
- \lambda_{13}^{\chi H} - \lambda_{14}^{\chi H} 
+ 2 \lambda_{4}^{\chi H} - 2 \lambda_{6}^{\chi H} 
+ \lambda_{8}^{\chi H} + \omega^2 \lambda_{9}^{\chi H}
\big) 
+ f_2^2 v_{\chi}^2 
\big(
\lambda_{1}^{\chi H} + \lambda_{8}^{\chi H} 
+ \omega^3 (\lambda_{10}^{\chi H} \nonumber \\
&&+ \lambda_{2}^{\chi H} + \lambda_{3}^{\chi H} + \lambda_{9}^{\chi H})
\big) 
+ \mu_{1}^2
\bigg)
=0,\nonumber \\
\frac{\partial V}{\partial \phi_1} &=& v_h^2 v_\Phi \left( k_1 \left( \lambda_{11}^{\Phi H} + \lambda_{12}^{\Phi H} - \lambda_{13}^{\Phi H} - \lambda_{14}^{\Phi H} + 2 \lambda_{4}^{\Phi H} - 2 \lambda_{7}^{\Phi H} + \lambda_{8}^{\Phi H} + \omega \left( \omega \lambda_{10}^{\Phi H} + \lambda_{9}^{\Phi H} \right) \right) \right.\nonumber \\
&& \left. + k_2 \left( \lambda_{11}^{\Phi H} + \lambda_{12}^{\Phi H} + \lambda_{13}^{\Phi H} + \lambda_{14}^{\Phi H} + 2 \left( \lambda_{4}^{\Phi H} + \lambda_{7}^{\Phi H} \right) + \lambda_{8}^{\Phi H} + \omega \left( \lambda_{10}^{\Phi H} + \omega \lambda_{9}^{\Phi H} \right) \right) \right) = 0,\nonumber \\
\frac{\partial V}{\partial \phi_1^\dagger} &=& v_h^2 v_\Phi \left( k_2 \left( \lambda_{11}^{\Phi H} + \lambda_{12}^{\Phi H} - \lambda_{13}^{\Phi H} - \lambda_{14}^{\Phi H} + 2 \lambda_{4}^{\Phi H} - 2 \lambda_{7}^{\Phi H} + \lambda_{8}^{\Phi H} + \omega \left( \omega \lambda_{10}^{\Phi H} + \lambda_{9}^{\Phi H} \right) \right) \right.\nonumber \\
&& \left.+ k_1 \left( \lambda_{11}^{\Phi H} + \lambda_{12}^{\Phi H} + \lambda_{13}^{\Phi H} + \lambda_{14}^{\Phi H} + 2 \left( \lambda_{4}^{\Phi H} + \lambda_{7}^{\Phi H} \right) + \lambda_{8}^{\Phi H} + \omega \left( \lambda_{10}^{\Phi H} + \omega \lambda_{9}^{\Phi H} \right) \right) \right)
=0,\nonumber \\
\frac{\partial V}{\partial \phi_2} &=& v_{\phi} \bigg(
k_1 k_2^2 v_{\phi}^2 
\big( 2 \lambda_{1}^{\phi} + \omega^2 \lambda_{2}^{\phi} + \omega^4 \lambda_{2}^{\phi} + 4 \lambda_{3}^{\phi} - 2 \lambda_{5}^{\phi} \big) 
+ k_2 \bigg( v_h^2 \big( \omega^2 \lambda_{10}^{\phi H} + \lambda_{11}^{\phi H} + \lambda_{12}^{\phi H} 
- \lambda_{13}^{\phi H} - \lambda_{14}^{\phi H} \nonumber \\
&&+ 2 \lambda_{4}^{\phi H} - 2 \lambda_{7}^{\phi H} 
+ \lambda_{8}^{\phi H} + \omega^4 \lambda_{9}^{\phi H} \big) 
+ f_1 f_2 v_{\chi}^2 \big( \omega^2 \lambda_{10}^{\phi \chi} + \lambda_{11}^{\phi \chi} + \lambda_{12}^{\phi \chi} 
- \lambda_{13}^{\phi \chi} - \lambda_{14}^{\phi \chi} + 2 \lambda_{4}^{\phi \chi} - 2 \lambda_{7}^{\phi \chi} \nonumber \\
&&+ \lambda_{8}^{\phi \chi} + \omega^4 \lambda_{9}^{\phi \chi} \big) \bigg) 
+ k_1 \bigg( 2 k_1^2 v_{\phi}^2 \big( \lambda_{1}^{\phi} + \omega^3 \lambda_{2}^{\phi} \big) 
+ f_2^2 v_{\chi}^2 \big( \lambda_{11}^{\phi \chi} - \lambda_{12}^{\phi \chi} - \lambda_{13}^{\phi \chi} 
+ \lambda_{14}^{\phi \chi} + \lambda_{1}^{\phi \chi} + \omega^2 \lambda_{2}^{\phi \chi} \nonumber \\
&&+ \omega^4 \lambda_{3}^{\phi \chi} \big) 
+ v_h^2 \big( 2 \lambda_{11}^{\phi H} - 2 \lambda_{12}^{\phi H} + 3 \lambda_{1}^{\phi H} 
+ \lambda_{8}^{\phi H} + \omega \big( \lambda_{2}^{\phi H} + \omega \big( \lambda_{2}^{\phi H} + \lambda_{3}^{\phi H} 
+ \omega \big( \lambda_{10}^{\phi H} + \lambda_{2}^{\phi H} + \lambda_{3}^{\phi H} \nonumber \\
&&+ \omega \lambda_{3}^{\phi H} + \lambda_{9}^{\phi H} \big) \big) \big) \big) 
+ f_1^2 v_{\chi}^2 \big( \lambda_{1}^{\phi \chi} + \lambda_{8}^{\phi \chi} + \omega^3 \big( \lambda_{10}^{\phi \chi} + \lambda_{2}^{\phi \chi} 
+ \lambda_{3}^{\phi \chi} + \lambda_{9}^{\phi \chi} \big) \big) 
+ \mu_{2}^2
\bigg)
= 0,\nonumber \\
\frac{\partial V}{\partial \phi_2^\dagger} &=& v_{\phi} \bigg( 2 k_1^3 v_{\phi}^2 \big( \lambda_1^{\phi} + \omega^3 \lambda_2^{\phi} \big) 
+ k_2 \bigg( v_h^2 \big( \omega^4 \lambda_{10}^{\phi H} + \lambda_{11}^{\phi H} + \lambda_{12}^{\phi H} 
+ \lambda_{13}^{\phi H} + \lambda_{14}^{\phi H} + 2 \big( \lambda_4^{\phi H} + \lambda_7^{\phi H} \big) 
+ \lambda_8^{\phi H} \nonumber \\
&&+ \omega^2 \lambda_9^{\phi H} \big) 
+ f_1 f_2 v_{\chi}^2 \big( \omega^4 \lambda_{10}^{\phi \chi} + \lambda_{11}^{\phi \chi} + \lambda_{12}^{\phi \chi} 
+ \lambda_{13}^{\phi \chi} + \lambda_{14}^{\phi \chi} + 2 \big( \lambda_4^{\phi \chi} + \lambda_7^{\phi \chi} \big) 
+ \lambda_8^{\phi \chi} + \omega^2 \lambda_9^{\phi \chi} \big) \bigg) \nonumber \\
&&+ k_1 \bigg( f_2^2 v_{\chi}^2 \big( \lambda_{11}^{\phi \chi} - \lambda_{12}^{\phi \chi} - \lambda_{13}^{\phi \chi} 
+ \lambda_{14}^{\phi \chi} + \lambda_1^{\phi \chi} + \omega^2 \lambda_2^{\phi \chi} + \omega^4 \lambda_3^{\phi \chi} \big) 
+ k_2^2 v_{\phi}^2 \big( 2 \lambda_1^{\phi} + \omega^2 \lambda_2^{\phi} + \omega^4 \lambda_2^{\phi} \nonumber \\
&&+ 4 \lambda_3^{\phi} 
+ 2 \lambda_5^{\phi} \big) 
+ v_h^2 \big( 2 \lambda_{11}^{\phi H} - 2 \lambda_{12}^{\phi H} + 3 \lambda_1^{\phi H} + \lambda_8^{\phi H} 
+ \omega \big( \lambda_2^{\phi H} + \omega \big( \lambda_2^{\phi H} \nonumber \\
&&+ \lambda_3^{\phi H} 
+ \omega \big( \lambda_{10}^{\phi H} + \lambda_2^{\phi H} + \lambda_3^{\phi H} + \omega \lambda_3^{\phi H} + \lambda_9^{\phi H} \big) \big) \big) \big) 
+ f_1^2 v_{\chi}^2 \big( \lambda_1^{\phi \chi} + \lambda_8^{\phi \chi} + \omega^3 \big( \lambda_{10}^{\phi \chi} 
+ \lambda_2^{\phi \chi} + \lambda_3^{\phi \chi} \nonumber \\
&&+ \lambda_9^{\phi \chi} \big) \big) 
+ \mu_2^2 \bigg)
 = 0,\nonumber \\
 \footnotesize
 \frac{\partial V}{\partial \phi_3} &=& v_{\phi} \bigg( 2 k_2^3 v_{\phi}^2 \big( \lambda_1^{\phi} + \omega^3 \lambda_2^{\phi} \big) + k_1 \bigg( v_h^2 \big( \omega^4 \lambda_{10}^{\phi H} + \lambda_{11}^{\phi H} + \lambda_{12}^{\phi H} + \lambda_{13}^{\phi H} + \lambda_{14}^{\phi H} + 2 \big( \lambda_4^{\phi H} + \lambda_7^{\phi H} \big) + \lambda_8^{\phi H} \nonumber \\
 &&+ \omega^2 \lambda_9^{\phi H} \big) + f_1 f_2 v_{\chi}^2 \big( \omega^4 \lambda_{10}^{\phi \chi} + \lambda_{11}^{\phi \chi} + \lambda_{12}^{\phi \chi} + \lambda_{13}^{\phi \chi} + \lambda_{14}^{\phi \chi} + 2 \big( \lambda_4^{\phi \chi} + \lambda_7^{\phi \chi} \big) + \lambda_8^{\phi \chi} + \omega^2 \lambda_9^{\phi \chi} \big) \bigg) \nonumber \\
&& + k_2 \bigg( f_1^2 v_{\chi}^2 \big( \lambda_{11}^{\phi \chi} - \lambda_{12}^{\phi \chi} + \lambda_{13}^{\phi \chi} - \lambda_{14}^{\phi \chi} + \lambda_1^{\phi \chi} + \omega^4 \lambda_2^{\phi \chi} + \omega^2 \lambda_3^{\phi \chi} \big) + k_1^2 v_{\phi}^2 \big( 2 \lambda_1^{\phi} + \omega^2 \lambda_2^{\phi} + \omega^4 \lambda_2^{\phi}\nonumber 
 \nonumber 
 \end{eqnarray}
 \newpage
 \begin{eqnarray}
 &&+ 4 \lambda_3^{\phi} + 2 \lambda_5^{\phi} \big) \bigg) + v_h^2 \big( 2 \lambda_{11}^{\phi H} - 2 \lambda_{12}^{\phi H} + 3 \lambda_1^{\phi H} + \lambda_8^{\phi H} + \omega \big( \lambda_3^{\phi H} + \omega \big( \lambda_2^{\phi H} + \lambda_3^{\phi H} + \omega \big( \lambda_{10}^{\phi H} 
 + \lambda_2^{\phi H} \nonumber \\
&& + \omega \lambda_2^{\phi H} + \lambda_3^{\phi H} + \lambda_9^{\phi H} \big) \big) \big) \big) + f_2^2 v_{\chi}^2 \big( \lambda_1^{\phi \chi} + \lambda_8^{\phi \chi} + \omega^3 \big( \lambda_{10}^{\phi \chi} + \lambda_2^{\phi \chi} + \lambda_3^{\phi \chi} + \lambda_9^{\phi \chi} \big) \big) 
+ \mu_2^2 \bigg)
=0,\nonumber \\
\frac{\partial V}{\partial \phi_3^\dagger} &=& v_{\phi} \bigg( k_1^2 k_2 v_{\phi}^2 \big( 2 \lambda_1^{\phi} + \omega^2 \lambda_2^{\phi} + \omega^4 \lambda_2^{\phi} + 4 \lambda_3^{\phi} - 2 \lambda_5^{\phi} \big) 
+ k_1 \bigg( v_h^2 \big( \omega^2 \lambda_{10}^{\phi H} + \lambda_{11}^{\phi H} + \lambda_{12}^{\phi H} 
- \lambda_{13}^{\phi H} - \lambda_{14}^{\phi H} \nonumber \\
&&+ 2 \lambda_4^{\phi H} 
- 2 \lambda_7^{\phi H} + \lambda_8^{\phi H} + \omega^4 \lambda_9^{\phi H} \big) 
+ f_1 f_2 v_{\chi}^2 \big( \omega^2 \lambda_{10}^{\phi \chi} + \lambda_{11}^{\phi \chi} + \lambda_{12}^{\phi \chi} 
- \lambda_{13}^{\phi \chi} - \lambda_{14}^{\phi \chi} + 2 \lambda_4^{\phi \chi} 
- 2 \lambda_7^{\phi \chi} \nonumber \\
&&+ \lambda_8^{\phi \chi} + \omega^4 \lambda_9^{\phi \chi} \big) \bigg) 
+ k_2 \bigg( 2 k_2^2 v_{\phi}^2 \big( \lambda_1^{\phi} + \omega^3 \lambda_2^{\phi} \big) 
+ f_1^2 v_{\chi}^2 \big( \lambda_{11}^{\phi \chi} - \lambda_{12}^{\phi \chi} + \lambda_{13}^{\phi \chi} 
- \lambda_{14}^{\phi \chi} + \lambda_1^{\phi \chi} + \omega^4 \lambda_2^{\phi \chi} \nonumber \\
&&+ \omega^2 \lambda_3^{\phi \chi} \big) 
+ v_h^2 \big( 2 \lambda_{11}^{\phi H} - 2 \lambda_{12}^{\phi H} + 3 \lambda_1^{\phi H} + \lambda_8^{\phi H} 
+ \omega \big( \lambda_3^{\phi H} + \omega \big( \lambda_2^{\phi H} + \lambda_3^{\phi H} 
+ \omega \big( \lambda_{10}^{\phi H} + \lambda_2^{\phi H} \nonumber \\
&&+ \omega \lambda_2^{\phi H} 
+ \lambda_3^{\phi H} + \lambda_9^{\phi H} \big) \big) \big) \big) 
+ f_2^2 v_{\chi}^2 \big( \lambda_1^{\phi \chi} + \lambda_8^{\phi \chi} + \omega^3 \big( \lambda_{10}^{\phi \chi} 
+ \lambda_2^{\phi \chi} + \lambda_3^{\phi \chi} + \lambda_9^{\phi \chi} \big) \big) 
+ \mu_2^2 \bigg)
=0,\nonumber \\
\frac{\partial V}{\partial \chi_1} &=&v_h^2 v_{\chi} \bigg( f_1 \big( \lambda_{11}^{\chi H} + \lambda_{12}^{\chi H} - \lambda_{13}^{\chi H} - \lambda_{14}^{\chi H} 
+ 2 \lambda_4^{\chi H} - 2 \lambda_7^{\chi H} + \lambda_8^{\chi H} + \omega \big( \omega \lambda_{10}^{\chi H} + \lambda_9^{\chi H} \big) \big)
+ f_2 \big( \lambda_{11}^{\chi H} \nonumber \\
&&+ \lambda_{12}^{\chi H} + \lambda_{13}^{\chi H} + \lambda_{14}^{\chi H} 
+ 2 \big( \lambda_4^{\chi H} + \lambda_7^{\chi H} \big) + \lambda_8^{\chi H} + \omega \big( \lambda_{10}^{\chi H} + \omega \lambda_9^{\chi H} \big) \big) \bigg)
=0,\nonumber \\
\frac{\partial V}{\partial \chi_1^\dagger} &=&
v_h^2 v_{\chi} \bigg( f_2 \big( \lambda_{11}^{\chi H} + \lambda_{12}^{\chi H} - \lambda_{13}^{\chi H} - \lambda_{14}^{\chi H} 
+ 2 \lambda_4^{\chi H} - 2 \lambda_7^{\chi H} + \lambda_8^{\chi H} + \omega \big( \omega \lambda_{10}^{\chi H} + \lambda_9^{\chi H} \big) \big)
+ f_1 \big( \lambda_{11}^{\chi H} \nonumber \\
&&+ \lambda_{12}^{\chi H} + \lambda_{13}^{\chi H} + \lambda_{14}^{\chi H} 
+ 2 \big( \lambda_4^{\chi H} + \lambda_7^{\chi H} \big) + \lambda_8^{\chi H} + \omega \big( \lambda_{10}^{\chi H} + \omega \lambda_9^{\chi H} \big) \big) \bigg)
=0,\nonumber \\
\frac{\partial V}{\partial \chi_2} &=&
v_{\chi} \bigg( 2 f_1^3 v_{\chi}^2 \left( \lambda_1^{\chi} + \omega^3 \lambda_2^{\chi} \right) 
+ f_2 \big( k_1 k_2 v_{\phi}^2 \big( \omega^4 \lambda_{10}^{\phi \chi} + \lambda_{11}^{\phi \chi} + \lambda_{12}^{\phi \chi} - \lambda_{13}^{\phi \chi} - \lambda_{14}^{\phi \chi} + 2 \lambda_4^{\phi \chi} - 2 \lambda_6^{\phi \chi} + \lambda_8^{\phi \chi} \nonumber \\
&&+ \omega^2 \lambda_9^{\phi \chi} \big) \big) 
+ v_h^2 \big( \omega^2 \lambda_{10}^{\chi H} + \lambda_{11}^{\chi H} + \lambda_{12}^{\chi H} - \lambda_{13}^{\chi H} - \lambda_{14}^{\chi H} + 2 \lambda_4^{\chi H} - 2 \lambda_7^{\chi H} + \lambda_8^{\chi H} + \omega^4 \lambda_9^{\chi H} \big) \bigg) \nonumber \\
&&+ f_1 \big( k_2^2 v_{\phi}^2 \big( \lambda_{11}^{\phi \chi} - \lambda_{12}^{\phi \chi} + \lambda_{13}^{\phi \chi} - \lambda_{14}^{\phi \chi} + \lambda_1^{\phi \chi} \big) \big) 
+ 2 f_2^2 v_{\chi}^2 \lambda_1^{\chi} 
+ k_2^2 v_{\phi}^2 \omega^4 \lambda_2^{\phi \chi} 
+ f_2^2 v_{\chi}^2 \omega^2 \lambda_2^{\chi} 
+ f_2^2 v_{\chi}^2 \omega^4 \lambda_2^{\chi} \nonumber \\
&&+ k_2^2 v_{\phi}^2 \omega^2 \lambda_3^{\phi \chi} 
+ 4 f_2^2 v_{\chi}^2 \lambda_3^{\chi} - 2 f_2^2 v_{\chi}^2 \lambda_5^{\chi} 
+ k_1^2 v_{\phi}^2 \big( \lambda_1^{\phi \chi} + \lambda_8^{\phi \chi} + \omega^3 \big( \lambda_{10}^{\phi \chi} + \lambda_2^{\phi \chi} + \lambda_3^{\phi \chi} + \lambda_9^{\phi \chi} \big) \big) \nonumber \\
&&+ v_h^2 \big( 2 \lambda_{11}^{\chi H} - 2 \lambda_{12}^{\chi H} + 3 \lambda_1^{\chi H} + \lambda_8^{\chi H} + \omega \big( \lambda_2^{\chi H} + \omega \big( \lambda_2^{\chi H} + \lambda_3^{\chi H} + \omega \big( \lambda_{10}^{\chi H} + \lambda_2^{\chi H} + \lambda_3^{\chi H} + \omega \lambda_3^{\chi H} \nonumber \\
&&+ \lambda_9^{\chi H} \big) \big) \big) \big) \big) + \mu_{3}^2 \bigg)
=0,\nonumber \\
\frac{\partial V}{\partial \chi_2^\dagger} &=&
v_{\chi} \bigg( f_1 f_2^2 v_{\chi}^2 \big( 2 \lambda_1^{\chi} + \omega^2 \lambda_2^{\chi} + \omega^4 \lambda_2^{\chi} + 4 \lambda_3^{\chi} + 2 \lambda_5^{\chi} \big) 
+ f_2 \big( k_1 k_2 v_{\phi}^2 \big( \omega^2 \lambda_{10}^{\phi \chi} + \lambda_{11}^{\phi \chi} + \lambda_{12}^{\phi \chi} + \lambda_{13}^{\phi \chi} \nonumber \\
&&+ \lambda_{14}^{\phi \chi} + 2 \big( \lambda_4^{\phi \chi} + \lambda_6^{\phi \chi} \big) + \lambda_8^{\phi \chi} + \omega^4 \lambda_9^{\phi \chi} \big) \big) 
+ v_h^2 \big( \omega^4 \lambda_{10}^{\chi H} + \lambda_{11}^{\chi H} + \lambda_{12}^{\chi H} + \lambda_{13}^{\chi H} + \lambda_{14}^{\chi H} + 2 \big( \lambda_4^{\chi H} \nonumber \\
&&+ \lambda_7^{\chi H} \big) + \lambda_8^{\chi H} + \omega^2 \lambda_9^{\chi H} \big) \bigg) 
+ f_1 \big( k_2^2 v_{\phi}^2 \big( \lambda_{11}^{\phi \chi} - \lambda_{12}^{\phi \chi} + \lambda_{13}^{\phi \chi} - \lambda_{14}^{\phi \chi} + \lambda_1^{\phi \chi} \big) \big) 
+ 2 f_1^2 v_{\chi}^2 \lambda_1^{\chi} \nonumber \\
&&+ \omega^2 \big( 2 f_1^2 v_{\chi}^2 \omega \lambda_2^{\chi} + k_2^2 v_{\phi}^2 \big( \omega^2 \lambda_2^{\phi \chi} + \lambda_3^{\phi \chi} \big) \big) 
+ k_1^2 v_{\phi}^2 \big( \lambda_1^{\phi \chi} + \lambda_8^{\phi \chi} + \omega^3 \big( \lambda_{10}^{\phi \chi} + \lambda_2^{\phi \chi} + \lambda_3^{\phi \chi} + \lambda_9^{\phi \chi} \big) \big) \nonumber \\
&&+ v_h^2 \big( 2 \lambda_{11}^{\chi H} - 2 \lambda_{12}^{\chi H} + 3 \lambda_1^{\chi H} + \lambda_8^{\chi H} + \omega \big( \lambda_2^{\chi H} + \omega \big( \lambda_2^{\chi H} + \lambda_3^{\chi H} + \omega \big( \lambda_{10}^{\chi H} + \lambda_2^{\chi H} + \lambda_3^{\chi H} + \omega \lambda_3^{\chi H} \nonumber \\
&&+ \lambda_9^{\chi H} \big) \big) \big) \big) \big) + \mu_{3}^2 \bigg)
=0,\nonumber 
\end{eqnarray}
\newpage
\begin{eqnarray}
\frac{\partial V}{\partial \chi_3} &=&
v_{\chi} \bigg( f_1^2 f_2 v_{\chi}^2 \big( 2 \lambda_1^{\chi} + \omega^2 \lambda_2^{\chi} + \omega^4 \lambda_2^{\chi} + 4 \lambda_3^{\chi} + 2 \lambda_5^{\chi} \big) 
+ f_1 \big( k_1 k_2 v_{\phi}^2 \big( \omega^2 \lambda_{10}^{\phi \chi} + \lambda_{11}^{\phi \chi} + \lambda_{12}^{\phi \chi} + \lambda_{13}^{\phi \chi} + \lambda_{14}^{\phi \chi} \nonumber \\
&&+ 2 \big( \lambda_4^{\phi \chi} + \lambda_6^{\phi \chi} \big) + \lambda_8^{\phi \chi} + \omega^4 \lambda_9^{\phi \chi} \big) \big) 
+ v_h^2 \big( \omega^4 \lambda_{10}^{\chi H} + \lambda_{11}^{\chi H} + \lambda_{12}^{\chi H} + \lambda_{13}^{\chi H} + \lambda_{14}^{\chi H} + 2 \big( \lambda_4^{\chi H} + \lambda_7^{\chi H} \big) \nonumber \\
&&+ \lambda_8^{\chi H} + \omega^2 \lambda_9^{\chi H} \big) \bigg) 
+ f_2 \big( k_1^2 v_{\phi}^2 \big( \lambda_{11}^{\phi \chi} - \lambda_{12}^{\phi \chi} - \lambda_{13}^{\phi \chi} + \lambda_{14}^{\phi \chi} + \lambda_1^{\phi \chi} \big) \big) 
+ 2 f_2^2 v_{\chi}^2 \lambda_1^{\chi} 
+ \omega^2 \big( 2 f_2^2 v_{\chi}^2 \omega \lambda_2^{\chi} \nonumber \\
&&+ k_1^2 v_{\phi}^2 \big( \lambda_2^{\phi \chi} + \omega^2 \lambda_3^{\phi \chi} \big) \big) 
+ k_2^2 v_{\phi}^2 \big( \lambda_1^{\phi \chi} + \lambda_8^{\phi \chi} + \omega^3 \big( \lambda_{10}^{\phi \chi} + \lambda_2^{\phi \chi} + \lambda_3^{\phi \chi} + \lambda_9^{\phi \chi} \big) \big) 
+ v_h^2 \big( 2 \lambda_{11}^{\chi H} - 2 \lambda_{12}^{\chi H} \nonumber \\
&&+ 3 \lambda_1^{\chi H} + \lambda_8^{\chi H} + \omega \big( \lambda_3^{\chi H} + \omega \big( \lambda_2^{\chi H} + \lambda_3^{\chi H} + \omega \big( \lambda_{10}^{\chi H} + \lambda_2^{\chi H} + \omega \lambda_2^{\chi H} + \lambda_3^{\chi H} + \lambda_9^{\chi H} \big) \big) \big) \big) \big) \nonumber \\
&& + \mu_{3}^2 \bigg)
=0,\nonumber \\
\frac{\partial V}{\partial \chi_3^\dagger} &=&
v_{\chi} \bigg( 2 f_2^3 v_{\chi}^2 \big( \lambda_1^{\chi} + \omega^3 \lambda_2^{\chi} \big) 
+ f_1 \big( k_1 k_2 v_{\phi}^2 \big( \omega^4 \lambda_{10}^{\phi \chi} + \lambda_{11}^{\phi \chi} + \lambda_{12}^{\phi \chi} - \lambda_{13}^{\phi \chi} - \lambda_{14}^{\phi \chi} + 2 \lambda_4^{\phi \chi} - 2 \lambda_6^{\phi \chi} + \lambda_8^{\phi \chi} \nonumber \\
&&+ \omega^2 \lambda_9^{\phi \chi} \big) \big) 
+ v_h^2 \big( \omega^2 \lambda_{10}^{\chi H} + \lambda_{11}^{\chi H} + \lambda_{12}^{\chi H} - \lambda_{13}^{\chi H} - \lambda_{14}^{\chi H} + 2 \lambda_4^{\chi H} - 2 \lambda_7^{\chi H} + \lambda_8^{\chi H} + \omega^4 \lambda_9^{\chi H} \big) \bigg) \nonumber \\
&&+ f_2 \big( k_1^2 v_{\phi}^2 \big( \lambda_{11}^{\phi \chi} - \lambda_{12}^{\phi \chi} - \lambda_{13}^{\phi \chi} + \lambda_{14}^{\phi \chi} + \lambda_1^{\phi \chi} \big) \big) 
+ 2 f_1^2 v_{\chi}^2 \lambda_1^{\chi} 
+ k_1^2 v_{\phi}^2 \omega^2 \lambda_2^{\phi \chi} 
+ f_1^2 v_{\chi}^2 \omega^2 \lambda_2^{\chi} 
+ f_1^2 v_{\chi}^2 \omega^4 \lambda_2^{\chi} \nonumber \\
&&+ k_1^2 v_{\phi}^2 \omega^4 \lambda_3^{\phi \chi} 
+ 4 f_1^2 v_{\chi}^2 \lambda_3^{\chi} - 2 f_1^2 v_{\chi}^2 \lambda_5^{\chi} 
+ k_2^2 v_{\phi}^2 \big( \lambda_1^{\phi \chi} + \lambda_8^{\phi \chi} + \omega^3 \big( \lambda_{10}^{\phi \chi} + \lambda_2^{\phi \chi} + \lambda_3^{\phi \chi} + \lambda_9^{\phi \chi} \big) \big) \nonumber \\
&&+ v_h^2 \big( 2 \lambda_{11}^{\chi H} - 2 \lambda_{12}^{\chi H} + 3 \lambda_1^{\chi H} + \lambda_8^{\chi H} + \omega \big( \lambda_3^{\chi H} + \omega \big( \lambda_2^{\chi H} + \lambda_3^{\chi H} + \omega \big( \lambda_{10}^{\chi H} + \lambda_2^{\chi H} + \omega \lambda_2^{\chi H} + \lambda_3^{\chi H} \nonumber \\
&&+ \lambda_9^{\chi H} \big) \big) \big) \big) \big) + \mu_{3}^2 \bigg)
=0.
\end{eqnarray}

\section{\label{A}{Analytical Expressions of Observable Parameters of Various Types of Models under NH and IH}}

In Sect.\ref{Section 2(c)}, we have given the analytical expressions of predicted observable parameters for Type 1 models under both NH and IH. In this section, we are giving the expressions for all other models under both the hierarchies.

\textbf{Type 2 (NH)}
\begin{eqnarray}
\label{ec}
a &=& \Delta m^2_{21},\nonumber\\
b &=& \frac{\Delta m^2_{31}}{2} + \frac{1}{2(3-4\cos^2\,\rho)^2}[\Delta m^4_{31}(3-4\cos^2\,\rho)^2(16 \cos^4\,\rho-16 \cos^2\,\rho-3 \nonumber\\&&
+8\cos\,\rho \sqrt{3-3\cos^2\,\rho}-24 \sin^2\,\theta_{13}\,(2\cos^2\,\rho+2\cos\,\rho\sqrt{3-3\cos^2\,\rho}-3) \nonumber \\&&
+36\sin^2\,\theta_{13}(2\cos^2\,\rho+2\cos\,\rho\sqrt{3-3\cos^2\,\rho}-3)) ]^\frac{1}{2},\nonumber
\end{eqnarray}
\begin{eqnarray}
c &=& \frac{\Delta m^2_{31}}{2} - \frac{1}{2(3-4\cos^2\,\rho)^2}[\Delta m^4_{31}(3-4\cos^2\,\rho)^2(16 \cos^4\,\rho-16 \cos^2\,\rho-3 \nonumber\\&&
+8\cos\,\rho \sqrt{3-3\cos^2\,\rho}-24 \sin^2\,\theta_{13}\,(2\cos^2\,\rho+2\cos\,\rho\sqrt{3-3\cos^2\,\rho}-3) \nonumber \\&&
+36\sin^2\,\theta_{13}(2\cos^2\,\rho+2\cos\,\rho\sqrt{3-3\cos^2\,\rho}-3)) ]^\frac{1}{2},\nonumber \\
\sin^2\,\theta_{12} &=& \frac{\Delta m^2_{31}}{2 \Delta m^2_{31}+(\cos\rho+\sqrt{3-3\cos^2\rho}) \sqrt{\frac{\Delta m^4_{31}(1-3\sin^2\,\theta_{13})^2(\cos\rho-\sqrt{3-3\cos^2\rho})^2}{(3-4\cos^2\,\rho)^2}}},\nonumber\\
\sin^2\,\theta_{23} &=& \frac{\Delta m^2_{31}-(\cos\rho-\sqrt{3-3\cos^2\rho})\sqrt{\frac{\Delta m^4_{31}(1-3\sin^2\,\theta_{13})^2(\cos\rho-\sqrt{3-3\cos^2\rho})^2}{(3-4\cos^2\rho)^2}}} {2 \Delta m^2_{31}+(\cos\rho+\sqrt{3-3\cos^2\rho}) \sqrt{\frac{\Delta m^4_{31}(1-3\sin^2\,\theta_{13})^2(\cos\rho-\sqrt{3-3\cos^2\rho})^2}{(3-4\cos^2\,\rho)^2}}},\nonumber\\
\delta &=& -arg \left[\frac{e^{-i\eta_1(2\sqrt{\frac{c}{b+c}}-(1+i\sqrt{3})\sqrt{\frac{b}{b+c}}e^{-i\rho})}}{2\sqrt{3}} \right].
\end{eqnarray}

\textbf{Parametrization Conditions}
\begin{eqnarray}
\label{ed}
\textbf{(A).} && \frac{1}{6} \left( 3 \sqrt{\frac{c}{b + c}} \cos(\eta_4 + \rho + \eta_1) - \sqrt{3} \left( 2 \sqrt{\frac{b}{b + c}} \sin(\eta_4 + \eta_1) + \sqrt{\frac{c}{b + c}} \sin(\eta_4 + \rho + \eta_1) \right) \right) = 0, \nonumber \\
\textbf{(B).} && \frac{1}{6} \left( 3 \cos(\eta_5 + \eta_1) + \sqrt{3} \sin(\eta_5 + \eta_1) \right) = 0, \nonumber \\
 \textbf{(C).}&& \frac{1}{6} \left( 3 \sqrt{\frac{b}{b + c}} \cos(\rho + \eta_2) + \sqrt{3} \left( -2 \sqrt{\frac{c}{b + c}} \sin(\eta_2) + \sqrt{\frac{b}{b + c}} \sin(\rho + \eta_2) \right) \right) = 0, \nonumber \\
\textbf{(D).} && - \frac{\sqrt{\frac{c}{b + c}} \sin(\eta_3) + \sqrt{\frac{b}{b + c}} \sin(\rho + \eta_3)}{\sqrt{3}} = 0, \nonumber \\
\textbf{(E).} && \frac{1}{6 \sqrt{3}} \left( -3 \sqrt{\frac{c}{b + c}} \cos(\eta_4 + \rho + \eta_2) + 3 \cos(\eta_5 + \eta_2) \sqrt{\frac{b + c + \sqrt{bc} \cos(\rho) + \sqrt{3} \sqrt{bc} \sin(\rho)}{b + c}} \right. \nonumber \\
&& \left. - \sqrt{3} \left( 2 \sqrt{\frac{b}{b + c}} \sin(\eta_4 + \eta_2) + \sqrt{\frac{b + c + \sqrt{bc} \cos(\rho) + \sqrt{3} \sqrt{bc} \sin(\rho)}{b + c}} \sin(\eta_5 + \eta_2) \right.\right. \nonumber \\
&& \left.\left. + \sqrt{\frac{c}{b + c}} \sin(\eta_4 + \rho + \eta_2) \right) \right) = 0.
\end{eqnarray}

\textbf{Type 3 (NH)}
\begin{eqnarray}
\label{ee}
a &=& \Delta m^2_{21},\nonumber \\
b &=& \frac{\Delta m^2_{31}}{2} + \frac{\sqrt{\Delta m^4_{31}\,\cos^2\,\rho\,(\cos^2\,\rho-9\, \sin^4\,\theta_{13}+6\,\sin^2\,\theta_{13}-1)}}{2\,\cos^2\,\rho},\nonumber\\
c &=& \frac{\Delta m^2_{31}}{2}-\frac{\sqrt{\Delta m^4_{31}\,\cos^2\,\rho\,(\cos^2\,\rho-9\, \sin^4\,\theta_{13}+6\,\sin^2\,\theta_{13}-1)}}{2\,\cos^2\,\rho},\nonumber\\
\sin^2\,\theta_{12} &=& \frac{\Delta m^2_{31}}{2\, \Delta m^2_{31}-\sqrt{\frac{\Delta m^4_{31}(-1+3\,\sin^2\,\theta_{13})^2}{\cos^2\,\rho}} \cos\,\rho},\nonumber\\
\sin^2\,\theta_{23} &=& \frac{2\,\Delta m^2_{31}- \sqrt{\frac{\Delta m^4_{31}(1-3\,\sin^2\,\theta_{13})^2}{\cos^2\,\rho}} (x+\sqrt{3-3\,\cos^2\,\rho})}{4\,\Delta m^2_{31}-2\sqrt{\frac{\Delta m^4_{31}(-1+3\,\sin^2\,\theta_{13})^2}{\cos^2\,\rho} \cos\,\rho}},\nonumber\\
\delta &=& -arg \left[ \frac{e^{-i \eta_1}(\sqrt{\frac{c}{b+c}}+\sqrt{\frac{b}{b+c}}e^{-i \rho})}{\sqrt{3}} \right].
\end{eqnarray}

\textbf{Parametrization Conditions}
\begin{eqnarray}
\label{ef}
\textbf{(A).}&& \sqrt{c}\,\sin\,(\rho+\eta_1+\eta_4)-\sqrt{b}\,\sin\,(\eta_1+\eta_4)=0,\nonumber\\
\textbf{(B).}&& \sin\,(\eta_1+\eta_5)=0,\nonumber\\
\textbf{(C).}&& \sqrt{3b}\,\sin\,(\rho+\eta_2)-2\sqrt{3c}\,\sin\,\eta_2-3\sqrt{b}\,\cos\,(\rho+\eta_2)=0,\nonumber\\
\textbf{(D).}&& \sqrt{3b}\,\sin\,(\rho+\eta_3)-2\sqrt{3c}\,\sin\,\eta_3-3\sqrt{b}\,\cos\,(\rho+\eta_3)=0,\nonumber\\
\textbf{(E).}&& \sqrt{b+c-2\sqrt{bc}\,\cos\rho}\,(\sqrt{3}\,\sin\,(\eta_2+\eta_5)-3\,\cos\,(\eta_2+\eta_5)) \nonumber \\&&
-3\sqrt{c}\,\cos\,(\rho+\eta_2+\eta_4)+\sqrt{3c}\,\sin\,(\rho+\eta_2+\eta_4) \nonumber \\
&&+2\sqrt{3b}\,\sin\,(\eta_2+\eta_4)=0.
\end{eqnarray}

\textbf{Type 4 (NH)}
\begin{eqnarray}
\label{eg}
a &=& \Delta m^2_{21},\nonumber \\
b &=& \frac{\Delta m^2_{31}}{2} + \frac{1}{2(3-4\cos^2\,\rho)^2}[\Delta m^4_{31}(3-4\cos^2\,\rho)^2(16 \cos^4\,\rho-16 \cos^2\,\rho-3 \nonumber\\&&
+8\cos\,\rho \sqrt{3-3\cos^2\,\rho}-24 \sin^2\,\theta_{13}\,(2\cos^2\,\rho+2\cos\,\rho\sqrt{3-3\cos^2\,\rho}-3) \nonumber \\&&
+36\sin^2\,\theta_{13}(2\cos^2\,\rho+2\cos\,\rho\sqrt{3-3\cos^2\,\rho}-3)) ]^\frac{1}{2},\nonumber
\end{eqnarray}
\begin{eqnarray}
c &=& \frac{\Delta m^2_{31}}{2} - \frac{1}{2(3-4\cos^2\,\rho)^2}[\Delta m^4_{31}(3-4\cos^2\,\rho)^2(16 \cos^4\,\rho-16 \cos^2\,\rho-3 \nonumber\\&&
+8\cos\,\rho \sqrt{3-3\cos^2\,\rho}-24 \sin^2\,\theta_{13}\,(2\cos^2\,\rho+2\cos\,\rho\sqrt{3-3\cos^2\,\rho}-3) \nonumber \\&&
+36\sin^2\,\theta_{13}(2\cos^2\,\rho+2\cos\,\rho\sqrt{3-3\cos^2\,\rho}-3)) ]^\frac{1}{2},\nonumber\\
\sin^2\,\theta_{12} &=& \frac{\Delta m^2_{31}}{2 \Delta m^2_{31}+(\cos\rho+\sqrt{3-3\cos^2\rho}) \sqrt{\frac{\Delta m^4_{31}(1-3\sin^2\,\theta_{13})^2(\cos\rho-\sqrt{3-3\cos^2\rho})^2}{(3-4\cos^2\,\rho)^2}}},\nonumber\\
\sin^2\,\theta_{23} &=& \frac{\Delta m^2_{31}+2\cos\,\rho\sqrt{\frac{\Delta m^4_{31}(1-3\sin^2\,\theta_{13})^2(\cos\rho-\sqrt{3-3\cos^2\rho})^2}{(3-4\cos^2\rho)^2}}} {2 \Delta m^2_{31}+(\cos\rho+\sqrt{3-3\cos^2\rho}) \sqrt{\frac{\Delta m^4_{31}(1-3\sin^2\,\theta_{13})^2(\cos\rho-\sqrt{3-3\cos^2\rho})^2}{(3-4\cos^2\,\rho)^2}}},\nonumber\\
\delta &=& -arg \left[\frac{e^{-i\eta_1(2\sqrt{\frac{c}{b+c}}-(1+i\sqrt{3})\sqrt{\frac{b}{b+c}}e^{-i\rho})}}{2\sqrt{3}} \right].
\end{eqnarray}

\textbf{Parametrization Conditions}
\begin{eqnarray}
\label{eh}
\textbf{(A).}&& 3\sqrt{c}\,\cos\,(\rho+\eta_1+\eta_4)-2\sqrt{3b}\,\sin\,(\eta_1+\eta_4)-\sqrt{3c}\,\sin\,(\rho+\eta_1+\eta_4)=0,\nonumber\\
\textbf{(B).}&& 3\cos\,(\eta_1+\eta_5)+\sqrt{3}\,\sin\,(\eta_1+\eta_5)=0,\nonumber\\
\textbf{(C).}&& \sqrt{c}\,\sin\,\eta_2+\sqrt{b}\,\sin\,(\rho+\eta_2) =0,\nonumber\\
\textbf{(D).}&& 3 \sqrt{b}\,\cos\,(\rho+\eta_3)-2\sqrt{3c}\,\sin\,\eta_3+\sqrt{3b}\,\sin\,(\rho+\eta_3)=0,\nonumber\\
\textbf{(E).}&& \sqrt{b+c+\sqrt{bc}\,(\cos\rho+\sqrt{3}\,\sin\rho)}\,\sin\,(\eta_2+\eta_5)+\sqrt{c}\,\sin\,(\rho+\eta_2+\eta_4) \nonumber \\&&
-\sqrt{b}\,\sin\,(\eta_2+\eta_4)=0.
\end{eqnarray}

\textbf{Type 5 (NH)} 
\begin{eqnarray}
a & = & \Delta m^2_{21}, \nonumber \\
b & = & \frac{\Delta m^2_{31} \cos^2 \rho + \sqrt{\Delta m^4_{31} \cos^2 \rho \left(-1 + 6 \sin^2 \theta_{13} - 9 \sin^4 \theta_{13} + \cos^2 \rho\right)}}{2 \cos^2 \rho}, \nonumber \\
c & = & \frac{1}{2} \left(\Delta m^2_{31} - \frac{\sqrt{\Delta m^4_{31} \cos^2 \rho \left(-1 + 6 \sin^2 \theta_{13} - 9 \sin^4 \theta_{13} + \cos^2 \rho\right)}}{\cos^2 \rho}\right), \nonumber 
\end{eqnarray}
\begin{eqnarray}
\sin^2\, \theta_{12} & = & \frac{\Delta m^2_{31}}{2 \Delta m^2_{31} - \sqrt{\frac{\Delta m^4_{31} \left(-1 + 3 \sin^2 \theta_{13}\right)^2}{\cos^2 \rho}} \cos \rho}, \nonumber \\
\sin^2\, \theta_{23} & = & \frac{2 \Delta m^2_{31} + \sqrt{\frac{\Delta m^4_{31} \left(1 - 3 \sin^2 \theta_{13}\right)^2}{\cos^2 \rho}} \left(-\cos \rho + \sqrt{3 - 3 \cos^2 \rho}\right)}{4 \Delta m^2_{31} - 2 \sqrt{\frac{\Delta m^4_{31} \left(-1 + 3 \sin^2 \theta_{13}\right)^2}{\cos^2 \rho}} \cos \rho}, \nonumber \\
\delta & = & -\arg\left[\frac{e^{-i \eta_1} \left(\sqrt{\frac{c}{b + c}} + \sqrt{\frac{b}{b + c}} e^{-i \rho}\right)}{\sqrt{3}}\right].
\end{eqnarray}

\textbf{Parametrization Conditions}
\begin{eqnarray}
\textbf{(A).}&& \frac{-\sqrt{\frac{b}{b + c}} \sin(\eta_4 + \eta_1) + \sqrt{\frac{c}{b + c}} \sin(\eta_4 + \rho + \eta_1)}{\sqrt{3}} = 0, \nonumber \\
\textbf{(B).}&& -\frac{\sin(\eta_5 + \eta_1)}{\sqrt{3}} = 0, \nonumber \\
\textbf{(C).}&& \frac{1}{6} \left( 3 \sqrt{\frac{c}{b + c}} \cos(\eta_1) - 3 \sqrt{\frac{b}{b + c}} \cos(\rho + \eta_1) 
+ \sqrt{3} \left( \sqrt{\frac{c}{b + c}} \sin(\eta_1) \right.\right.\nonumber \\
&& \left.\left. + \sqrt{\frac{b}{b + c}} \sin(\rho + \eta_1) \right) \right) = 0, \nonumber \\
\textbf{(D).}&& \frac{1}{6} \left( -3 \sqrt{\frac{c}{b + c}} \cos(\eta_2) + 3 \sqrt{\frac{b}{b + c}} \cos(\rho + \eta_2) 
+ \sqrt{3} \left( \sqrt{\frac{c}{b + c}} \sin(\eta_2) \right.\right.\nonumber \\
&& \left.\left.+ \sqrt{\frac{b}{b + c}} \sin(\rho + \eta_2) \right) \right) = 0, \nonumber \\
\textbf{(E).}&& \frac{1}{6} \left( \sqrt{3} \sqrt{\frac{b}{b + c}} \cos(\eta_4 + \eta_2) 
+ \sqrt{3} \sqrt{\frac{c}{b + c}} \cos(\eta_4 + \rho + \eta_2) 
+ \sqrt{\frac{b}{b + c}} \sin(\eta_4 + \eta_2)\right. \nonumber \\
&& \left.+ 2 \sqrt{\frac{b + c - 2 \sqrt{b c} \cos(\rho)}{b + c}} \sin(\eta_5 + \eta_2) 
- \sqrt{\frac{c}{b + c}} \sin(\eta_4 + \rho + \eta_2) \right) = 0.
\end{eqnarray}

\textbf{Type 6 (NH)}
\begin{eqnarray}
a & = & \Delta m^2_{21}, \nonumber \\
b &=& \frac{1}{2} \Delta m^2_{31} + \frac{1}{(3 - 4 \cos^2 \rho)^2} \left[ \Delta m^4_{31} (3 - 4 \cos^2 \rho)^2 ( -3 + 8 \cos \rho ( -2 \cos \rho + 2 \cos^3 \rho 
\right. \nonumber \\
&& \left.+ \sqrt{3 - 3 \cos^2 \rho} )
- 24 \sin^2 \theta_{13} ( -3 + 2 \cos \rho ( \cos \rho + \sqrt{3 - 3 \cos^2 \rho} ) ) + 36 \sin^2 \theta_{13}^2 ( -3 \right. \nonumber \\
&& \left.+ 2 \cos \rho ( \cos \rho 
+ \sqrt{3 - 3 \cos^2 \rho} ) ) ) \right]^{\frac{1}{2}}, \nonumber 
\end{eqnarray}
\begin{eqnarray}
c &=& \frac{1}{2} \Delta m^2_{31} - \frac{1}{(3 - 4 \cos^2 \rho)^2} [ \Delta m^4_{31} (3 - 4 \cos^2 \rho)^2 ( -3 + 8 \cos \rho ( -2 \cos \rho + 2 \cos^3 \rho \nonumber \\
&& + \sqrt{3 - 3 \cos^2 \rho} ) 
- 24 \sin^2 \theta_{13} ( -3 + 2 \cos \rho ( \cos \rho + \sqrt{3 - 3 \cos^2 \rho} ) ) + 36 \sin^2 \theta_{13}^2 ( -3 \nonumber \\
&& + 2 \cos \rho ( \cos \rho 
+ \sqrt{3 - 3 \cos^2 \rho} ) ) ) ]^{\frac{1}{2}},\nonumber \\
\sin^2 \theta_{12} & = &\frac{\Delta m^2_{31}}{2 \Delta m^2_{31} + (\cos \rho + \sqrt{3 - 3 \cos^2 \rho}) \sqrt{-\frac{(\Delta m^2_{31} - 3 \Delta m^2_{31} \sin^2 \theta_{13})^2 (-3 + 2 \cos^2 \rho + 2 \cos \rho \sqrt{3 - 3 \cos^2 \rho})}{(3 - 4 \cos^2 \rho)^2}}},\nonumber \\
\sin^2 \theta_{23} & = &\frac{\Delta m^2_{31} + \left( - \cos \rho + \sqrt{3 - 3 \cos^2 \rho} \right) \sqrt{-\left( \frac{(\Delta m^2_{31} - 3 \Delta m^2_{31} \sin^2 \theta_{13})^2 (-3 + 2 \cos^2 \rho + 2 \cos \rho \sqrt{3 - 3 \cos^2 \rho})}{(3 - 4 \cos^2 \rho)^2} \right)}}{2 \Delta m^2_{31} + \left( \cos \rho + \sqrt{3 - 3 \cos^2 \rho} \right) \sqrt{-\left( \frac{(\Delta m^2_{31} - 3 \Delta m^2_{31} \sin^2 \theta_{13})^2 (-3 + 2 \cos^2 \rho + 2 \cos \rho \sqrt{3 - 3 \cos^2 \rho})}{(3 - 4 \cos^2 \rho)^2} \right)}}, \nonumber \\
\delta &=&-arg\left[ \frac{i e^{-i (\rho + \eta_1)} \left( (i + \sqrt{3}) \sqrt{\frac{b}{b + c}} - (-i + \sqrt{3}) \sqrt{\frac{c}{b + c}} e^{i \rho} \right)}{2 \sqrt{3}}\right],\nonumber \\
\end{eqnarray}

\textbf{Parametrization Conditions}
\begin{eqnarray}
\textbf{(A).}&& \frac{1}{6} \bigg( -3 \sqrt{\frac{b}{b + c}} \cos(\eta_4 + \eta_1) - 3 \sqrt{\frac{c}{b + c}} \cos(\eta_4 + \rho + \eta_1) + \sqrt{3} ( \sqrt{\frac{b}{b + c}} \sin(\eta_4 + \eta_1) \nonumber \\
&&- \sqrt{\frac{c}{b + c}} \sin(\eta_4 + \rho + \eta_1) ) \bigg)
 = 0, \nonumber \\
\textbf{(B).}&&\frac{\sin(\eta_5 + \eta_1)}{\sqrt{3}}
=0,\nonumber 
\end{eqnarray}
\begin{eqnarray}
\textbf{(C).}&& \frac{1}{6} \left( 3 \sqrt{\frac{c}{b + c}} \cos(\eta_2) - 3 \sqrt{\frac{b}{b + c}} \cos(\rho + \eta_2) 
+ \sqrt{3} \left( \sqrt{\frac{c}{b + c}} \sin(\eta_2) \right.\right.\nonumber \\
&&\left.\left.+ \sqrt{\frac{b}{b + c}} \sin(\rho + \eta_2) \right) \right) = 0, \nonumber \\
\textbf{(D).}&& -\frac{1}{\sqrt{3}} \left( \sqrt{\frac{c}{b + c}} \sin(\eta_3) + \sqrt{\frac{b}{b + c}} \sin(\rho + \eta_3) \right) = 0, \nonumber \\
\textbf{(E).}&& \frac{1}{6} \left( \sqrt{3} \sqrt{\frac{b}{b + c}} \cos(\eta_4 + \eta_2) 
+ \sqrt{3} \sqrt{\frac{c}{b + c}} \cos(\eta_4 + \rho + \eta_2) 
+ \sqrt{\frac{b}{b + c}} \sin(\eta_4 + \eta_2) \right. \nonumber \\
&& \left. + 2 \sqrt{\frac{b + c + \sqrt{b c} \cos(\rho) + \sqrt{3} \sqrt{b c} \sin(\rho)}{b + c}} \sin(\eta_5 + \eta_2) \right.\nonumber \\
&&\left.- \sqrt{\frac{c}{b + c}} \sin(\eta_4 + \rho + \eta_2) \right) = 0.
\end{eqnarray}

\textbf{Type 7 (NH)}
\begin{eqnarray}
a &=& \Delta m^2_{21}, \nonumber \\
b &=& \frac{\Delta m^2_{31} \cos^2 \rho + [\Delta m^4_{31} \cos^2 \rho (-1 + 6 \sin^2 \theta_{13} - 9 \sin^4 \theta_{13} + \cos^2 \rho)]^{1/2}}{2 \cos^2 \rho}, \nonumber \\
c &=& \frac{1}{2} \left( \Delta m^2_{31} - \frac{[\Delta m^4_{31} \cos^2 \rho (-1 + 6 \sin^2 \theta_{13} - 9 \sin^4 \theta_{13} + \cos^2 \rho)]^{1/2}}{\cos^2 \rho} \right),\nonumber \\
\sin^2 \theta_{12} &=& \frac{\Delta m^2_{31}}{2 \Delta m^2_{31} - \sqrt{\frac{\Delta m^4_{31} (-1 + 3 \sin^2 \theta_{13})^2}{\cos^2 \rho}} \cos \rho},\nonumber \\
\sin^2 \theta_{23} &=& \frac{2 \Delta m^2_{31} - \sqrt{\frac{\Delta m^4_{31} (1 - 3 \sin^2 \theta_{13})^2}{\cos^2 \rho}} (\cos \rho + \sqrt{3 - 3 \cos^2 \rho})}{4 \Delta m^2_{31} - 2 \sqrt{\frac{\Delta m^4_{31} (1 - 3 \sin^2 \theta_{13})^2}{\cos^2 \rho}} \cos \rho},\,\nonumber \\
\delta &=& -\arg \left[\frac{e^{-i \eta_1} \left( \sqrt{\frac{c}{b + c}} + \sqrt{\frac{b}{b + c}} e^{-i \rho} \right)}{\sqrt{3}} \right].
\end{eqnarray}

\textbf{Parametrization Conditions}
\begin{eqnarray}
\textbf{(A).} && \frac{-\sqrt{\frac{b}{b + c}} \sin(\eta_4 + \eta_1) + \sqrt{\frac{c}{b + c}} \sin(\eta_4 + \rho + \eta_1)}{\sqrt{3}} = 0, \nonumber \\
\textbf{(B).} && -\frac{\sin(\eta_5 + \eta_1)}{\sqrt{3}} = 0, \nonumber \\
\textbf{(C).} && \frac{1}{6} \left( -3 \sqrt{\frac{c}{b + c}} \cos(\eta_2) + 3 \sqrt{\frac{b}{b + c}} \cos(\rho + \eta_2) 
+ \sqrt{3} \left( \sqrt{\frac{c}{b + c}} \sin(\eta_2) \right.\right.\nonumber \\
&&\left.\left.+ \sqrt{\frac{b}{b + c}} \sin(\rho + \eta_2) \right) \right) = 0, \nonumber \\
\textbf{(D).} && \frac{1}{6} \left( 3 \sqrt{\frac{c}{b + c}} \cos(\eta_3) - 3 \sqrt{\frac{b}{b + c}} \cos(\rho + \eta_3) 
+ \sqrt{3} \left( \sqrt{\frac{c}{b + c}} \sin(\eta_3) \right.\right.\nonumber \\
&&\left.\left. + \sqrt{\frac{b}{b + c}} \sin(\rho + \eta_3) \right) \right) = 0, \nonumber \\
\textbf{(E).} && \frac{1}{6} \left( -\sqrt{3} \sqrt{\frac{b}{b + c}} \cos(\eta_4 + \eta_2) 
- \sqrt{3} \sqrt{\frac{c}{b + c}} \cos(\eta_4 + \rho + \eta_2) 
+ \sqrt{\frac{b}{b + c}} \sin(\eta_4 + \eta_2)\right. \nonumber \\
&& \left.+ 2 \sqrt{\frac{b + c - 2 \sqrt{b c} \cos(\rho)}{b + c}} \sin(\eta_5 + \eta_2) 
- \sqrt{\frac{c}{b + c}} \sin(\eta_4 + \rho + \eta_2) \right) = 0.
\end{eqnarray}

\textbf{Type 8 (NH)}
\begin{eqnarray}
a &=& \Delta m^2_{21}, \nonumber \\
b &=& \frac{1}{2 (3 - 4 \cos^2 \rho)^2} \bigg( \Delta m^2_{31} (3 - 4 \cos^2 \rho)^2 + 
\big[ \Delta m^4_{31} (3 - 4 \cos^2 \rho)^2 \big( -3 - 16 \cos^2 \rho + 16 \cos^4 \rho \nonumber \\
&&+ 8 \cos \rho \sqrt{3 - 3 \cos^2 \rho} - 
24 \sin^2 \theta_{13} (-3 + 2 \cos^2 \rho + 2 \cos \rho \sqrt{3 - 3 \cos^2 \rho}) + 
36 \sin^4 \theta_{13} (-3 \nonumber \\
&&+ 2 \cos^2 \rho + 2 \cos \rho \sqrt{3 - 3 \cos^2 \rho}) \big) \big]^{1/2} \bigg), \nonumber \\
c &=& \frac{1}{2 (3 - 4 \cos^2 \rho)^2} \bigg( \Delta m^2_{31} (3 - 4 \cos^2 \rho)^2 - 
\big[ \Delta m^4_{31} (3 - 4 \cos^2 \rho)^2 \big( -3 - 16 \cos^2 \rho + 16 \cos^4 \rho \nonumber \\
&& + 8 \cos \rho \sqrt{3 - 3 \cos^2 \rho} - 
24 \sin^2 \theta_{13} (-3 + 2 \cos^2 \rho + 2 \cos \rho \sqrt{3 - 3 \cos^2 \rho}) + 
36 \sin^4 \theta_{13} (-3 \nonumber \\
&&+ 2 \cos^2 \rho + 2 \cos \rho \sqrt{3 - 3 \cos^2 \rho}) \big) \big]^{1/2} \bigg),\nonumber \\
\sin^2 \theta_{12} &=& \frac{\Delta m^2_{31}}{2 \Delta m^2_{31} + (\cos \rho + \sqrt{3 - 3 \cos^2 \rho}) \sqrt{-\frac{(\Delta m^2_{31} - 3 \Delta m^2_{31} \sin^2 \theta_{13})^2 (-3 + 2 \cos^2 \rho + 2 \cos \rho \sqrt{3 - 3 \cos^2 \rho})}{(3 - 4 \cos^2 \rho)^2}}}, \nonumber \\
\sin^2 \theta_{23} &=& \frac{\Delta m^2_{31} + 2 \cos \rho \sqrt{-\frac{(\Delta m^2_{31} - 3 \Delta m^2_{31} \sin^2 \theta_{13})^2 (-3 + 2 \cos^2 \rho + 2 \cos \rho \sqrt{3 - 3 \cos^2 \rho})}{(3 - 4 \cos^2 \rho)^2}}}{2 \Delta m^2_{31} + (\cos \rho + \sqrt{3 - 3 \cos^2 \rho}) \sqrt{-\frac{(\Delta m^2_{31} - 3 \Delta m^2_{31} \sin^2 \theta_{13})^2 (-3 + 2 \cos^2 \rho + 2 \cos \rho \sqrt{3 - 3 \cos^2 \rho})}{(3 - 4 \cos^2 \rho)^2}}}, \nonumber \\
\delta &=& -arg \left[\frac{ i e^{-i \rho + \eta_1)} \left( (i + \sqrt{3}) \sqrt{\frac{b}{b + c}} - (-i + \sqrt{3}) \sqrt{\frac{c}{b + c}} e^{i \rho} \right) }{2 \sqrt{3}} \right].
\end{eqnarray}

\textbf{Parametrization Conditions}
\begin{eqnarray}
\textbf{(A).} && \frac{1}{6} \left( -3 \sqrt{\frac{b}{b + c}} \cos(\eta_4 + \eta_1) 
- 3 \sqrt{\frac{c}{b + c}} \cos(\eta_4 + \rho + \eta_1) 
+ \sqrt{3} \left( \sqrt{\frac{b}{b + c}} \sin(\eta_4 + \eta_1) \right.\right. \nonumber \\
&&\left.\left. - \sqrt{\frac{c}{b + c}} \sin(\eta_4 + \rho + \eta_1) \right) \right) = 0, \nonumber \\
\textbf{(B).} && -\frac{\sin(\eta_5 + \eta_1)}{\sqrt{3}} = 0, \nonumber \\
\textbf{(C).} && -\frac{\sqrt{\frac{c}{b + c}} \sin(\eta_2) 
+ \sqrt{\frac{b}{b + c}} \sin(\rho + \eta_2)}{\sqrt{3}} = 0, \nonumber 
\end{eqnarray}
\begin{eqnarray}
\textbf{(D).} && \frac{1}{6} \left( 3 \sqrt{\frac{c}{b + c}} \cos(\eta_3) 
- 3 \sqrt{\frac{b}{b + c}} \cos(\rho + \eta_3) 
+ \sqrt{3} \left( \sqrt{\frac{c}{b + c}} \sin(\eta_3) \right. \right.\nonumber \\
&&\left. \left.+ \sqrt{\frac{b}{b + c}} \sin(\rho + \eta_3) \right) \right) = 0, \nonumber \\
\textbf{(E).} && \frac{1}{3} \left( -\sqrt{\frac{b}{b + c}} \sin(\eta_4 + \eta_2) 
+ \sqrt{\frac{b + c + \sqrt{b c} \cos(\rho) 
+ \sqrt{3} \sqrt{b c} \sin(\rho)}{b + c}} \sin(\eta_5 + \eta_2)\right. \nonumber \\
&&\left. + \sqrt{\frac{c}{b + c}} \sin(\eta_4 + \rho + \eta_2) \right) = 0.
\end{eqnarray}

\textbf{Type 9 (NH)}
\begin{eqnarray}
a &=& \Delta m^2_{21}, \nonumber \\
b &=& \frac{\Delta m^2_{31} \cos^2 \rho + \sqrt{\Delta m^4_{31} \cos^2 \rho (-1 + 6 \sin^2 \theta_{13} - 9 \sin^4 \theta_{13} + \cos^2 \rho)}}{2 \cos^4 \rho}, \nonumber \\
c &=& \frac{1}{2} \left( \Delta m^2_{31} - \frac{\sqrt{\Delta m^4_{31} \cos^2 \rho (-1 + 6 \sin^2 \theta_{13} - 9 \sin^4 \theta_{13} + \cos^2 \rho)}}{\cos^2 \rho} \right),\nonumber \\
\sin^2 \theta_{12} &=& \frac{\Delta m^2_{31}}{2 \Delta m^2_{31} - \sqrt{\frac{(\Delta m^2_{31})^2 (-1 + 3 \sin^2 \theta_{13})^2}{\cos^2 \rho}} \cos \rho},\nonumber \\
\sin^2 \theta_{23} & =& \frac{2 \Delta m^2_{31} + \sqrt{\frac{\Delta m^4_{31} (1 - 3 \sin^2 \theta_{13})^2}{\cos^2 \rho}} (-\cos \rho + \sqrt{3 - 3 \cos^2 \rho})}{4 \Delta m^2_{31} - 2 \sqrt{\frac{\Delta m^4_{31} (-1 + 3 \sin^2 \theta_{13})^2}{\cos^2 \rho}} \cos \rho},\nonumber \\
\delta &=& -\arg\left[ \frac{ \left( e^{-i \eta_1} \left( \sqrt{\frac{c}{b + c}} + \sqrt{\frac{b}{b + c}} e^{-i \arccos(\cos \rho)} \right) \right) }{\sqrt{3}}\right].
\end{eqnarray}

\textbf{Parametrization Conditions}
\begin{eqnarray}
\textbf{(A).} && \frac{-\sqrt{\frac{b}{b + c}} \sin(\eta_4 + \eta_1) + \sqrt{\frac{c}{b + c}} \sin(\eta_4 + \rho + \eta_1)}{\sqrt{3}} = 0, \nonumber \\
\textbf{(B).} && -\frac{\sin(\eta_5 + \eta_1)}{\sqrt{3}} = 0, \nonumber 
\end{eqnarray}
\begin{eqnarray}
\textbf{(C).} && \frac{1}{6} \left( -3 \sqrt{\frac{c}{b + c}} \cos(\eta_2) 
+ \sqrt{3} \left( \sqrt{\frac{c}{b + c}} \sin(\eta_2) 
- 2 \sqrt{\frac{b}{b + c}} \sin(\rho + \eta_2) \right) \right) = 0, \nonumber \\
\textbf{(D).} && \frac{1}{6} \left( 3 \sqrt{\frac{c}{b + c}} \cos(\eta_3) 
+ \sqrt{3} \left( \sqrt{\frac{c}{b + c}} \sin(\eta_3) 
- 2 \sqrt{\frac{b}{b + c}} \sin(\rho + \eta_3) \right) \right) = 0, \nonumber \\
\textbf{(E).}&& \frac{1}{6 \sqrt{3}} \left( 
-3 \sqrt{\frac{b}{b + c}} \cos(\eta_4 + \eta_2) 
- 3 \sqrt{\frac{b + c - 2 \sqrt{b c} \cos(\rho)}{b + c}} \cos(\eta_5 + \eta_2) \right. \nonumber \\
&& \left.+ \sqrt{3} \left( 
\sqrt{\frac{b}{b + c}} \sin(\eta_4 + \eta_2) 
- \sqrt{\frac{b + c - 2 \sqrt{b c} \cos(\rho)}{b + c}} \sin(\eta_5 + \eta_2) \right. \right. \nonumber \\
&& \left. \left. + 2 \sqrt{\frac{c}{b + c}} \sin(\eta_4 + \rho + \eta_2) 
\right) 
\right) = 0.
\end{eqnarray}

\textbf{Type 10 (NH)}
\begin{eqnarray}
a &=& \Delta m^2_{21}, \nonumber \\
b &=& \frac{1}{2 (3 - 4 \cos^2 \rho)^2} \left[ \Delta m^4_{31} (3 - 4 \cos^2 \rho)^2 + \left( \Delta m^2_{31} (3 - 4 \cos^2 \rho)^2 \left( -3 - 16 \cos^2 \rho + 16 \cos^4 \rho \right.\right.\right.   \nonumber \\
&&\left.\left.\left.+ 8 \cos \rho \sqrt{3 - 3 \cos^2 \rho} - 24 \sin^2 \theta_{13} \left( -3 + 2 \cos^2 \rho + 2 \cos \rho \sqrt{3 - 3 \cos^2 \rho} \right) + 36 \sin^2 \theta_{13}^2 \left( -3 \right.\right.\right. \right.\nonumber \\
 &&\left.\left.\left.\left. + 2 \cos^2 \rho + 2 \cos \rho \sqrt{3 - 3 \cos^2 \rho} \right) \right) \right) \right]^{1/2}, \nonumber \\
c &=& \frac{1}{2 (3 - 4 \cos^2 \rho)^2} \left[ \Delta m^4_{31} (3 - 4 \cos^2 \rho)^2 - \left( \Delta m^2_{31} (3 - 4 \cos^2 \rho)^2 \left( -3 - 16 \cos^2 \rho + 16 \cos^4 \rho \right.\right.\right.        \nonumber \\
&&\left.\left.\left.+ 8 \cos \rho \sqrt{3 - 3 \cos^2 \rho} - 24 \sin^2 \theta_{13} \left( -3 + 2 \cos^2 \rho + 2 \cos \rho \sqrt{3 - 3 \cos^2 \rho} \right) + 36 \sin^2 \theta_{13}^2 \left( -3  \right.\right.\right. \right.\nonumber \\
&&\left.\left.\left.\left.+ 2 \cos^2 \rho + 2 \cos \rho \sqrt{3 - 3 \cos^2 \rho} \right) \right) \right) \right]^{1/2},\nonumber \\
\sin^2 \theta_{12} &=& \frac{\Delta m^2_{31}}{2 \Delta m^2_{31} + ( \cos \rho + \sqrt{3 - 3 \cos^2 \rho} ) \sqrt{ - \frac{( \Delta m^2_{31} - 3 \Delta m^2_{31} \sin^2 \theta_{13} )^2 ( -3 + 2 \cos^2 \rho + 2 \cos \rho \sqrt{3 - 3 \cos^2 \rho} )}{(3 - 4 \cos^2 \rho)^2} } },\nonumber \\
\sin^2 \theta_{23} &=& \frac{\Delta m^2_{31} + (-\cos \rho + \sqrt{3 - 3 \cos^2 \rho}) \sqrt{-\frac{(\Delta m^2_{31} - 3 \Delta m^2_{31} \sin^2 \theta_{13})^2 (-3 + 2 \cos^2 \rho + 2 \cos \rho \sqrt{3 - 3 \cos^2 \rho})}{(3 - 4 \cos^2 \rho)^2}}}{2 \Delta m^2_{31} + (\cos \rho + \sqrt{3 - 3 \cos^2 \rho}) \sqrt{-\frac{(\Delta m^2_{31} - 3 \Delta m^2_{31} \sin^2 \theta_{13})^2 (-3 + 2 \cos^2 \rho + 2 \cos \rho \sqrt{3 - 3 \cos^2 \rho})}{(3 - 4 \cos^2 \rho)^2}}},\nonumber \\
\delta &=& -\arg \bigg[ \frac{e^{-i \eta_1} \big( 
  i (i + \sqrt{3}) \sqrt{\frac{c}{b + c}} + 
   2 \sqrt{\frac{b}{b + c}} e^{-i \rho} \big)}{2 \sqrt{3}} \bigg].
\end{eqnarray}

\textbf{Parametrization Conditions}
\begin{eqnarray}
\textbf{(A).} && \frac{1}{6} \left( 3 \sqrt{\frac{b}{b + c}} \cos(\eta_4 + \eta_1) 
+ \sqrt{3} \left( \sqrt{\frac{b}{b + c}} \sin(\eta_4 + \eta_1) 
+ 2 \sqrt{\frac{c}{b + c}} \sin(\eta_4 + \rho + \eta_1) \right) \right) = 0, \nonumber \\
\textbf{(B).} && \frac{1}{6} \left( -3 \cos(\eta_5 + \eta_1) 
+ \sqrt{3} \sin(\eta_5 + \eta_1) \right) = 0,  \nonumber   \\
\textbf{(C).} && \frac{1}{6} \left( -3 \sqrt{\frac{c}{b + c}} \cos(\eta_2) 
+ \sqrt{3} \left( \sqrt{\frac{c}{b + c}} \sin(\eta_2) 
- 2 \sqrt{\frac{b}{b + c}} \sin(\rho + \eta_2) \right) \right) = 0,  \nonumber   \\
\textbf{(D).} && -\frac{1}{\sqrt{3}} \left( \sqrt{\frac{c}{b + c}} \sin(\eta_3) 
+ \sqrt{\frac{b}{b + c}} \sin(\rho + \eta_3) \right) = 0,   \nonumber   \\
\textbf{(E).} && \frac{1}{6 \sqrt{3}} \left( 
-3 \sqrt{\frac{b}{b + c}} \cos(\eta_4 + \eta_2) 
- 3 \cos(\eta_5 + \eta_2) 
\sqrt{\frac{b + c + \sqrt{b c} \cos(\rho) 
+ \sqrt{3} \sqrt{b c} \sin(\rho)}{b + c}} \right. \nonumber \\
&& \qquad + \sqrt{3} \left( 
\sqrt{\frac{b}{b + c}} \sin(\eta_4 + \eta_2) 
- \sqrt{\frac{b + c + \sqrt{b c} \cos(\rho) 
+ \sqrt{3} \sqrt{b c} \sin(\rho)}{b + c}} 
\sin(\eta_5 + \eta_2) \right. \nonumber \\
&& \qquad \left. + 2 \sqrt{\frac{c}{b + c}} \sin(\eta_4 + \rho + \eta_2) \right) \Bigg) = 0.
\end{eqnarray}

\textbf{Type 11 (NH)}
\begin{eqnarray}
a &=& \Delta m^2_{21}, \nonumber \\
b &=& \frac{\Delta m^2_{31} \cos^2 \rho + \sqrt{\Delta m^4_{31} \cos^2 \rho (-1 + 6 \sin^2 \theta_{13} - 9 \sin^4 \theta_{13} + \cos^2 \rho)}}{2 \cos^2 \rho}, \nonumber \\
c &=& \frac{1}{2} \left( \Delta m^2_{31} - \frac{\sqrt{\Delta m^4_{31} \cos^2 \rho (-1 + 6 \sin^2 \theta_{13} - 9 \sin^4 \theta_{13} + \cos^2 \rho)}}{\cos^2 \rho} \right).,\nonumber \\
\sin^2 \theta_{12} &=& \frac{\Delta m^2_{31}}{2 \Delta m^2_{31} - \sqrt{\frac{\Delta m^4_{31} (-1 + 3 \sin^2 \theta_{13})^2}{\cos^2 \rho}} \cos \rho},\nonumber \\
\sin^2 \theta_{23} &=& \frac{2 \Delta m^2_{31} - \sqrt{\frac{\Delta m^4_{31} (1 - 3 \sin^2 \theta_{13})^2}{\cos^2 \rho}} (\cos \rho + \sqrt{3 - 3 \cos^2 \rho})}{4 \Delta m^2_{31} - 2 \sqrt{\frac{\Delta m^4_{31} (1 - 3 \sin^2 \theta_{13})^2}{\cos^2 \rho}} \cos \rho},\nonumber \\
\delta &=& -arg \left[ \frac{e^{-i \eta_1} \left(\sqrt{\frac{c}{b + c}} + \sqrt{\frac{b}{b + c}} e^{-i \rho}\right)}{\sqrt{3}} \right].
\end{eqnarray}

\textbf{Parametrization Conditions}
\begin{eqnarray}
\textbf{(A).} && \frac{-\sqrt{\frac{b}{b+c}} \sin(\eta_4 + \eta_1) + \sqrt{\frac{c}{b+c}} \sin(\eta_4 + \rho + \eta_1)}{\sqrt{3}} = 0, \nonumber   \\
\textbf{(B).} && -\frac{\sin(\eta_5 + \eta_1)}{\sqrt{3}} = 0,   \nonumber   \\
\textbf{(C).} && \frac{1}{6} \left(3 \sqrt{\frac{c}{b+c}} \cos(\eta_2) + \sqrt{3} \left(\sqrt{\frac{c}{b+c}} \sin(\eta_2) - 2 \sqrt{\frac{b}{b+c}} \sin(\rho + \eta_2) \right)\right) = 0,   \nonumber   \\
\textbf{(D).} && \frac{1}{6} \left(-3 \sqrt{\frac{c}{b+c}} \cos(\eta_3) + \sqrt{3} \left(\sqrt{\frac{c}{b+c}} \sin(\eta_3) - 2 \sqrt{\frac{b}{b+c}} \sin(\rho + \eta_3) \right)\right) = 0,  \nonumber     \\
\textbf{(E).} && \frac{1}{6 \sqrt{3}} \left(3 \sqrt{\frac{b}{b+c}} \cos(\eta_4 + \eta_2) + 3 \sqrt{\frac{b+c - 2 \sqrt{b c} \cos(\rho)}{b+c}} \cos(\eta_5 + \eta_2) \right. \nonumber \\
&& \left. \left. + \sqrt{3} \left(\sqrt{\frac{b}{b+c}} \sin(\eta_4 + \eta_2) - \sqrt{\frac{b+c - 2 \sqrt{b c} \cos(\rho)}{b+c}} \sin(\eta_5 + \eta_2)  \right. \right.\right. \nonumber \\
&&\left.\left.+ 2 \sqrt{\frac{c}{b+c}} \sin(\eta_4 + \rho + \eta_2)\right)\right) = 0.
\end{eqnarray}

\vspace{5pt}
\textbf{Type 12 (NH)}
\begin{eqnarray}
a &=& \Delta m^2_{21}, \nonumber \\
b &=& \frac{1}{2 (3 - 4 \cos^2 \rho)^2} \Big( \Delta m^2_{31} (3 - 4 \cos^2 \rho)^2 
+ \Big[ \Delta m^4_{31} (3 - 4 \cos^2 \rho)^2 
\big( -3 - 16 \cos^2 \rho + 16 \cos^4 \rho \nonumber \\
&&+ 8 \cos \rho \sqrt{3 - 3 \cos^2 \rho} 
 - 24 \sin^2 \theta_{13} (-3 + 2 \cos^2 \rho + 2 \cos \rho \sqrt{3 - 3 \cos^2 \rho}) 
+ 36 \sin^4 \theta_{13} (-3  \nonumber \\
&&+ 2 \cos^2 \rho + 2 \cos \rho \sqrt{3 - 3 \cos^2 \rho}) 
\big) \Big]^{\frac{1}{2}} \Big), \nonumber \\
c &=& \frac{1}{2 (3 - 4 \cos^2 \rho)^2} \Big( \Delta m^2_{31} (3 - 4 \cos^2 \rho)^2 
- \Big[ \Delta m^4_{31} (3 - 4 \cos^2 \rho)^2 
\big( -3 - 16 \cos^2 \rho + 16 \cos^4 \rho \nonumber \\
&&+ 8 \cos \rho \sqrt{3 - 3 \cos^2 \rho} 
 - 24 \sin^2 \theta_{13} (-3 + 2 \cos^2 \rho + 2 \cos \rho \sqrt{3 - 3 \cos^2 \rho}) 
+ 36 \sin^4 \theta_{13} (-3 \nonumber \\
&&+ 2 \cos^2 \rho + 2 \cos \rho \sqrt{3 - 3 \cos^2 \rho}) 
\big) \Big]^{\frac{1}{2}} \Big),\nonumber \\
\sin^2 \theta_{12} &=& \frac{\Delta m^2_{31}}{2 \Delta m^2_{31} + (\cos \rho + [3 - 3 \cos^2 \rho]^{1/2}) [-\frac{((\Delta m^2_{31} - 3 \Delta m^2_{31} \sin^2 \theta_{13})^2 (-3 + 2 \cos^2 \rho + 2 \cos \rho [3 - 3 \cos^2 \rho]^{1/2}))}{(3 - 4 \cos^2 \rho)^2}]^{1/2}},\nonumber 
\end{eqnarray}
\begin{eqnarray}
\sin^2 \theta_{23} &=& \frac{\Delta m^2_{31} + 2 \cos \rho \sqrt{-\frac{(\Delta m^2_{31} - 3 \Delta m^2_{31} \sin^2 \theta_{13})^2 (-3 + 2 \cos^2 \rho + 2 \cos \rho \sqrt{3 - 3 \cos^2 \rho})}{(3 - 4 \cos^2 \rho)^2}}}{2 \Delta m^2_{31} + (\cos \rho + \sqrt{3 - 3 \cos^2 \rho}) \sqrt{-\frac{(\Delta m^2_{31} - 3 \Delta m^2_{31} \sin^2 \theta_{13})^2 (-3 + 2 \cos^2 \rho + 2 \cos \rho \sqrt{3 - 3 \cos^2 \rho})}{(3 - 4 \cos^2 \rho)^2}}} \nonumber  \\
\delta &=&-arg \left[ \frac{e^{-i \eta_1} \big( 
i (i + \sqrt{3}) \sqrt{\frac{c}{b + c}} + 
2 \sqrt{\frac{b}{b + c}} e^{-i  \rho} \big)}{42 \sqrt{3}} \right].
\end{eqnarray}

\textbf{Parametrization Conditions}
\begin{eqnarray}
 \textbf{(A).}&& \frac{1}{6} \left( 3 \sqrt{\frac{b}{b + c}} \cos(\eta_4 + \eta_1) + 
\sqrt{3} \left( \sqrt{\frac{b}{b + c}} \sin(\eta_4 + \eta_1) + 
2 \sqrt{\frac{c}{b + c}} \sin(\eta_4 + \rho + \eta_1) \right) \right) = 0, \nonumber\\
\textbf{(B).} && \frac{1}{6} \left( -3 \cos(\eta_5 + \eta_1) + 
\sqrt{3} \sin(\eta_5 + \eta_1) \right) = 0, \nonumber\\[10pt]
\textbf{(C).} && -\frac{1}{\sqrt{3}} \left( 
\sqrt{\frac{c}{b + c}} \sin(\eta_2) + 
\sqrt{\frac{b}{b + c}} \sin(\rho + \eta_2) \right) = 0, \nonumber \\[10pt]
\textbf{(D).} && \frac{1}{6} \left( -3 \sqrt{\frac{c}{b + c}} \cos(\eta_3) + 
\sqrt{3} \left( \sqrt{\frac{c}{b + c}} \sin(\eta_3) - 
2 \sqrt{\frac{b}{b + c}} \sin(\rho + \eta_3) \right) \right) = 0, \nonumber \\[10pt]
\textbf{(E).} && \frac{1}{3} \left( 
-\sqrt{\frac{b}{b + c}} \sin(\eta_4 + \eta_2) + 
\sqrt{\frac{b + c + \sqrt{b c} \cos(\rho) + \sqrt{3} \sqrt{b c} \sin(\rho)}{b + c}} \sin(\eta_5 + \eta_2) \right.\nonumber \\
&&\left. + \sqrt{\frac{c}{b + c}} \sin(\eta_4 + \rho + \eta_2) \right) = 0.
\end{eqnarray}

\textbf{Type 2 (IH)}
\begin{eqnarray}
\label{ek}
a &=& \Delta m^2_{21}-\Delta m^2_{31},\nonumber\\
b &=& -\frac{\Delta m^2_{31}}{2} + \frac{1}{2(3-4\cos^2\,\rho)^2}[\Delta m^4_{31}(3-4\cos^2\,\rho)^2(16 \cos^4\,\rho-16 \cos^2\,\rho-3 \nonumber\\&&
-8\cos\,\rho \sqrt{3-3\cos^2\,\rho}+24 \sin^2\,\theta_{13}\,(3-2\cos^2\,\rho+2\cos\,\rho\sqrt{3-3\cos^2\,\rho}) \nonumber \\&&
-36\sin^2\,\theta_{13}(3-2\cos^2\,\rho+2\cos\,\rho\sqrt{3-3\cos^2\,\rho})) ]^\frac{1}{2},\nonumber\\
c &=& -\frac{\Delta m^2_{31}}{2} - \frac{1}{2(3-4\cos^2\,\rho)^2}[\Delta m^4_{31}(3-4\cos^2\,\rho)^2(16 \cos^4\,\rho-16 \cos^2\,\rho-3 \nonumber\\&&
-8\cos\,\rho \sqrt{3-3\cos^2\,\rho}+24 \sin^2\,\theta_{13}\,(3-2\cos^2\,\rho+2\cos\,\rho\sqrt{3-3\cos^2\,\rho}) \nonumber \\&&
-36\sin^2\,\theta_{13}(3-2\cos^2\,\rho+2\cos\,\rho\sqrt{3-3\cos^2\,\rho})) ]^\frac{1}{2},\nonumber\\
\sin^2\,\theta_{12} &=& \frac{\Delta m^2_{31}}{2 \Delta m^2_{31}+(\cos\rho-\sqrt{3-3\cos^2\rho}) \sqrt{\frac{\Delta m^4_{31}(1-3\sin^2\,\theta_{13})^2(\cos\rho+\sqrt{3-3\cos^2\rho})^2}{(3-4\cos^2\,\rho)^2}}},\nonumber\\
\sin^2\,\theta_{23} &=& \frac{\Delta m^2_{31}+2\cos\,\rho\sqrt{\frac{\Delta m^4_{31}(1-3\sin^2\,\theta_{13})^2(\cos\rho+\sqrt{3-3\cos^2\rho})^2}{(3-4\cos^2\rho)^2}}} {2 \Delta m^2_{31}+(\cos\rho-\sqrt{3-3\cos^2\rho}) \sqrt{\frac{\Delta m^4_{31}(1-3\sin^2\,\theta_{13})^2(\cos\rho+\sqrt{3-3\cos^2\rho})^2}{(3-4\cos^2\,\rho)^2}}},\nonumber\\
\delta &=& -arg \left[ e^{-i\eta_1} \left( \sqrt{\frac{b}{3(b+c)}} +\frac{1}{6}(\sqrt{3}-3i)\sqrt{\frac{c}{b+c}}e^{-i\rho} \right) \right].
\end{eqnarray}

\textbf{Parametrization Conditions:}
\begin{eqnarray}
\label{el}
\textbf{(A).}&&3\sqrt{b}\,\cos\,(\rho+\eta_1+\eta_4)-2\sqrt{3c}\,\sin\,(\eta_1+\eta_4)+\sqrt{3b}\,\sin\,(\rho+\eta_1+\eta_4)=0,\nonumber\\
\textbf{(B).} && \sqrt{3}\,\sin\,(\eta_1+\eta_5)-3\,\cos\,(\eta_1+\eta_5)=0,\nonumber\\
\textbf{(C).} && \sqrt{c}\,\sin\,(\rho+\eta_2)-\sqrt{b}\,\sin\,\eta_2 =0,\nonumber\\
\textbf{(D).} && 3\sqrt{c}\,\cos\,(\rho+\eta_3)-2\sqrt{3\,b}\,\sin\,\eta_3-\sqrt{3\,c}\,\sin\,(\rho+\eta_3)=0,\nonumber\\
\textbf{(E).} && \sqrt{b+c-\sqrt{b\,c}\,\cos\rho+\sqrt{3\,b\,c}\,\sin\rho}\,\sin\,(\eta_2+\eta_5)-\sqrt{c}\,\sin\,(\eta_2+\eta_4) \nonumber \\&&
-\sqrt{b}\,\sin\,(\rho+\eta_2+\eta_4)=0.
\end{eqnarray}

\textbf{Type 3 (IH)}
\begin{eqnarray}
\label{em}
a &=& \Delta m^2_{21}-\Delta m^2_{31},\nonumber \\
b &=&- \frac{\Delta m^2_{31}}{2} + \frac{\sqrt{\Delta m^4_{31}\,\cos^2\,\rho\,(\cos^2\,\rho-9\, \sin^4\,\theta_{13}+6\,\sin^2\,\theta_{13}-1)}}{2\,\cos^2\,\rho},\nonumber\\
c &=& -\frac{\Delta m^2_{31}}{2}-\frac{\sqrt{\Delta m^4_{31}\,\cos^2\,\rho\,(\cos^2\,\rho-9\, \sin^4\,\theta_{13}+6\,\sin^2\,\theta_{13}-1)}}{2\,\cos^2\,\rho},\nonumber\\
\sin^2\,\theta_{12} &=& \frac{\Delta m^2_{31}}{2\, \Delta m^2_{31}-\sqrt{\frac{\Delta m^4_{31}(-1+3\,\sin^2\,\theta_{13})^2}{\cos^2\,\rho}} \cos\,\rho},\nonumber\\
\sin^2\,\theta_{23} &=& \frac{2\,\Delta m^2_{31}- \sqrt{\frac{\Delta m^4_{31}(1-3\,\sin^2\,\theta_{13})^2}{\cos^2\,\rho}} (\cos\rho+\sqrt{3-3\,\cos^2\,\rho})}{4\,\Delta m^2_{31}-2\sqrt{\frac{\Delta m^4_{31}(-1+3\,\sin^2\,\theta_{13})^2}{\cos^2\,\rho}} \cos\,\rho},\nonumber\\
\delta &=& -arg \left[ \frac{e^{-i \eta_1}(\sqrt{\frac{b}{b+c}}-\sqrt{\frac{c}{b+c}}\,e^{-i \rho})}{\sqrt{3}} \right].
\end{eqnarray}

\textbf{Parametrization Conditions:}
\begin{eqnarray}
\label{en}
\textbf{(A).} &&\sqrt{b}\,sin\,(\rho+\eta_1+\eta_4)-\sqrt{c}\,\sin\,(\eta_1+\eta_4)=0,\nonumber\\
\textbf{(B).} && \sin\,(\eta_1+\eta_5)=0,\nonumber\\
\textbf{(C).} && 3\sqrt{c}\,\cos\,(\rho+\eta_2)-2\sqrt{3\,b}\,\sin\,\eta_2-\sqrt{3\,c}\,\sin\,(\rho+\eta_2)=0,\nonumber\\
\textbf{(D).} && 3\sqrt{c}\,\cos\,(\rho+\eta_3)+2\sqrt{3\,b}\,\sin\,\eta_3+\sqrt{3\,c}\,\sin\,(\rho+\eta_3)=0,\nonumber\\
\textbf{(E).} && \sqrt{b+c+2\sqrt{b\,c}\,\cos\rho}\,(3\,\cos\,(\eta_2+\eta_5)+\sqrt{3}\,\sin\,(\eta_2+\eta_5))\nonumber \\&&
+\sqrt{b}\,(3\,\cos\,(\rho+\eta_2+\eta_4)-\sqrt{3}\,sin\,(\rho+\eta_2+\eta_4))\nonumber \\
&&+2\sqrt{3\,c}\,\sin\,(\eta_2+\eta_4)=0.
\end{eqnarray}

\textbf{Type 4 (IH)}
\begin{eqnarray}
\label{eo}
a &=& \Delta m^2_{21}-\Delta m^2_{31},\nonumber\\
b &=& -\frac{\Delta m^2_{31}}{2} + \frac{1}{2(3-4\cos^2\,\rho)^2}[\Delta m^4_{31}(3-4\cos^2\,\rho)^2(16 \cos^4\,\rho-16 \cos^2\,\rho-3 \nonumber\\&&
-8\cos\,\rho \sqrt{3-3\cos^2\,\rho}+24 \sin^2\,\theta_{13}\,(3-2\cos^2\,\rho+2\cos\,\rho\sqrt{3-3\cos^2\,\rho}) \nonumber \\&&
-36\sin^2\,\theta_{13}(3-2\cos^2\,\rho+2\cos\,\rho\sqrt{3-3\cos^2\,\rho})) ]^\frac{1}{2},\nonumber \\
c &=& -\frac{\Delta m^2_{31}}{2} - \frac{1}{2(3-4\cos^2\,\rho)^2}[\Delta m^4_{31}(3-4\cos^2\,\rho)^2(16 \cos^4\,\rho-16 \cos^2\,\rho-3 \nonumber\\&&
-8\cos\,\rho \sqrt{3-3\cos^2\,\rho}+24 \sin^2\,\theta_{13}\,(3-2\cos^2\,\rho+2\cos\,\rho\sqrt{3-3\cos^2\,\rho}) \nonumber \\&&
-36\sin^2\,\theta_{13}(3-2\cos^2\,\rho+2\cos\,\rho\sqrt{3-3\cos^2\,\rho})) ]^\frac{1}{2},\nonumber \\
\sin^2\,\theta_{12} &=& \frac{\Delta m^2_{31}}{2 \Delta m^2_{31}+(\cos\rho-\sqrt{3-3\cos^2\rho}) \sqrt{\frac{\Delta m^4_{31}(1-3\sin^2\,\theta_{13})^2(\cos\rho+\sqrt{3-3\cos^2\rho})^2}{(3-4\cos^2\,\rho)^2}}},\nonumber \\
\sin^2\,\theta_{23} &=& \frac{\Delta m^2_{31}-(\cos\rho+\sqrt{3-3\,\cos^2\rho})\sqrt{\frac{\Delta m^4_{31}(1-3\sin^2\,\theta_{13})^2(\cos\rho+\sqrt{3-3\cos^2\rho})^2}{(3-4\cos^2\rho)^2}}} {2 \Delta m^2_{31}+(\cos\rho-\sqrt{3-3\cos^2\rho}) \sqrt{\frac{\Delta m^4_{31}(1-3\sin^2\,\theta_{13})^2(\cos\rho+\sqrt{3-3\cos^2\rho})^2}{(3-4\cos^2\,\rho)^2}}},\nonumber \\
\delta &=& -arg \left[ e^{-i\eta_1} \left( \sqrt{\frac{b}{3(b+c)}} +\frac{1}{6}(\sqrt{3}-3i)\sqrt{\frac{c}{b+c}}e^{-i\rho} \right) \right].
\end{eqnarray}

\textbf{Parametrization Condtions:}
\begin{eqnarray}
\label{ep}
\textbf{(A).} &&3\sqrt{b}\,\cos\,(\rho+\eta_1+\eta_4)-2\sqrt{3c}\,\sin\,(\eta_1+\eta_4)+\sqrt{3b}\,\sin\,(\rho+\eta_1+\eta_4)=0,\nonumber\\
\textbf{(B).} && \sqrt{3}\,\sin\,(\eta_1+\eta_5)-3\,\cos\,(\eta_1+\eta_5)=0,\nonumber\\
\textbf{(C).} && 3\sqrt{c}\,\cos\,(\rho+\eta_2)-2\sqrt{3\,b}\,\sin\,\eta_2-\sqrt{3\,c}\,\sin\,(\rho+\eta_2)=0,\nonumber\\
\textbf{(D).} && \sqrt{c}\,\sin\,(\rho+\eta_3)-\sqrt{b}\,\sin\,\eta_3 =0,\nonumber\\
\textbf{(E).} && \sqrt{b+c-\sqrt{b\,c}\,\cos\rho+\sqrt{3\,b\,c}\,\sin\rho}\,(\sqrt{3}\,\sin\,(\eta_2+\eta_5)+3\,\cos\,(\eta_2+\eta_4)) \nonumber \\&&
+\sqrt{b}\,(3\,\cos\,(\rho+\eta_2+\eta_4)-\sqrt{3}\,\sin\,(\rho+\eta_2+\eta_4)) \nonumber \\
&&+2\,\sqrt{3\,c}\,\sin\,(\eta_2+\eta_4)=0.
\end{eqnarray}

\textbf{Type 5 (IH)}
\begin{eqnarray}
a &=& \Delta m^2_{21} - \Delta m^2_{31}, \nonumber \\
b &=& \frac{-\Delta m^2_{31} \cos^2 \rho + \sqrt{\Delta m^4_{31} \cos^2 \rho (-1 + 6 \sin^2 \theta_{13} - 9 \sin^4 \theta_{13} + \cos^2 \rho)}}{2 \cos^2 \rho}, \nonumber \\
c &=& -\frac{\Delta m^2_{31} \cos^2 \rho + \sqrt{\Delta m^4_{31} \cos^2 \rho (-1 + 6 \sin^2 \theta_{13} - 9 \sin^4 \theta_{13} + \cos^2 \rho)}}{2 \cos^2 \rho}.,\nonumber \\
\sin^2 \theta_{12} &=& \frac{\Delta m^2_{31}}{2 \Delta m^2_{31} - \sqrt{\frac{\Delta m^4_{31} (-1 + 3 \sin^2 \theta_{13})^2}{\cos^2 \rho}} \cos \rho}, \nonumber \\
\sin^2 \theta_{23} &=& \frac{2 \Delta m^2_{31} + \sqrt{\frac{\Delta m^4_{31} (1 - 3 \sin^2 \theta_{13})^2}{\cos^2 \rho}} (-\cos \rho + \sqrt{3 - 3 \cos^2 \rho})}{4 \Delta m^2_{31} - 2 \sqrt{\frac{\Delta m^4_{31} (-1 + 3 \sin^2 \theta_{13})^2}{\cos^2 \rho}} \cos \rho}, \nonumber \\
\delta &=& -arg \left[\frac{e^{-i \eta_1} \left( \sqrt{\frac{b}{b + c}} - \sqrt{\frac{c}{b + c}} e^{-i  \rho} \right)}{\sqrt{3}} \right].
\end{eqnarray}

\textbf{Parametrization Conditions}
\begin{eqnarray}
\textbf{(A).} && -\frac{1}{\sqrt{3}} \left( 
\sqrt{\frac{c}{b+c}} \sin(\eta_4 + \eta_1) + 
\sqrt{\frac{b}{b+c}} \sin(\eta_4 + \rho + \eta_1) 
\right) = 0, \nonumber \\
\textbf{(B).} && -\frac{\sin(\eta_5 + \eta_1)}{\sqrt{3}} = 0, \nonumber \\
\textbf{(C).} && \frac{1}{6} \left( 
3 \sqrt{\frac{b}{b+c}} \cos(\eta_2) + 
3 \sqrt{\frac{c}{b+c}} \cos(\rho + \eta_2) + 
\sqrt{3} \left( 
\sqrt{\frac{b}{b+c}} \sin(\eta_2) \right.\right. \nonumber \\
&& \left.\left. - \sqrt{\frac{c}{b+c}} \sin(\rho + \eta_2) 
\right) \right) = 0, \nonumber \\
\textbf{(D).} && \frac{1}{6} \left( 
-3 \sqrt{\frac{b}{b+c}} \cos(\eta_3) - 
3 \sqrt{\frac{c}{b+c}} \cos(\rho + \eta_3) + 
\sqrt{3} \left( 
\sqrt{\frac{b}{b+c}} \sin(\eta_3) \right.\right. \nonumber \\
&& \left.\left.- \sqrt{\frac{c}{b+c}} \sin(\rho + \eta_3) 
\right) \right) = 0, \nonumber \\
\textbf{(E).} && \frac{1}{6} \left( 
\sqrt{3} \sqrt{\frac{c}{b+c}} \cos(\eta_4 + \eta_2) - 
\sqrt{3} \sqrt{\frac{b}{b+c}} \cos(\eta_4 + \rho + \eta_2) + 
\sqrt{\frac{c}{b+c}} \sin(\eta_4 + \eta_2) \right.\nonumber \\
&& \left. + 2 \sqrt{\frac{b+c + 2 \sqrt{bc} \cos(\rho)}{b+c}} \sin(\eta_5 + \eta_2) + 
\sqrt{\frac{b}{b+c}} \sin(\eta_4 + \rho + \eta_2) 
\right) = 0. 
\end{eqnarray}

\textbf{Type 6 (IH)}
\begin{eqnarray}
a &=& \Delta m^2_{21} - \Delta m^2_{31}, \nonumber \\
b &=& \frac{1}{2 (3 - 4 \cos^2 \rho)^2} \left[ - \Delta m^2_{31} (3 - 4 \cos^2 \rho)^2 + \left( \Delta m^4_{31} (3 - 4 \cos^2 \rho)^2 \left( -3 - 16 \cos^2 \rho + 16 \cos^4 \rho \right.\right.\right. \nonumber \\
&&\left.\left.\left.- 8 \cos \rho \sqrt{3 - 3 \cos^2 \rho} + 24 \sin^2 \theta_{13} (3 - 2 \cos^2 \rho + 2 \cos \rho \sqrt{3 - 3 \cos^2 \rho}) - 36 \sin^4 \theta_{13} (3 \right.\right.\right. \nonumber \\
&& \left.\left.\left. - 2 \cos^2 \rho + 2 \cos \rho \sqrt{3 - 3 \cos^2 \rho}) \right) \right) \right]^{1/2}, \nonumber \\
c &=& -\frac{1}{2 (3 - 4 \cos^2 \rho)^2} \left[ \Delta m^2_{31} (3 - 4 \cos^2 \rho)^2 + \left( \Delta m^4_{31} (3 - 4 \cos^2 \rho)^2 \left( -3 - 16 \cos^2 \rho + 16 \cos^4 \rho \right.\right.\right.\nonumber \\
&&\left.\left.\left. - 8 \cos \rho \sqrt{3 - 3 \cos^2 \rho} + 24 \sin^2 \theta_{13} (3 - 2 \cos^2 \rho + 2 \cos \rho \sqrt{3 - 3 \cos^2 \rho}) - 36 \sin^4 \theta_{13} (3 \right.\right.\right. \nonumber \\
&& \left.\left.\left.- 2 \cos^2 \rho + 2 \cos \rho \sqrt{3 - 3 \cos^2 \rho}) \right) \right) \right]^{1/2}, \nonumber \\
\sin^2 \theta_{12} &=& \frac{\Delta m^2_{31}}{2 \Delta m^2_{31} + (\cos \rho - \sqrt{3 - 3 \cos^2 \rho}) \sqrt{ \frac{ (\Delta m^2_{31} - 3 \Delta m^2_{31} \sin^2 \theta_{13})^2 (3 - 2 \cos^2 \rho + 2 \cos \rho \sqrt{3 - 3 \cos^2 \rho}) }{(3 - 4 \cos^2 \rho)^2} } }, \nonumber \\
\sin^2 \theta_{23} &=& \frac{\Delta m^2_{31} + 2 \cos \rho \sqrt{\frac{(\Delta m^2_{31} - 3 \Delta m^2_{31} \sin^2 \theta_{13})^2 (3 - 2 \cos^2 \rho + 2 \cos \rho \sqrt{3 - 3 \cos^2 \rho})}{(3 - 4 \cos^2 \rho)^2}}}{2 \Delta m^2_{31} + (\cos \rho - \sqrt{3 - 3 \cos^2 \rho}) \sqrt{\frac{(\Delta m^2_{31} - 3 \Delta m^2_{31} \sin^2 \theta_{13})^2 (3 - 2 \cos^2 \rho + 2 \cos \rho \sqrt{3 - 3 \cos^2 \rho})}{(3 - 4 \cos^2 \rho)^2}}}, \nonumber \\
\delta &=&-arg \left[ \frac{e^{-i (\rho + \eta_1)} \left( (1 + i \sqrt{3}) \sqrt{\frac{c}{b + c}} + i (i + \sqrt{3}) \sqrt{\frac{b}{b + c}} e^{i \rho} \right)}{2 \sqrt{3}} \right].
\end{eqnarray}

\textbf{Parametrization Conditions}
\begin{eqnarray}
\textbf{(A).} && \frac{1}{6} \left( 3 \sqrt{\frac{c}{b + c}} \cos(\eta_4 + \eta_1) - 3 \sqrt{\frac{b}{b + c}} \cos(\eta_4 + \rho + \eta_1) + \sqrt{3} \left( \sqrt{\frac{c}{b + c}} \sin(\eta_4 + \eta_1) \right.\right. \nonumber \\
&& \left.\left.+ \sqrt{\frac{b}{b + c}} \sin(\eta_4 + \rho + \eta_1) \right) \right) = 0, \nonumber \\
\textbf{(B).} && -\frac{\sin(\eta_5 + \eta_1)}{\sqrt{3}} = 0, \nonumber \\
\textbf{(C).} && \frac{-\sqrt{\frac{b}{b + c}} \sin(\eta_1) + \sqrt{\frac{c}{b + c}} \sin(\rho + \eta_1)}{\sqrt{3}} = 0, \nonumber 
\end{eqnarray}
\begin{eqnarray}
\textbf{(D).} && \frac{1}{6} \left( -3 \sqrt{\frac{b}{b + c}} \cos(\eta_1) - 3 \sqrt{\frac{c}{b + c}} \cos(\rho + \eta_1) + \sqrt{3} \left( \sqrt{\frac{b}{b + c}} \sin(\eta_1) \right.\right. \nonumber \\
&& \left.\left. - \sqrt{\frac{c}{b + c}} \sin(\rho + \eta_1) \right) \right) = 0, \nonumber \\
\textbf{(E).} && \frac{1}{3} \left( -\sqrt{\frac{c}{b + c}} \sin(\eta_4 + \eta_2) + \sqrt{\frac{b + c - \sqrt{b c} \cos(\rho) + \sqrt{3} \sqrt{b c} \sin(\rho)}{b + c}} \sin(\eta_5 + \eta_2) \right. \nonumber \\
&& \left. - \sqrt{\frac{b}{b + c}} \sin(\eta_4 + \rho + \eta_2) \right) = 0.
\end{eqnarray}

\textbf{Type 7 (IH)}
\begin{eqnarray}
a &=& \Delta m^2_{21} - \Delta m^2_{31}, \nonumber \\
b &=& \frac{-\Delta m^2_{31} \cos^2 \rho + \sqrt{ [\Delta m^4_{31} \cos^4 \rho (-1 + 6 \sin^2 \theta_{13} - 9 \sin^4 \theta_{13} + \cos^2 \rho)] }}{2 \cos^2 \rho}, \nonumber \\
c &=& -\frac{(\Delta m^2_{31} \cos^2 \rho + \sqrt{ [\Delta m^4_{31} \cos^4 \rho (-1 + 6 \sin^2 \theta_{13} - 9 \sin^4 \theta_{13} + \cos^2 \rho)] }) )}{2 \cos^2 \rho}, \nonumber \\
\sin^2\,\theta_{12} &=& \frac{\Delta m^2_{31}}{2 \Delta m^2_{31} - \sqrt{ \left( \frac{\Delta m^4_{31} (-1 + 3 \sin^2 \theta_{13})^2}{\cos^2 \rho} \right) } \cos \rho} \nonumber \\
\sin^2\,\theta_{23} &=& \frac{2 \Delta m^2_{31} - \sqrt{\frac{\Delta m^4_{31} (1 - 3 \sin^2 \theta_{13})^2}{\cos^2 \rho}} (\cos \rho + \sqrt{3 - 3 \cos^2 \rho})}{4 \Delta m^2_{31} - 2 \sqrt{\frac{\Delta m^4_{31} (-1 + 3 \sin^2 \theta_{13})^2}{\cos^2 \rho}} \cos \rho}, \nonumber \\
\delta &=& -arg \left[ \frac{e^{-i \eta_1} \left( \sqrt{\frac{b}{b + c}} - \sqrt{\frac{c}{b + c}} e^{-i \rho} \right)}{\sqrt{3}} \right].
\end{eqnarray}

\textbf{Parametrization Conditions}
\begin{eqnarray}
\textbf{(A).} && -\frac{\sqrt{\frac{c}{b + c}} \sin(\eta_4 + \eta_1) + \sqrt{\frac{b}{b + c}} \sin(\eta_4 + \rho + \eta_1)}{\sqrt{3}} = 0, \nonumber \\
\textbf{(B).} && -\frac{\sin(\eta_5 + \eta_1)}{\sqrt{3}} = 0, \nonumber 
\end{eqnarray}
\begin{eqnarray}
\textbf{(C).} && \frac{1}{6} \left( -3 \sqrt{\frac{b}{b + c}} \cos(\eta_2) - 3 \sqrt{\frac{c}{b + c}} \cos(\rho + \eta_2) + \sqrt{3} \left( \sqrt{\frac{b}{b + c}} \sin(\eta_2) \right.\right.\nonumber \\
&& \left.\left. - \sqrt{\frac{c}{b + c}} \sin(\rho + \eta_2) \right) \right) = 0, \nonumber \\
\textbf{(D).} && \frac{1}{6} \left( 3 \sqrt{\frac{b}{b + c}} \cos(\eta_3) + 3 \sqrt{\frac{c}{b + c}} \cos(\rho + \eta_3) + \sqrt{3} \left( \sqrt{\frac{b}{b + c}} \sin(\eta_3) \right.\right.\nonumber \\
&& \left.\left. - \sqrt{\frac{c}{b + c}} \sin(\rho + \eta_3) \right) \right) = 0, \nonumber \\
\textbf{(E).} && \frac{1}{6} \left( -\sqrt{3} \sqrt{\frac{c}{b + c}} \cos(\eta_4 + \eta_2) + \sqrt{3} \sqrt{\frac{b}{b + c}} \cos(\eta_4 + \rho + \eta_2) + \sqrt{\frac{c}{b + c}} \sin(\eta_4 + \eta_2) \right.\nonumber \\
&& \left. + 2 \sqrt{\frac{b + c + 2 \sqrt{b c} \cos(\rho)}{b + c}} \sin(\eta_5 + \eta_2) + \sqrt{\frac{b}{b + c}} \sin(\eta_4 + \rho + \eta_2) \right) = 0.
\end{eqnarray}

\textbf{Type 8 (IH)}
\begin{eqnarray}
a &=& \Delta m^2_{21} - \Delta m^2_{31}, \nonumber \\
b &=& \frac{1}{2(3 - 4 \cos^2 \rho)^2} \Big[ -\Delta m^2_{31} (3 - 4 \cos^2 \rho)^2 + \Big( \Delta m^4_{31} (3 - 4 \cos^2 \rho)^2 \Big( -3 - 16 \cos^2 \rho + 16 \cos^4 \rho \nonumber \\
&&- 8 \cos \rho [3 - 3 \cos^2 \rho]^{1/2}  + 24 \sin^2 \theta_{13} (3 - 2 \cos^2 \rho + 2 \cos \rho [3 - 3 \cos^2 \rho]^{1/2}) - 36 \sin^4 \theta_{13} (3\nonumber \\
&& - 2 \cos^2 \rho + 2 \cos \rho [3 - 3 \cos^2 \rho]^{1/2}) \Big) \Big)^{1/2} \Big], \nonumber \\
c &=& -\frac{1}{2(3 - 4 \cos^2 \rho)^2} \Big[ \Delta m^2_{31} (3 - 4 \cos^2 \rho)^2 + \Big( \Delta m^4_{31} (3 - 4 \cos^2 \rho)^2 \Big( -3 - 16 \cos^2 \rho + 16 \cos^4 \rho \nonumber \\
&&- 8 \cos \rho [3 - 3 \cos^2 \rho]^{1/2}  + 24 \sin^2 \theta_{13} (3 - 2 \cos^2 \rho + 2 \cos \rho [3 - 3 \cos^2 \rho]^{1/2}) - 36 \sin^4 \theta_{13} (3 \nonumber \\
&& - 2 \cos^2 \rho + 2 \cos \rho [3 - 3 \cos^2 \rho]^{1/2}) \Big) \Big)^{1/2} \Big],\nonumber \\
\sin^2\,\theta_{12} &=& \frac{\Delta m^2_{31}}{2 \Delta m^2_{31} + (\cos \rho - (3 - 3 \cos^2 \rho)^{1/2}) 
\Bigg[\frac{(\Delta m^2_{31} - 3 \Delta m^2_{31} \sin^2 \theta_{13})^2 
(3 - 2 \cos^2 \rho + 2 \cos \rho (3 - 3 \cos^2 \rho)^{1/2})}{(3 - 4 \cos^2 \rho)^2}\Bigg]^{1/2}},\nonumber \\
\sin^2\,\theta_{23} &=& \frac{\Delta m^2_{31} - (\cos \rho + (3 - 3 \cos^2 \rho)^{1/2}) 
\bigg[\frac{(\Delta m^2_{31} - 3 \Delta m^2_{31} \sin^2 \theta_{13})^2 
(3 - 2 \cos^2 \rho + 2 \cos \rho (3 - 3 \cos^2 \rho)^{1/2})}{(3 - 4 \cos^2 \rho)^2}\bigg]^{1/2}}
{2 \Delta m^2_{31} + (\cos \rho - (3 - 3 \cos^2 \rho)^{1/2}) 
\bigg[\frac{(\Delta m^2_{31} - 3 \Delta m^2_{31} \sin^2 \theta_{13})^2 
(3 - 2 \cos^2 \rho + 2 \cos \rho (3 - 3 \cos^2 \rho)^{1/2})}{(3 - 4 \cos^2 \rho)^2}\bigg]^{1/2}},
\nonumber \\
\delta &=&-arg \left[ \frac{e^{-i (\rho + \eta_{1})} \big( (1 + i \sqrt{3}) \big[ \frac{c}{b + c} \big]^{1/2} 
+ i (i + \sqrt{3}) \big[ \frac{b}{b + c} \big]^{1/2} e^{i \rho} \big)}{2 \sqrt{3}} \right].
\end{eqnarray}

\textbf{Parametrization Conditions}
\begin{eqnarray}
\textbf{(A).} && \frac{1}{6} \bigg( 
3 \sqrt{\frac{c}{b + c}} \cos(\eta_4 + \eta_1) - 
3 \sqrt{\frac{b}{b + c}} \cos(\eta_4 + \rho + \eta_1) + 
\sqrt{3} \bigg( \sqrt{\frac{c}{b + c}} \sin(\eta_4 + \eta_1) \nonumber \\
&&+ \sqrt{\frac{b}{b + c}} \sin(\eta_4 + \rho + \eta_1) \bigg) \bigg) = 0, \nonumber \\
\textbf{(B).} && -\frac{\sin(\eta_5 + \eta_1)}{\sqrt{3}} = 0, \nonumber \\
\textbf{(C).} && \frac{1}{6} \bigg( 
-3 \sqrt{\frac{b}{b + c}} \cos(\eta_1) - 
3 \sqrt{\frac{c}{b + c}} \cos(\rho + \eta_1) + 
\sqrt{3} \bigg( \sqrt{\frac{b}{b + c}} \sin(\eta_1) \nonumber \\
&& - \sqrt{\frac{c}{b + c}} \sin(\rho + \eta_1) \bigg) \bigg) = 0, \nonumber \\
\textbf{(D).} && \frac{-\sqrt{\frac{b}{b + c}} \sin(\eta_3) + 
\sqrt{\frac{c}{b + c}} \sin(\rho + \eta_3)}{\sqrt{3}} = 0, \nonumber \\
\textbf{(E).} && \frac{1}{6} \bigg( 
-\sqrt{3} \sqrt{\frac{c}{b + c}} \cos(\eta_4 + \eta_2) + 
\sqrt{3} \sqrt{\frac{b}{b + c}} \cos(\eta_4 + \rho + \eta_2) + 
\sqrt{\frac{c}{b + c}} \sin(\eta_4 + \eta_2) \nonumber \\
&&+ 2 \sqrt{\frac{b + c - \sqrt{b c} \cos(\rho) + \sqrt{3} \sqrt{b c} \sin(\rho)}{b + c}} \sin(\eta_5 + \eta_2) + 
\sqrt{\frac{b}{b + c}} \sin(\eta_4 \nonumber \\
&& + \rho + \eta_2) \bigg) = 0.
\end{eqnarray}

\textbf{Type 9 (IH)}
\begin{eqnarray}
a &=& \Delta m^2_{21} - \Delta m^2_{31}, \nonumber \\
b &=& \frac{-\Delta m^2_{31} \cos^2 \rho + [\Delta m^4_{31} \cos^2 \rho (-1 + 6 \sin^2 \theta_{13} - 9 \sin^4 \theta_{13} + \cos^2 \rho)]^{1/2}}{2 \cos^2 \rho}, \nonumber \\
c &=& -\frac{\Delta m^2_{31} \cos^2 \rho + [\Delta m^4_{31} \cos^2 \rho (-1 + 6 \sin^2 \theta_{13} - 9 \sin^4 \theta_{13} + \cos^2 \rho)]^{1/2}}{2 \cos^2 \rho},\nonumber \\
\sin^2\,\theta_{12} &=& \frac{\Delta m^2_{31}}{2 \Delta m^2_{31} - [(\Delta m^4_{31} (-1 + 3 \sin^2 \theta_{13})^2) / (\cos^2 \rho)]^{1/2} \cos \rho}, \nonumber \\
\sin^2\,\theta_{223} &=& \frac{2 \Delta m^2_{31} + \Bigg[ \frac{\Delta m^4_{31} (1 - 3 \sin^2 \theta_{13})^2}{\cos^2 \rho} \Bigg]^{1/2} (-\cos \rho + [3 - 3 \cos^2 \rho]^{1/2})}{4 \Delta m^2_{31} - 2 \Bigg[ \frac{\Delta m^4_{31} (-1 + 3 \sin^2 \theta_{13})^2}{\cos^2 \rho} \Bigg]^{1/2} \cos \rho}, \nonumber \\
\delta &=& -arg \left[ \frac{e^{-i \eta_1} \Bigg( \Big( \frac{b}{b + c} \Big)^{1/2} - \Big( \frac{c}{b + c} \Big)^{1/2} e^{-i \rho} \Bigg)}{\sqrt{3}}\right].
\end{eqnarray}

\textbf{Parametrization Conditions}
\begin{eqnarray}
\textbf{(A).} && -\frac{1}{\sqrt{3}} \left( 
   \sqrt{\frac{c}{b + c}} \sin(\eta_4 + \eta_1) + 
   \sqrt{\frac{b}{b + c}} \sin(\eta_4 + \rho + \eta_1) 
\right) = 0, \nonumber \\
\textbf{(B).} && -\frac{\sin(\eta_5 + \eta_1)}{\sqrt{3}} = 0, \nonumber \\
\textbf{(C).} && \frac{1}{6} \left( 
   -3 \sqrt{\frac{b}{b + c}} \cos(\eta_2) + 
   \sqrt{3} \left( 
      \sqrt{\frac{b}{b + c}} \sin(\eta_2) + 
      2 \sqrt{\frac{c}{b + c}} \sin(\rho + \eta_2) 
   \right) 
\right) = 0, \nonumber \\
\textbf{(D).} && \frac{1}{6} \left( 
   3 \sqrt{\frac{b}{b + c}} \cos(\eta_3) + 
   \sqrt{3} \left( 
      \sqrt{\frac{b}{b + c}} \sin(\eta_3) + 
      2 \sqrt{\frac{c}{b + c}} \sin(\rho + \eta_3) 
   \right) 
\right) = 0, \nonumber \\
\textbf{(E).} && \frac{1}{6 \sqrt{3}} \left( 
   -3 \sqrt{\frac{c}{b + c}} \cos(\eta_4 + \eta_2) - 
   3 \sqrt{\frac{b + c + 2 \sqrt{b c} \cos(\rho)}{b + c}} \cos(\eta_5 + \eta_2) \right. \nonumber \\ 
   && \left. + \sqrt{3} \left( 
      \sqrt{\frac{c}{b + c}} \sin(\eta_4 + \eta_2) - 
      \sqrt{\frac{b + c + 2 \sqrt{b c} \cos(\rho)}{b + c}} \sin(\eta_5 + \eta_2) \right.\right. \nonumber \\
    &&\left. \left.  - 2 \sqrt{\frac{b}{b + c}} \sin(\eta_4 + \rho + \eta_2) 
   \right) 
\right) = 0.
\end{eqnarray}

\textbf{Type 10 (IH)}
\begin{eqnarray}
a &=& \Delta m^2_{21} - \Delta m^2_{31}, \nonumber \\
b &=& \frac{1}{2 (3 - 4 \cos^2 \rho)^2} \left[ - \Delta m^2_{31} (3 - 4 \cos^2 \rho)^2 + \left( \Delta m^4_{31} (3 - 4 \cos^2 \rho)^2 \left( -3 - 16 \cos^2 \rho + 16 \cos^4 \rho \right.\right.\right. \nonumber \\
&& \left.\left.\left. - 8 \cos \rho \sqrt{3 - 3 \cos^2 \rho} + 24 \sin^2 \theta_{13} (3 - 2 \cos^2 \rho + 2 \cos \rho \sqrt{3 - 3 \cos^2 \rho}\right) - 36 \sin^4 \theta_{13} (3 \right.\right.\nonumber \\
&& \left.\left.- 2 \cos^2 \rho + 2 \cos \rho \sqrt{3 - 3 \cos^2 \rho}) \right) \right]^{\frac{1}{2}}, \nonumber \\
c &=& -\frac{1}{2 (3 - 4 \cos^2 \rho)^2} \left[ \Delta m^2_{31} (3 - 4 \cos^2 \rho)^2 + \left( \Delta m^4_{31} (3 - 4 \cos^2 \rho)^2 \left( -3 - 16 \cos^2 \rho + 16 \cos^4 \rho \right.\right.\right. \nonumber \\
&& \left.\left.\left. - 8 \cos \rho \sqrt{3 - 3 \cos^2 \rho} + 24 \sin^2 \theta_{13} (3  - 2 \cos^2 \rho + 2 \cos \rho \sqrt{3 - 3 \cos^2 \rho} \right) - 36 \sin^4 \theta_{13} (3 
\right.\right.\nonumber \\
&& \left.\left. - 2 \cos^2 \rho + 2 \cos \rho \sqrt{3 - 3 \cos^2 \rho}) \right) \right]^{\frac{1}{2}}, \nonumber 
\end{eqnarray}
\begin{eqnarray}
\sin^2\,\theta_{12} &=& \frac{\Delta m^2_{31}}{2 \Delta m^2_{31} + (\cos \rho - \sqrt{3 - 3 \cos^2 \rho}) \sqrt{ \frac{(\Delta m^2_{31} - 3 \Delta m^2_{31} \sin^2 \theta_{13})^2 (3 - 2 \cos^2 \rho + 2 \cos \rho \sqrt{3 - 3 \cos^2 \rho})}{(3 - 4 \cos^2 \rho)^2}}}, \nonumber \\
\sin^2\,\theta_{23} &=& \frac{\Delta m^2_{31} + 2 \cos \rho \sqrt{\frac{(\Delta m^2_{31} - 3 \Delta m^2_{31} \sin^2 \theta_{13})^2 (3 - 2 \cos^2 \rho + 2 \cos \rho \sqrt{3 - 3 \cos^2 \rho})}{(3 - 4 \cos^2 \rho)^2}}}{2 \Delta m^2_{31} + (\cos \rho - \sqrt{3 - 3 \cos^2 \rho}) \sqrt{\frac{(\Delta m^2_{31} - 3 \Delta m^2_{31} \sin^2 \theta_{13})^2 (3 - 2 \cos^2 \rho + 2 \cos \rho \sqrt{3 - 3 \cos^2 \rho})}{(3 - 4 \cos^2 \rho)^2}}}, \nonumber \\
\delta &=&-arg \left[ \frac{e^{-i \eta_1} \left[ (-1 - i \sqrt{3}) \sqrt{\frac{b}{b + c}} - 2 \sqrt{\frac{c}{b + c}} e^{-i \rho} \right]}{2 \sqrt{3}} \right].
\end{eqnarray}

\textbf{Parametrization Conditions}
\begin{eqnarray}
\textbf{(A).} && \frac{1}{6} \left( -3 \sqrt{\frac{c}{b + c}} \cos(\eta_4 + \eta_1) + \sqrt{3} \left( \sqrt{\frac{c}{b + c}} \sin(\eta_4 + \eta_1) - 2 \sqrt{\frac{b}{b + c}} \sin(\eta_4 + \rho + \eta_1) \right) \right) = 0, \nonumber \\
\textbf{(B).} && \frac{1}{6} \left( 3 \cos(\eta_5 + \eta_1) + \sqrt{3} \sin(\eta_5 + \eta_1) \right) = 0, \nonumber \\
\textbf{(C).} && \frac{-\sqrt{\frac{b}{b + c}} \sin(\eta_1) + \sqrt{\frac{c}{b + c}} \sin(\rho + \eta_2)}{\sqrt{3}} = 0, \nonumber \\
\textbf{(D).} && \frac{1}{6} \left( 3 \sqrt{\frac{b}{b + c}} \cos(\eta_3) + \sqrt{3} \left( \sqrt{\frac{b}{b + c}} \sin(\eta_3) + 2 \sqrt{\frac{c}{b + c}} \sin(\rho + \eta_3) \right) \right) = 0, \nonumber \\
\textbf{(E).} && \frac{1}{3} \left( -\sqrt{\frac{c}{b + c}} \sin(\eta_4 + \eta_2) + \sqrt{\frac{b + c - \sqrt{b c} \cos(\rho) + \sqrt{3} \sqrt{b c} \sin(\rho)}{b + c}} \sin(\eta_5 + \eta_2) \right. \nonumber \\
&& \left.- \sqrt{\frac{b}{b + c}} \sin(\eta_4 + \rho + \eta_2) \right) = 0.
\end{eqnarray}

\textbf{Type 11 (IH)}
\begin{eqnarray}
a &=& \Delta m^2_{21} - \Delta m^2_{31}, \nonumber \\
b &=& \frac{-\Delta m^2_{31} \cos^2 \rho + \sqrt{ [\Delta m^4_{31} \cos^4 \rho (-1 + 6 \sin^2 \theta_{13} - 9 \sin^4 \theta_{13} + \cos^2 \rho)] }}{2 \cos^2 \rho}, \nonumber \\
c &=& - \left( \frac{\Delta m^2_{31} \cos^2 \rho + \sqrt{ [\Delta m^4_{31} \cos^4 \rho (-1 + 6 \sin^2 \theta_{13} - 9 \sin^4 \theta_{13} + \cos^2 \rho)] }}{2 \cos^2 \rho} \right), \nonumber 
\end{eqnarray}
\begin{eqnarray}
\sin^2\,\theta_{12} &=& \frac{\Delta m^2_{31}}{2 \Delta m^2_{31} - \sqrt{\frac{(\Delta m^2_{31})^2 (-1 + 3 \sin^2 \theta_{13})^2}{\cos^2 \rho}} \cos \rho} \nonumber \\
\sin^2\,\theta_{23} &=& \frac{2 \Delta m^2_{31} - \sqrt{\left[\frac{\Delta m^4_{31} (1 - 3 \sin^2 \theta_{13})^2}{\cos^2 \rho}\right]} \left( \cos \rho + \sqrt{3 - 3 \cos^2 \rho} \right)}{4 \Delta m^2_{31} - 2 \sqrt{\left[\frac{\Delta m^4_{31} (-1 + 3 \sin^2 \theta_{13})^2}{\cos^2 \rho}\right]} \cos \rho}, \nonumber \\
\delta &=& -arg \left[\frac{e^{-i \eta_1} \left( \sqrt{\frac{b}{b + c}} - \sqrt{\frac{c}{b + c}} e^{-i \rho} \right)}{\sqrt{3}} \right].
\end{eqnarray}

\textbf{Parametrization Conditions}
\begin{eqnarray}
\textbf{(A).} && -\frac{\left(\sqrt{\frac{c}{b+c}} \sin(\eta_4 + \eta_1) + \sqrt{\frac{b}{b+c}} \sin(\eta_4 + \rho + \eta_1)\right)}{\sqrt{3}}=0, \nonumber \\
\textbf{(B).} && -\frac{\sin(\eta_5 + \eta_1)}{\sqrt{3}}=0, \nonumber \\
\textbf{(C).} && \frac{1}{6} \left( 3 \sqrt{\frac{b}{b+c}} \cos(\eta_2) + \sqrt{3} \left(\sqrt{\frac{b}{b+c}} \sin(\eta_2) + 2 \sqrt{\frac{c}{b+c}} \sin(\rho + \eta_2)\right)\right)=0, \nonumber \\
\textbf{(D).} && \frac{1}{6} \left( -3 \sqrt{\frac{b}{b+c}} \cos(\eta_3) + \sqrt{3} \left(\sqrt{\frac{b}{b+c}} \sin(\eta_3) + 2 \sqrt{\frac{c}{b+c}} \sin(\rho + \eta_3)\right)\right)=0, \nonumber \\
\textbf{(E).} && \frac{1}{6\sqrt{3}} \left( 3 \sqrt{\frac{c}{b+c}} \cos(\eta_4 + \eta_2) + 3 \sqrt{\frac{b+c+2\sqrt{bc} \cos(\rho)}{b+c}} \cos(\eta_5 + \eta_2) \right. \nonumber \\
&& \left. + \sqrt{3} \left(\sqrt{\frac{c}{b+c}} \sin(\eta_4 + \eta_2) - \sqrt{\frac{b+c+2\sqrt{bc} \cos(\rho)}{b+c}} \sin(\eta_5 + \eta_2) \right.\right.\nonumber \\
 && \left.\left.- 2 \sqrt{\frac{b}{b+c}} \sin(\eta_4 + \rho + \eta_2)\right)\right)=0.
\end{eqnarray}

\textbf{Type 12 (IH)}
\begin{eqnarray}
a &=& \Delta m^2_{21} - \Delta m^2_{31}, \nonumber \\
b &=& \frac{1}{2 (3 - 4 \cos^2 \rho)^2} \Big[ - \Delta m^2_{31} (3 - 4 \cos^2 \rho)^2 + \Big( \Delta m^4_{31} (3 - 4 \cos^2 \rho)^2 \Big( -3 - 16 \cos^2 \rho + 16 \cos^4 \rho \nonumber \\
&&- 8 \cos \rho \sqrt{3 - 3 \cos^2 \rho} +  24 \sin^2 \theta_{13} (3 - 2 \cos^2 \rho + 2 \cos \rho \sqrt{3 - 3 \cos^2 \rho}) - 36 \sin^2 \theta_{13}^2 (3 \nonumber \\
&& - 2 \cos^2 \rho + 2 \cos \rho \sqrt{3 - 3 \cos^2 \rho}) \Big) \Big) \Big]^{1/2}, \nonumber 
\end{eqnarray}
\begin{eqnarray}
c &=& -\frac{1}{2 (3 - 4 \cos^2 \rho)^2} \Big[ \Delta m^2_{31} (3 - 4 \cos^2 \rho)^2 + \Big( \Delta m^4_{31} (3 - 4 \cos^2 \rho)^2 \Big( -3 - 16 \cos^2 \rho + 16 \cos^4 \rho \nonumber \\
&&- 8 \cos \rho \sqrt{3 - 3 \cos^2 \rho} +  24 \sin^2 \theta_{13} (3 - 2 \cos^2 \rho + 2 \cos \rho \sqrt{3 - 3 \cos^2 \rho}) - 36 \sin^2 \theta_{13}^2 (3 \nonumber \\
&&- 2 \cos^2 \rho + 2 \cos \rho \sqrt{3 - 3 \cos^2 \rho}) \Big) \Big) \Big]^{1/2}, \nonumber \\
\sin^2\,\theta_{12} &=& \frac{\Delta m^2_{31}}{2 \Delta m^2_{31} + (\cos \rho - \sqrt{3 - 3 \cos^2 \rho}) \sqrt{ \frac{(\Delta m^2_{31} - 3 \Delta m^2_{31} \sin^2 \theta_{13})^2 (3 - 2 \cos^2 \rho + 2 \cos \rho \sqrt{3 - 3 \cos^2 \rho})}{(3 - 4 \cos^2 \rho)^2} }}, \nonumber \\
\sin^2\,\theta_{23} &=& \frac{\Delta m^2_{31} - ( \cos \rho + \sqrt{3 - 3 \cos^2 \rho} ) \sqrt{ \frac{( \Delta m^2_{31} - 3 \Delta m^2_{31} \sin^2 \theta_{13} )^2 ( 3 - 2 \cos^2 \rho + 2 \cos \rho \sqrt{3 - 3 \cos^2 \rho} )}{( 3 - 4 \cos^2 \rho )^2} } }{ 2 \Delta m^2_{31} + ( \cos \rho - \sqrt{3 - 3 \cos^2 \rho} ) \sqrt{ \frac{( \Delta m^2_{31} - 3 \Delta m^2_{31} \sin^2 \theta_{13} )^2 ( 3 - 2 \cos^2 \rho + 2 \cos \rho \sqrt{3 - 3 \cos^2 \rho} )}{( 3 - 4 \cos^2 \rho )^2} } }, \nonumber \\
\delta &=& -arg \left[ \frac{e^{-i \eta_1} \left[(-1 - i \sqrt{3}) \sqrt{\frac{b}{b + c}} - 2 \sqrt{\frac{c}{b + c}} e^{-i \rho} \right]}{2 \sqrt{3}} \right]. \nonumber \\
\end{eqnarray}

\textbf{Parametrization Conditions}
\begin{eqnarray}
\textbf{(A).} && \frac{1}{6} \left( -3 \sqrt{\frac{c}{b+c}} \cos(\eta_4 + \eta_1) + \sqrt{3} \left( \sqrt{\frac{c}{b+c}} \sin(\eta_4 + \eta_1) - 2 \sqrt{\frac{b}{b+c}} \sin(\eta_4 + \rho + \eta_1) \right) \right) = 0, \nonumber \\
\textbf{(B).} && \frac{1}{6} \left( 3 \cos(\eta_5 + \eta_1) + \sqrt{3} \sin(\eta_5 + \eta_1) \right) = 0, \nonumber \\
\textbf{(C).} && \frac{1}{6} \left( 3 \sqrt{\frac{b}{b+c}} \cos(\eta_2) + \sqrt{3} \left( \sqrt{\frac{b}{b+c}} \sin(\eta_2) + 2 \sqrt{\frac{c}{b+c}} \sin(\rho + \eta_2) \right) \right) = 0, \nonumber \\
\textbf{(D).} && \frac{-\sqrt{\frac{b}{b+c}} \sin(\eta_3) + \sqrt{\frac{c}{b+c}} \sin(\rho + \eta_3)}{\sqrt{3}} = 0, \nonumber \\
\textbf{(E).} && \frac{1}{6 \sqrt{3}} \left( 3 \sqrt{\frac{c}{b+c}} \cos(\eta_4 + \eta_2) + 3 \cos(\eta_5 + \eta_2) \sqrt{\frac{b+c - \sqrt{b c} \cos(\rho) + \sqrt{3} \sqrt{b c} \sin(\rho)}{b+c}} \right. \nonumber \\
& & \left. + \sqrt{3} \left( \sqrt{\frac{c}{b+c}} \sin(\eta_4 + \eta_2) - \sqrt{\frac{b+c - \sqrt{b c} \cos(\rho) + \sqrt{3} \sqrt{b c} \sin(\rho)}{b+c}} \sin(\eta_5 + \eta_2) \right. \right. \nonumber \\
& & \left. \left. - 2 \sqrt{\frac{b}{b+c}} \sin(\eta_4 + \rho + \eta_2) \right) \right) = 0.
\end{eqnarray}

\bibliography{paper2.bib}

\end{document}